\documentclass [10pt,twocolumn]{IEEEtran}
\usepackage{algorithm,algorithmic,amsbsy,amsmath,amssymb,epsfig,bbm,mathrsfs,multirow,amsthm,bbm}
\usepackage[T1]{fontenc}
\usepackage[latin9]{inputenc}
\usepackage{amsmath}
\usepackage{fixltx2e}
\usepackage{algorithm}
\usepackage{array,multirow,graphicx}
\usepackage{color}
\usepackage{cite}
\usepackage{float}
\usepackage{makecell}
\usepackage[x11names,dvipsnames,table]{xcolor} 
\usepackage{colortbl}
\usepackage{multirow}
\usepackage{subcaption} 
\usepackage{lipsum}
\usepackage{url}

\definecolor{mygray}{gray}{0.6}
\definecolor{myblue}{rgb}{0.8,0.85,1} 
\usepackage{array}
\newcolumntype{L}[1]{>{\raggedright\let\newline\\\arraybackslash\hspace{0pt}}m{#1}}
\newcolumntype{C}[1]{>{\centering\let\newline\\\arraybackslash\hspace{0pt}}m{#1}}
\newcolumntype{R}[1]{>{\raggedleft\let\newline\\\arraybackslash\hspace{0pt}}m{#1}}

\usepackage{array, tabularx, boldline}
\usepackage{eurosym}
\usepackage{amstext} 
\DeclareRobustCommand{\officialeuro}{%
  \ifmmode\expandafter\text\fi
  {\fontencoding{U}\fontfamily{eurosym}\selectfont e}}

\usepackage{array, tabularx, boldline}
\usepackage{graphicx}
\usepackage{cellspace}
\DeclareMathOperator{\E}{\mathbb{E}}

\begin{document}
\title{Radio Resource Management in Joint Radar and Communication: A Comprehensive Survey}
\author{
}

\author{Nguyen Cong Luong, Xiao Lu, \textit{Member, IEEE}, Dinh Thai Hoang, \textit{Member, IEEE}, Dusit Niyato, \textit{Fellow, IEEE}, and Dong In Kim, \textit{Fellow, IEEE}
\thanks{N.~C.~Luong  is  with  Faculty  of  Information  Technology,  PHENIKAA University, Hanoi 12116, Vietnam, and is with PHENIKAA Research and Technology Institute (PRATI), A\&A Green Phoenix Group JSC,  No.167 Hoang Ngan, Trung Hoa, Cau Giay, Hanoi 11313, Vietnam. Email: luong.nguyencong@phenikaa-uni.edu.vn.}
\thanks{X.~Lu is with the Department of Electrical and Computer Engineering, University of Alberta, Canada. Email: lu9@ualberta.ca.}
\thanks{D.~T.~Hoang is with the Faculty of Engineering and Information Technology, University of Technology Sydney, Australia. Email: hoang.dinh@uts.edu.au.}
\thanks{D.~Niyato is with School of Computer Science and Engineering, Nanyang Technological University, Singapore. E-mail: dniyato@ntu.edu.sg.}
\thanks{D.~I.~Kim is with the Department of Electrical and Computer Engineering,Sungkyunkwan University, South Korea. Email: dikim@skku.ac.kr.}
}

\maketitle
\begin{abstract}

Joint radar and communication (JRC) has recently attracted substantial attention. The first reason is that JRC allows individual radar and communication systems to share spectrum bands and thus improves the spectrum utilization. The second reason is that JRC enables a single hardware platform, e.g., an autonomous vehicle or a UAV, to simultaneously perform the communication function and the radar function. As a result, JRC is able to improve the efficiency of resources, i.e., spectrum and energy, reduce the system size, and minimize the system cost. However, there are several challenges to be solved for the JRC design. In particular, sharing the spectrum imposes the interference caused by the systems, and sharing the hardware platform and energy resource complicates the design of the JRC transmitter and compromises the performance of each function. To address the challenges, several resource management approaches have been recently proposed, and this paper presents a comprehensive literature review on resource management for JRC. First, we give fundamental concepts of JRC, important performance metrics used in JRC systems, and applications of the JRC systems. Then, we review and analyze resource management approaches, i.e., spectrum sharing, power allocation, and interference management, for JRC. In addition, we present security issues to JRC and provide a discussion of countermeasures to the security issues. Finally, we highlight important challenges in the JRC design and discuss future research directions related to JRC.

\end{abstract}

\section{Introduction}
\label{sec:intro}

Frequency spectrum is becoming increasingly congested due to the rapid growth of wireless devices and mobile services. As a result, the price of the available wireless spectrum has experienced a sharp rise during recent years~\cite{liu2020joint}. As reported in \cite{Sean_5G_auction}, telecommunications companies in South Korea paid a total of \$3.3 billion for two bands, i.e., $3.5$ GHz and $28$ GHz, of 5G network. In the UK, as reported in \cite{auction_result}, mobile network operators were required to pay the total of $\pounds1.3$ billion for the $2.5$ GHz band (used for 4G network) and $3.4$ GHz (used for 5G network). The number of active IoT devices is expected to reach $24.1$ billion in 2030 that requires extra spectral resources~\cite{IoT_growth}. As a consequence, the mobile network operators need to seek opportunities to reuse or share spectrum of other applications and systems~\cite{liu2020joint}. Otherwise, as presented in~\cite{griffiths2014radar}, radar systems have the huge chunks of spectrum available at radar frequencies, i.e., ranging from $3-30$ MHz band to $110-300$ GHz  band. This enables the spectrum sharing between the radar systems and communication systems that leads to a convergence trend of the radar and the communication, namely \textit{joint radar and communication (JRC)}~\cite{feng2020joint}, \cite{chiriyath2017radar}.~In general, there are two main categories of JRC~\cite{liu2020joint}: coexisting radar and communication (CRC) and dual function radar-communication (DFRC). In particular, CRC allows individual radar and communication systems to share the spectrum. CRC can be found in several existing scenarios such as sharing spectrum between the airborne early warning radar systems and TDD-LTE system, spectrum between vessel traffic service radars and WLAN networks, e.g., using IEEE 802.11a/h/j standards. On the contrary to CRC, DFRC enables a single hardware platform, e.g., an autonomous vehicle or a UAV, to simultaneously perform the communication function and the radar function. A prototype of DFRC based on the IEEE 802.11a/g for autonomous vehicular forward collision detection has been implemented in~\cite{daniels2017forward}. With the frequency of $4.89$ GHz, the bandwidth of $20$ MHz, and the transmit power gain of $10$ dBm, the radar function of DFRC can provide a meter-level accuracy for single-target detection up to $30$ m. In addition, the communication function of DFRC is directly integrable into the dedicated short-range communication (DSRC) protocol (IEEE 802.11p) that enables the vehicle to implement the vehicle-to-vehicle and vehicle-to-roadside communications.


As such, JRC is able to improve the efficiency of resources, i.e., spectrum and energy, reduce the system size, and minimize the system cost. These benefits enable JRC as a promising technology for several military applications. In particular, JRC can be used in shipborne systems to perform a radar scanning activity while transmitting data to its allies in the sky. Also, JRC has been used in airborne systems to enhance electronic warfare by communicating and detecting objects at long distances simultaneously. Moreover, JRC has been used to allow ground-based systems such as tanks and reconnaissance vehicles to efficiently communicate and detect objects at short distances, e.g., battlefields. Apart from the military applications, JRC is a promising technology for civilian applications supporting logistics automation markets such as autonomous vehicle systems and flying wireless mesh networks. The market is expected to reach~\$$81$ and \$$290$ billion in 2030 and 2040, respectively~\cite{logistic}. Further discussions of important applications of JRC are provided in Section~~\ref{apps_JRC}.

\subsection{Challenges of Designing JRC Systems}

However, there are several challenges to be solved for the JRC design. First, JRC such as CRC enables the spectrum sharing between the radar system and the communication system. This imposes the interference caused by both the systems that can significantly degrade the performance. For example, as the Aegis combat system, i.e., an American integrated naval weapons system, shares the S-band, i.e., the $3.5$ GHz band, with a cellular system including $100$ base stations, the miss detection probability of the system can be up to $95$\%~\cite{khawar2014mathematical}. This raises the interference management and power allocation issues for JRC. Second, JRC such as DFRC allows the communication function and radar function to share a single hardware platform, spectrum and energy resources. This complicates the design of the JRC transmitter and compromises the performance of each function. The design of the JRC becomes more challenging as DFRC is applied to autonomous vehicles in which the JRC systems are densely deployed in urban environments and the radar functions need to detect multiple objects in short ranges on the order of a few tens of meters. To address the issues, several resource management approaches including spectrum sharing with waveform design, time sharing, spatial beamforming, and power allocation have been recently proposed for JRC.

\subsection{Contributions of the Paper}

There are several surveys and tutorials on JRC that are given in~\cite{feng2020joint},~\cite{Labib2017},~\cite{ma2019joint}, and~ \cite{gameiro2018research}. In particular, the authors in~\cite{feng2020joint} and \cite{Labib2017} highlight the applications of JRC and review the state-of-the-art for JRC systems. The authors in \cite{ma2019joint} provide an overview of DFRC used particularly for autonomous vehicles. This work can be considered to be a good tutorial that provides basic concepts of DFRC and explains spectrum sharing strategies for DFRC. The authors in \cite{gameiro2018research} discuss research challenges, trends, and applications of JRC. The existing surveys/tutorials are generally covering all issues in JRC. However, the existing surveys and tutorials have the following limitations:
\begin{itemize}
\item Basic concepts and important performance metrics related to resource management in JRC systems are not sufficiently provided. 
\item Resource management issues such as power allocation, security issues, and countermeasures for JRC are not well investigated and discussed.
\item Many state-of-the-art technologies for both DFRC and CRC are not thoughtfully updated and reviewed.  
\item Many emerging research topics as well as new issues introduced recently are not comprehensively discussed.  
\end{itemize}

This motivates us to have a comprehensive survey on JRC. In particular, our survey pays special attention to ``resource management'' for JRC. The survey has the following contributions:
\begin{itemize}
\item We provide fundamental knowledge of JRC that elaborates basic concepts of JRC and important performance metrics used for resource management. In addition, we discuss application scenarios of JRC in practice. 
\item We review and discuss a number of spectrum sharing approaches for JRC. The spectrum sharing approaches are based on communication signal, radar signal, time sharing, and antenna allocation. We furthermore analyze and compare the advantages and disadvantages of the approaches.  
\item We review, discuss, and analyze power allocation and interference management approaches for JRC.  
\item We present security issues to JRC and provide a discussion of countermeasures to the security issues.
\item We highlight challenges and discuss potential research directions related to JRC.
\end{itemize}

A comparison of contribution between our work and the relevant surveys are shown in Table~\ref{tab:table_comparison}.

\begin{table}[h!]
\scriptsize
  \caption{\small {A comparison of contribution between relevant surveys and our survey}}
  \label{tab:table_comparison}\centering
  \begin{tabularx}{8.7cm}{|Sl|X|}
    \hline
  \cellcolor{mygray} \textbf{Refs.} &   \cellcolor{mygray} \textbf{Contribution}\\   
            \hline
\cite{feng2020joint}, \cite{Labib2017}  & Survey on applications of JRC\\
  \hline
  \cite{ma2019joint} &   Tutorial on DFRC systems particularly for autonomous vehicle systems and spectrum sharing strategies for DFRC\\
  \hline
  \cite{gameiro2018research}  &  Discussions of  research challenges, trends, and applications of JRC.\\
  \hline
   \textbf{Our work} &Tutorial on JRC and comprehensive survey of resource management issues in JRC including spectrum sharing, power allocation, and interference management in JRC, discussion on security issues in JRC systems, discussions on challenges and potential research directions related to JRC.\\
  \hline
  \end{tabularx}
\end{table}

\begin{figure*}[t!]
\centering
\includegraphics[scale =0.5]{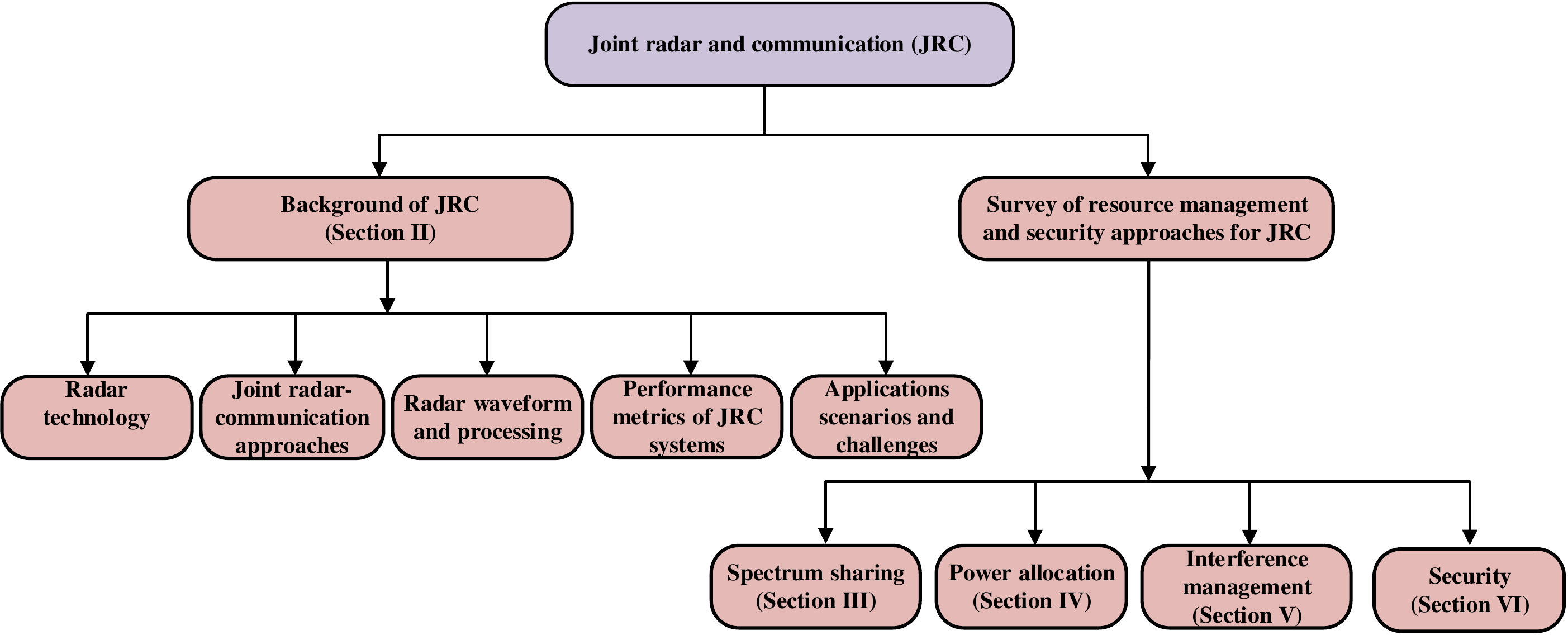}
 \caption{\small Classification of related studies to be discussed in this survey.}
 \label{classification}
\end{figure*}

For the reader's convenience, we classify the related studies according to the resource management issues, i.e., spectrum sharing, power allocation, and interference management, that is shown in Fig.~\ref{classification}. As such, the readers who are interested in or working on the related issues will benefit greatly from our insightful reviews and indepth discussions of existing approaches, remaining/open issues, and potential solutions. For this, the rest of this paper is organized as follows. Section~\ref{sec:radar-commu} introduces the basic concepts of JRC as well as important performance metrics and applications of JRC. Section~\ref{sec:radar-commu-mul} reviews the spectrum sharing approaches for JRC. Section~\ref{power_all_sec} discusses power allocation approaches for JRC. Section~ \ref{interference_man} presents interference cancellation approaches for JRC.~Section~\ref{sec:security} presents security issues to JRC and discusses the countermeasures. Section~\ref{sec:challenges_open_issues} highlights important challenges and potential research directions. Section~\ref{sec:conclusions} concludes the paper. The list of abbreviations commonly appeared in this paper is given in Table~\ref{tab:table_abb}.

\begin{table}[h!]
\scriptsize
  \caption{\small List of common abbreviations used in this paper}
  \label{tab:table_abb}\centering
  \begin{tabularx}{8.7cm}{|Sl|X|}
    \hline
  \cellcolor{mygray} \textbf{Abbreviation} &   \cellcolor{mygray} \textbf{Description} \\   
            \hline
 CPM/CRB&Continuous Phase Modulation/Crame\'r-Rao bound\\
  \hline
CSI& Channel State Information\\
   \hline
   CRC& Coexisting Radar and Communication\\
      \hline
DFT&Discrete Fourier Transform\\
  \hline
DFRC &Dual Function Radar-Communication\\
  \hline
FFT/IFFT&Fast Fourier Transform/Inverse Fast Fourier Transform\\
  \hline
  FH/LFM&Frequency Hopping/Linear Frequency Modulation\\
   \hline
FMCW&Frequency Modulated Continuous Wave\\
          \hline
JRC&Joint Radar and Communication\\
 \hline
PRI& Pulse Repetition Interval\\
  \hline
PAPR&Peak-to-Average Power Ratio\\
  \hline
OFDM&Orthogonal Frequency-Division Multiplexing\\
 \hline
MI& Mutual Information\\
 \hline
RCS& Radar Cross Section\\
\hline
SINR& Signal-to-Interference-Plus-Noise Ratio\\
  \hline
  \end{tabularx}
\end{table}

\section{Fundamental Background of Joint Radar and Communication Systems} 
\label{sec:radar-commu}

In this section, we present the fundamental background of JRC. To understand JRC, the basic knowledge of radar technology is necessary. Thus, we first provide some fundamental background of radar technologies. We then present and discuss the approaches for the integration of radar technologies into conventional data communication systems, i.e., the JRC systems. After that, we introduce and discuss performance metrics and applications of JRC systems.

\subsection{Radar Technology}

Radar (acronym for RAdio Detection And Ranging) is an electrical detection system that uses radio waves to determine target objects. In practice, radar systems have a long history and they have been widely implemented on many military and civilian applications~\cite{SkolnikRADAR2001}. In this section, we will study some basic concepts, architecture and typical parameters used in radar systems.

\subsubsection{Basic Concepts and Applications}

Radar (acronym for RAdio Detection And Ranging) is an electrical detection system that uses radio waves to determine target objects. The basic operation principle of a radar system is transmitting the radio waves to the air and then observing the received signals (reflected from the target objects) to determine characteristics of the  objects such as distance, directions, velocities, shapes and even materials~\cite{SkolnikRADAR2001}. The radar systems can hence find a number of applications in both military missions (e.g., to detect aircraft, ships, spy  unmanned aerial vehicle (UAV), and spacecraft) and civilian (e.g., robots, autonomous vehicles, and terrain exploration).

\subsubsection{Architecture and Main Components}

\begin{figure}[!]
	\centering
	\includegraphics[width=\linewidth]{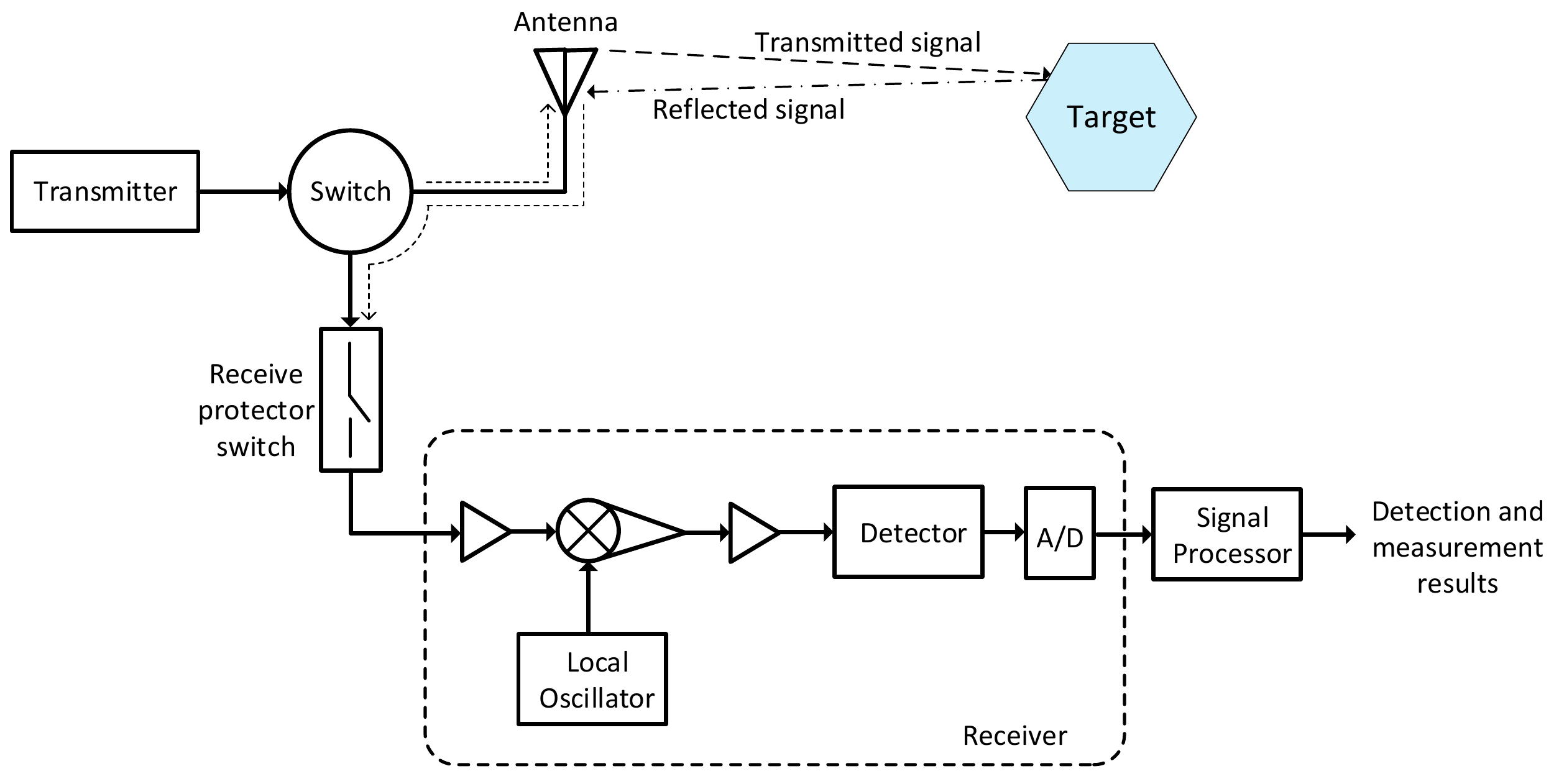}
	\caption{Main components of a radar system.}
	\label{fig_Radar}
\end{figure}

Figure~\ref{fig_Radar} illustrates the main components of a typical radar system which consist of (1) Transmitter, (2) Receiver, (3) Switch and Antenna and (4) Controller. Both transmitter and receiver components are connected to the switch and all of them are controlled by the controller as illustrated in Fig.~\ref{fig_Radar}. The main processes can be expressed as follows:
\begin{itemize}
	\item First, the transmitter generates the radio frequency (RF) signals and sends these signals (typically direct to a target) out through the antenna. The transmitted signals in the form of electromagnetic (EM) waves will be then propagated to the target objects through the environment, e.g., through the air. 
	\item Then, when the EM waves hit the target objects' surfaces, they will be reflected or scattered to the surrounding environment. 
	\item After that, if the radar system can receive the reflected signals (scattered signals or echo signals), the radar receiver will process and analyze these signals to determine the properties of the target objects. 
\end{itemize}
To avoid interference between the transmitted and reflected signals, the switch will be used. Note that radar receivers are usually, but not always, located at the same device with the transmitters. Thus, in the case if the transmitters and receivers are separated in different devices, the switch is not required.  

When the received signals are passed to the receiver, they first go through a low-noise amplifier in order to amplify a very low-power signal without significantly decreasing the received signal-to-noise (SNR) ratio. The amplified signals will be then shifted to an intermediate frequency by the intermediate frequency (IF) amplifier with the aim of extracting signals that have frequencies close to that of the transmitted signals. After that the detector device will be used to extract information from the modulated received signal. Note that in conventional radar systems, the detectors are often combined with analog-to-digital converter (ADC) and signal processors to create favorable conditions for analyzing and presenting the results. However, these components (i.e., ADC and signal processor) are not compulsory to implement on the radar systems.

\subsubsection{Target Identification and Radar Range}

\begin{figure}[!]
	\centering
	\includegraphics[scale=0.6]{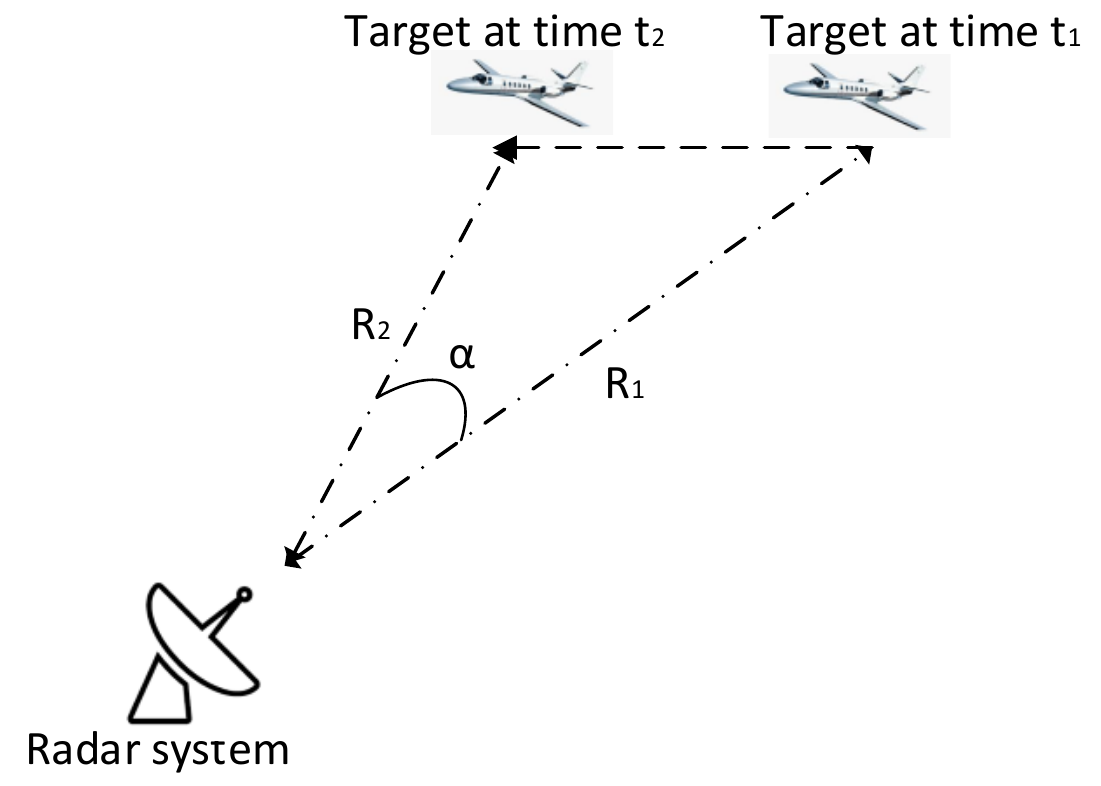}
	\caption{Target velocity.}
	\label{fig_Radar_velocity}
\end{figure}

\paragraph{Radar parameter} One of the most important goals of a radar system is to detect and determine the distance $R$ between the target object and the system. To calculate the distance $R$, we can measure the round-trip travel time of transmitted signals $\Delta T$ (i.e., the time from the signals transmitted from the system to the time that the system receives the reflected signals) and use the following equation~\cite{SkolnikRADAR2001}: 
\begin{equation}
\label{eq_Ra_Dis}
R = \frac{c \Delta T}{2} ,
\end{equation}
where $c$ is the speed of light ($c \approx 3 \times 10^8$ m/s). This equation implies that the distance between the radar system and the target object is directly proportional to the travel time of received signals. This principle is used in almost all radar systems to determine the distance of the target objects. Alternatively, this equation can be also used to determine the velocity of the target object. Specifically, if we denote $R_1$ and $R_2$ respectively are the distance from the target to the system at time $t_1$ and $t_2$, the target velocity can be estimated as follows:
\begin{equation}
v = \frac{\sqrt{R^2_1+R^2_2 - 2 R_1 R_2 \cos \alpha}}{|t_1-t_2|},
\end{equation}
where $\alpha$ is the angle between the radar antenna and the target at two different time $t_1$ and $t_2$ (as illustrated in Fig.~\ref{fig_Radar_velocity}) and $|.|$ is the absolute value function.

\paragraph{Radar range} Another equation that is also very important for radar systems to determine characteristics of target objects is radar range equation which can be expressed as follows~\cite{SkolnikRADAR2001}:  
\begin{equation}
\label{eq_Ra_ra}
P_r = \frac{P_t G_t G_r \lambda^2 \sigma F^4}{(4\pi)^3 R^4} ,
\end{equation}
where $P_t$ and $P_r$ are signal transmission power from the radar system and received signal power (reflected from the target object) at the system, respectively. $G_t$ and $G_r$ are the transmitting and receiving antenna gains of the radar system, respectively.\footnote{In cases if the radar system uses only one antenna for both transmitting and receiving signal as illustrated in Fig.~\ref{fig_Radar}, then $G_t=G_r$.} $\sigma$ is the scattering coefficient of the target object and $F$ is the pattern propagation factor. $\lambda$ is the wavelength of carrier frequency. $R$ is the distance between the target object and the system which can be calculated from~(\ref{eq_Ra_Dis}). From~(\ref{eq_Ra_ra}), by observing the received signal power at the radar system, we can infer some characteristics of the target object. For example, given the scattering coefficient, we can infer some features of target object (e.g., material and shapes). These information is especially important for military applications.

\paragraph{Range resolution} This is also an important metric used in radar systems to evaluate the system performance, especially related to radar operation efficiency. This metric is to show the ability of a radar system to differentiate between two or more targets that are very close in either range or bearing. There are three main factors impacting to the range resolution at different levels, i.e., the efficiency of the receiver and indicator, the types and sizes of targets and the width of the transmitted pulse. For a high accuracy system, it should be able to differentiate between the targets separated by one-half the pulse width time $\tau$, and thus its range resolution can be theoretically calculated by~\cite{Radar_Range_Resolution}:
\begin{equation}
\Delta R = \frac{c \tau}{2}, 
\label{eq_2R}
\end{equation}
where $\Delta R$ is the range resolution as a distance between the two targets in unit of meters. Here, we would like to note that for pulse compression systems, the range resolution will be calculated based on the bandwidth $B_p$ of the transmitted pulse (not by the pulse width as in~(\ref{eq_2R})) as follows:
\begin{equation}
\Delta R = \frac{c}{2 B_p}. 
\label{eq_3R}
\end{equation}

From~(\ref{eq_3R}), we can observe that the higher bandwidth of the transmitter pulse is, the better range resolution we can obtain.

\paragraph{Velocity resolution} Velocity resource is another parameter of a radar system, which defines the minimal radial velocity between two objects moving at the same range in order to make the radar system can differentiate discrete signals reflected from them. Given a time duration $T_c$ of the chirp, the velocity resource $\Delta V$ can be determined as follows~\cite{Campbell2012Surface}:
\begin{equation}
\Delta V = \frac{c}{2 f_c T_c},
\label{eq_RVR}
\end{equation}
where $f_c$ is the carrier center frequency. Here, we can observe that in order to enhance the velocity resolution for the radar system (i.e., make this parameter as small as possible), we can increase the chirp dispersion $T_c$.

\paragraph{Ambiguity function} An ambiguity function is a two-dimensional function of propagation time delay $\beta$ and Doppler frequency $f_D$ used to express the distortion of received signals due to the Doppler effect. Thus, the ambiguity function describes the propagation delay and Doppler relationship of the signals, and it can be defined as follows~\cite{Weiss2019mmwave}:
\begin{equation}
\mathcal{X}(\beta,f_D) = \int_{-\infty}^{+\infty}s(t)s^*(t-\beta) e^{i2\pi r f_D} dt,
\label{ambiguity_function}
\end{equation}
where $s(t)$ is a given complex baseband pulse signal and $s^*(t)$ is its complex conjugate with imaginary unit $i$. In radar systems, the ambiguity function is very useful for analyzing and designing the radar waveform due to its properties such as constant energy and symmetry with respect to the origin.

\subsection{Joint Radar-Communication Approaches}

We present three main approaches for integration between radar and communication systems. The key technologies together with advantages and disadvantages of each approach are then discussed.

\subsubsection{Frequency-division based Approaches}

This is the most simple approach for JRC systems. Specifically, to use both radar and communication functions, at the same time, they will be allocated to operate at separated antennas and transmitted at different frequencies. In this way, both functions can work fully independently and be easy to integrate into any existing systems. However, this approach requires the system to be equipped with separated antennas and frequencies which may not be cost-effective in implementing in civilian applications due to limited spectrum availability.

\subsubsection{Time-division based Approaches}

This is also a simple solution for combining both functions, i.e., radar and communication, into one system. The key idea of this approach is using a switch to choose and control the operation of these two functions. In particular, for this approach, the switch will take responsibility to control operations for communication and radar functions separately. For example, if the system needs to detect a target object, the radar function will be activated. Otherwise, if the system wants to transmit data, the communication function will be used. It is important to note that for this approach, even if we have two separated antennas to serve for these two functions, the functions should not work concurrently due to severe interference if they transmit at the same frequency. 

One of the biggest advantages of this approach is simplicity and ease of implementation. Both functions can be efficiently deployed and integrated into any system just by using a simple switch and without requiring re-designing radar and communication waveforms. However, this approach also has a few disadvantages. First, only one function can operate at a time, and thus the most important issue is to determine the appropriate working time for these functions to be activated. Some research works~\cite{chiriyath2017radar} propose solutions to fairly allocate working time for both functions (e.g., they work in a round-robin fashion). However, these solutions are not appropriate to implement on real-time systems when demands on radar and data communications are dynamic and uncertain. Deep reinforcement learning has been recently introduced to quickly find optimal decisions in a real-time manner~\cite{HieuiRDRCIEEE}, but its performance is much dependent on the accuracy of sensors, e.g., road friction sensor, weather station instrument and speedometer. More related research works are reviewed later in Section~\ref{sec:radar-commu-mul}.


\subsubsection{Dual function
		radar communication systems}

This is the most popular approach used in JRC systems, especially in autonomous systems mainly due to its outstanding features, e.g., low-cost and spectrum usage optimization. The core idea of this approach is integrating both functions on the same signals to transmit. To do so, there are two solutions, i.e., communication waveform-based and radar waveform-based solutions. The first solution (i.e., communication waveform) is based on the idea of embedding radar signals on the data communication waveform signals, while the second solution (i.e., radar waveform-based) is to embed data on the radar waveform signals.

\paragraph{Radar waveform-based}

The first approach to integrate the data communication functions into an existing radar system is to modify the radar waveform such that it can include digitally modulated data symbols.

In particular, for a conventional frequency-modulated continuous-wave waveform radar (FMCW) system, the transmitter can periodically transmit $M$ FMCW pulses with durations $T_p$ as follows~\cite{Ma2020Joint}: 
\begin{equation}
s_m (t) = e^{j2\pi f_c t + j \pi \gamma t^2} , \forall t \in [m T_{\text{PRI}}, m T_{\text{PRI}} + T_p], 
\label{eq_RWB}
\end{equation}
where $T_{\text{PRI}}$ is the pulse repetition interval and it is rather larger than $T_p$. In addition, in~(\ref{eq_RWB}), $f_c$ and $\gamma$, respectively, express the carrier frequency and the frequency modulation rate. Now, if we want to embed the data symbols on the radar signals, we can replace the $m$-th pulse $s_m (t)$ in~(\ref{eq_RWB}) by $s_m (t) e^{j \phi_m}$ where $\phi_m$ encapsulates the information message in the form of continuous phase modulation as introduced in~\cite{Roberton2003Integrated} or differential QPSK modulation as presented in~\cite{Sahin2017Anovel}. 

Another well-known method which also allows to transmit data based on the radar signals is based on the frequency modulation scheme~\cite{Saddik2007Ultra}. Specifically, the transmitter can use a positive frequency modulation rate $\gamma$ to transmit bits ``1'', while negative values can be used to transmit bits ``0''. Although the principle of this method is pretty simple and easy to decode information at the receiver, its communication rate is very low and its performance much depends on the pulse repetition interval (PRI).

\paragraph{Communication waveform-based} The idea of this solution is integrating radar function into the current conventional data communication waveform signals. The most effective communications waveform technique is using the OFDM signaling. The main reason is that OFDM is commonly used in both radar and data communication systems, and thus it can be more flexible and adaptable for the combination of both functions. In particular, in~(\ref{eq_OFDM}), we show the transmitted signal for an OFDM waveform radar in which $\{x_{m,n}\}$ are complex weights of the transmitted signals. Thus, if the data is embedded to the radar signals to transmit, we can replace complex weights $\{x_{m,n}\}$ by data symbols (e.g., bits ``1'' and ``0''). For the dual-function OFDM signals, the transmit information can cause a high degree of sidelobes after matched filtering, and thus we can allocate each transmitted symbol to a separated subscarrier to avoid this problem~\cite{sturm2011waveform}.

Table~\ref{tab_comparisons} provides some comparisons in terms of the main features, advantages and disadvantages of the aforementioned approaches.

\begin{table*}[!h]
	\centering
	\caption{JRC Approaches Comparisons} 
	\centering 
	\begin{tabular}{|c|c|c|c|}
		\hline
		\textbf{Approaches} & Frequency-sharing & Time-sharing & Signal-sharing \\
		\hline
		\hline
		\textbf{Features} & Two functions operate concurrently in  & Two functions operate sequentially  & Two functions operate concurrently \\
		& two separated frequency ranges & based on a switch & in the same frequency range \\
		\hline
		\textbf{Advantages} & Easy to implement and integrate & Easy to implement and integrate & High performance \\
		\hline
		\textbf{Disadvantages} & Low performance due to & Low performance due to & High complexity for designing \\	
		& sharing frequency & sharing operation time & hardware components \\	
		\hline				
	\end{tabular}
	\label{tab_comparisons}
\end{table*}

\subsection{Radar Waveform and Processing}

There are two typical waveforms used in JRC systems, i.e., Frequency-Modulated Continuous Wave (FMCW) and Orthogonal Frequency-Division Multiplexing (OFDM).

\paragraph{FMCW radar waveform}
\label{FMCW_tutorial}

\begin{figure}[!]
	\centering
	\includegraphics[width=\linewidth]{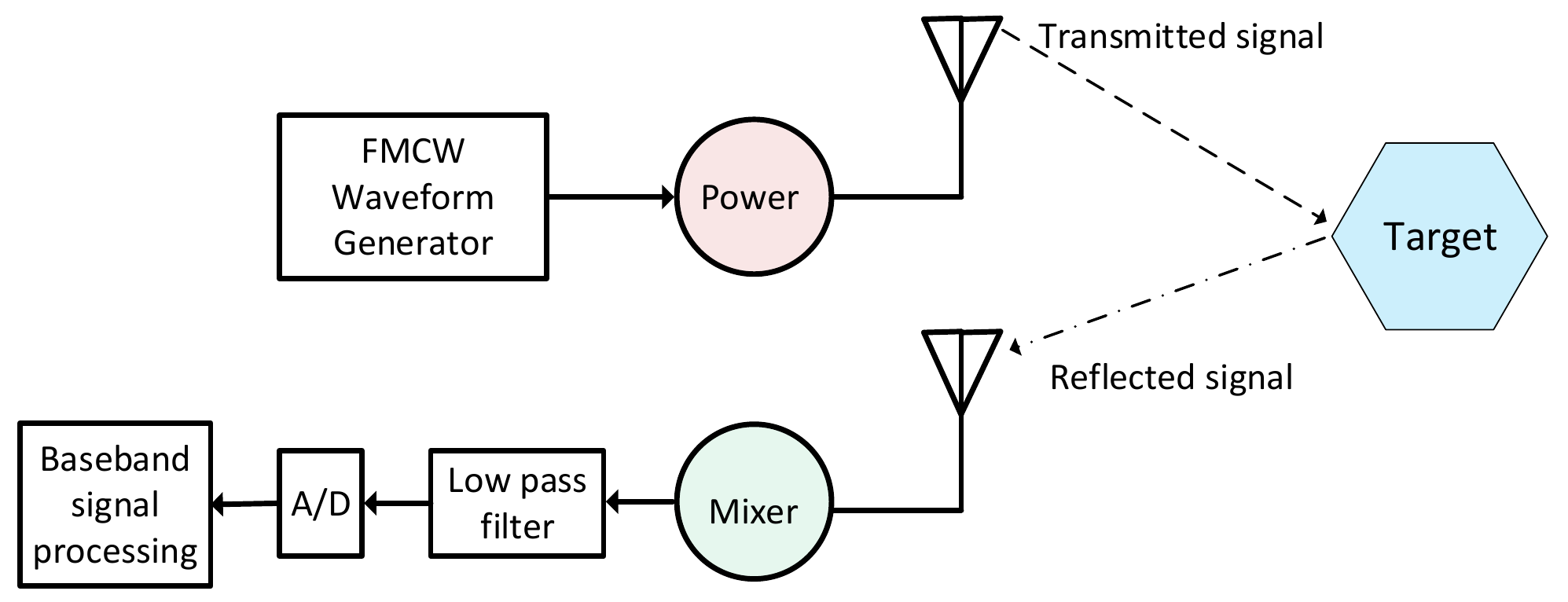}
	\caption{Block diagram of an FMCW system.}
	\label{fig_FMCW}
\end{figure}

FMCW is a type of linear frequency modulation (LFM) or chirp modulation in which the frequency increases or decreases with a so-called \textit{chirp rate}~\cite{wang2018joint}. Fig.~\ref{fig_FMCW} shows a general FMCW system with main components including a receiver, a transmitter, a analog-to-digital converter (ADC) and a mixer. In this figure, we can see that the modulated signal first is generated by an FMCW waveform generator before sending out through the transmitter's antenna. The received signal at the receiver's antenna will be then multiplied with the transmitted signal in the time domain before sending to the low pass filter (LPF) for further processing.

According to~\cite{WinklerRange2007}, the transmitted signal of an FMCW radar system can be modeled as:
\begin{equation}
s_{\text{T}}(t) = A_{\text{T}} \cos \Big( 2 \pi f_c t + 2 \pi \int_{0}^{t} f_{\text{T}} (\tau) d\tau \Big) ,
\end{equation}
where $A_{\text{T}}$ represents the transmitted signal amplitude. Moreover, $f_{\text{T}} (\tau)$ and $f_c$, respectively, express the transmit frequency and the carrier frequency of the system.

Let $f_D$ denote the Doppler frequency. The time delay and the receiving frequency can be expressed, respectively, as follows:
\begin{equation}
t_{\text{d}} = 2 \frac{R_0+vt}{c} \phantom{10} \text{and} \phantom{10} f_{\text{R}}(t) = \frac{B}{T} (t- t_{\text{d}}) + f_D ,
\end{equation}
where $R_0$ is the range at $t=0$, $B$ is the bandwidth, $v$ is the target velocity, $c$ is the speed of light and $T$ is the time duration. The received signal can be described as:
\begin{equation}
\begin{aligned}
& s_{\text{R}} (t)  = A_{\text{R}} \cos \Big( 2 \pi f_c (t-t_{\text{d}}) + 2 \pi \int_{0}^{t} f_{\text{R}}(\tau) d\tau		\Big)  \\
& = A_{\text{R}} \cos \Big\{ 2 \pi \Big( f_c (t - t_{\text{d}}) + \frac{B}{T} \big( \frac{1}{2} t^2 - t_d t	\big) + f_{\text{D}} t \Big) \Big\} ,
\end{aligned}
\end{equation}
where $A_R$ represents the received signal amplitude which depends on some factors such as transmission power, distance between the radar system and the target and antenna gains. 

Then, the transmitted signal and the received signal will be mixed by multiplication in the time domain before sending to the LPF. After that, we can obtain the intermediate frequency (IF) signal $S_{\text{IF}} (t)$ at the output of LPF as follows:
\begin{equation}
s_{\text{IF}} (t) = \frac{1}{2} \cos \Bigg( 2 \pi f_c \frac{2 R_0}{c} + 2 \pi \Big( \frac{2 R_0}{c} \frac{B}{T} + \frac{2 f_c v}{c}  \Big) t	\Bigg) .
\end{equation}

In a similar way, we can obtain the IF signal $S_{\text{IF}} (t)$ at the output of the LPF as follows:
\begin{equation}
s_{\text{IF}} (t) = \frac{1}{2} \cos \Bigg( 2 \pi f_c \frac{2 R_0}{c} + 2 \pi \Big( - \frac{2 R_0}{c} \frac{B}{T} + \frac{2 f_c v}{c}  \Big) t	\Bigg) .
\end{equation}

\begin{figure}[!]
	\centering
	\includegraphics[width=\linewidth]{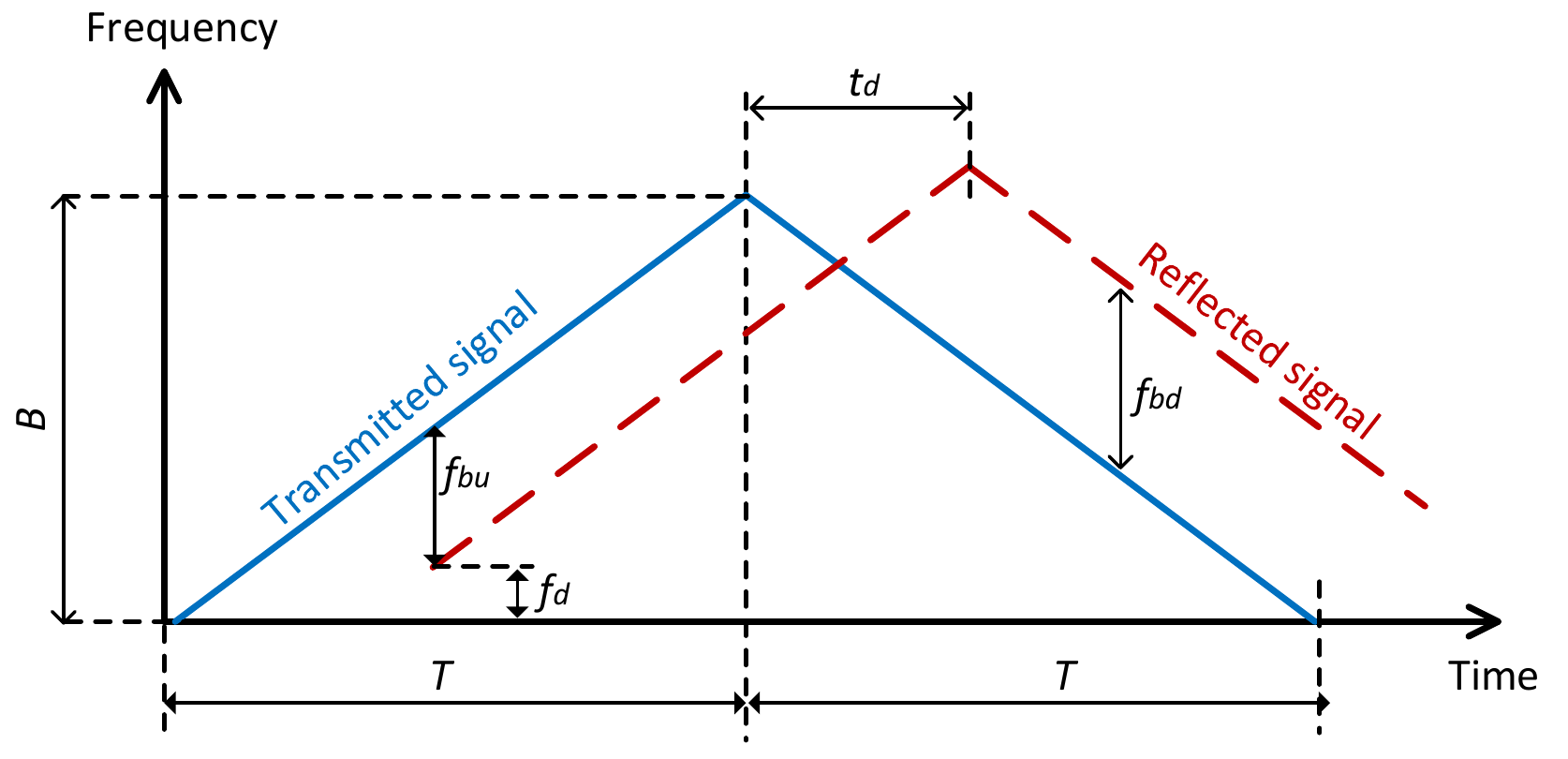}
	\caption{An illustration of FMCW radar system's waveform.}
	\label{fig_FMCW_rec}
\end{figure}

Finally, the up ramp beat frequency $f_{\text{bu}}$ and down ramp beat frequency $f_{\text{bd}}$ (as illustrated in Fig.~\ref{fig_FMCW_rec}) can be obtained in the following way:
\begin{align}
f_{\text{bu}} = \frac{2 R_0}{c} \frac{B}{T} + \frac{2 f_c v}{c} , \\
f_{\text{bd}} = - \frac{2 R_0}{c} \frac{B}{T} + \frac{2 f_c v}{c} .
\end{align}

\emph{
Based on these parameters, we then can derive the $v$ and $R_0$ of the target objective as follows:
\begin{equation}
	v = \frac{(f_{\text{bu}}+f_{\text{bd}}) c}{4 f_c}
\end{equation}
\begin{equation}
	R_0 = \frac{(f_{\text{bu}}-f_{\text{bd}}) cT}{4 B}
\end{equation}
}

\paragraph{OFDM radar waveform}
\label{OFDM_basic}

In this section, we consider a JRC system using standard OFDM modulation with cyclic prefix to to prevent inter-symbol interference when an OFDM signal is transmitted. In the following, we denote $T$ and $T_{\text{cp}}$, respectively, to be the data symbol durations and cyclic prefix. I this way, the OFDM symbol duration $T_{\text{o}}$ can be determined by $T_{\text{o}} = T_{\text{cp}} + T$. If we consider a maximum delay $\tau_{\max}$ for the system, we then can select the cyclic prefix $T_{\text{cp}} = C \frac{T}{M}$. Here, $C$ can be defined by $C= \lceil \frac{\tau_{\max}}{T/M} \rceil$, where $\lceil . \rceil$ is the ceiling function and $M$ is the number of subcarriers in the OFDM sysmbol. As a result, we can derive the OFDM frame duration by $T_f^{\text{OFDM}} = N T_{\text{o}}$, where $N$ is the number of OFDM symbols in the frame. 

According to~\cite{ma2019joint}, we can derive the continuous-time OFDM transmitted signal as follows:
\begin{equation}
\label{eq_OFDM}
s(t) = \sum_{n=0}^{N-1} \sum_{m=0}^{M-1} x_{m,n} \text{rect} (t - n T_{\text{o}}) e^{j 2 \pi m \Delta f(t - T_{\text{cp}} - n T_{\text{o}} ) }	,
\end{equation}
where $\text{rect} (t)$ is a step function that equals one when $t \in [0, T_{\text{o}}]$ and zero otherwise. Then, we can derive the received signal on the time-frequency selective channel as follows:
\begin{equation}
\begin{aligned}
y(t) &= \int_{}^{} h(t,\tau) s(t-\tau) d \tau  \\
&= \sum_{p=0}^{P-1} h_p s(t -\tau_p) e^{j 2 \pi \nu_p t} ,
\end{aligned}
\label{eq_OFDM_tra}
\end{equation}

Note that in~(\ref{eq_OFDM_tra}), we ignore the noise on the channel for presentation convenience and $h(t,\tau)$ is the radar channel model, which can be determined as follows:
\begin{equation}
h(t,\tau) = \sum_{p=0}^{P-1} h_p \Delta (t -\tau_p) e^{j 2 \pi \nu_p t},
\end{equation}
where $h_p$ is the complex channel gain, $\nu_p = \frac{2 v_p f_c}{c}$ and $\tau_p = \frac{2 r_p}{c}$ denotes a round-trip Doppler shift and delay, respectively. To determine the Doppler shift and delay, an ambiguity function can be used.

\begin{figure}[!]
	\centering
	\includegraphics[width=\linewidth]{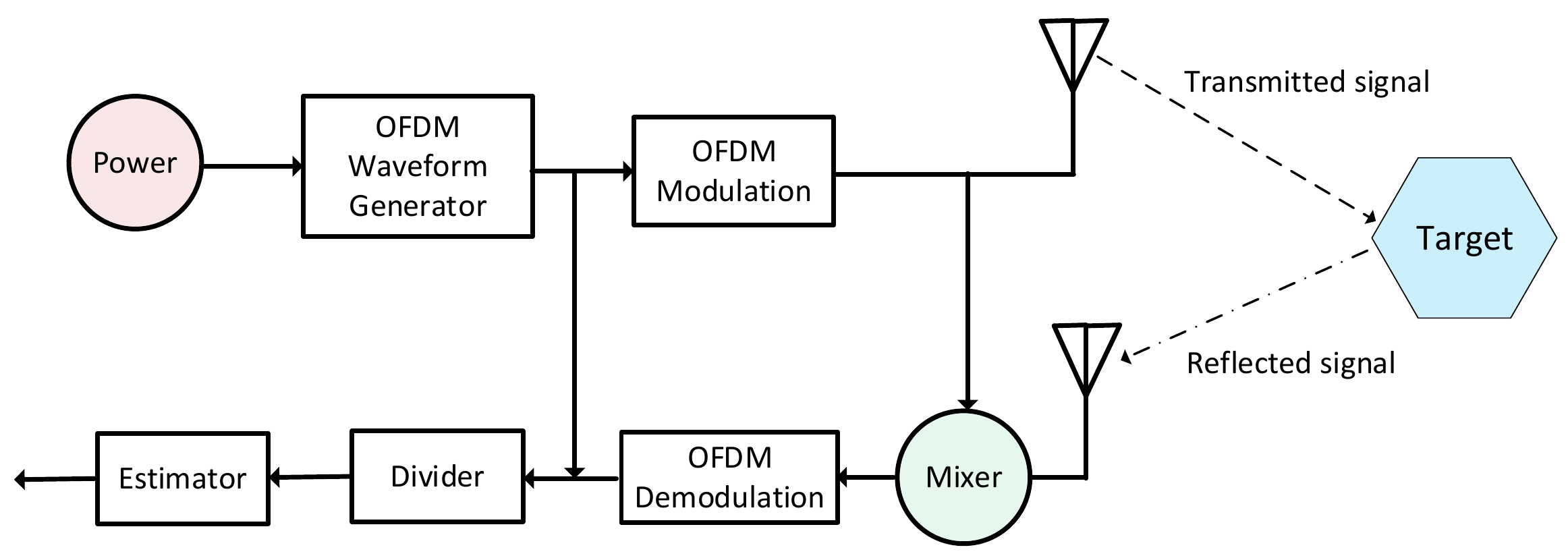}
	\caption{OFDM radar signal modulation.}
	\label{fig_OFDM_rec}
\end{figure}

\subsection{Performance Metrics of JRC Systems}

The main aim of JRC systems is to simultaneously perform both data and radar communication. Thus, there are two key metrics which are usually used to evaluate the performance for these functions in JRC systems, i.e., data communication rate and radar estimation rate .

\subsubsection{Data Communication Rate}

The communication rate indicates the number of data bits that we can transmit to the receiver. Typically, from~\cite{Huang2012Decentralized}, given the transmit power $P^\text{tr}$, the transmission rate can be determined as follows:
\begin{equation}
\label{eq_Shannon}
r_{\text{data}} = \kappa W \log_2 (1 + \frac{P^\text{tr}}{P_0}) ,
\end{equation}
where $P_0$ is the ratio between the noise power $N_0$ and the channel (power) gain efficiency $h$, i.e., $P_0 = \frac{N_0}{h}$. In addition, in~\ref{eq_Shannon}, $W$ and $\kappa \in [0,1]$, respectively, express the channel bandwidth and efficiency of data transmission. From~(\ref{eq_Shannon}), there are few factors that affect performance of data transmission rate.

\begin{itemize}
	\item \textit{Bandwidth} ($W$): This parameter is impacted by the communication approach of JRC system. For example, for time-sharing and signal-sharing approaches, the JRC system can utilize all bandwidth for both radar and data communication functions. However, for the frequency-sharing approach, the JRC system needs to trade-off between radar and data communication activities. Given a fixed amount of allocated bandwidth, the more bandwidth allocated for data communication activities, the less the system has for radar activities and vice versa. 
	\item \textit{Transmit power} ($P^\text{tr}$): This parameter has significant influence to the communication approach of JRC systems. Specifically, for frequency sharing and signal sharing approaches, when the JRC system has to perform both functions concurrently given that it is supplied by only one energy source, it needs to control the power allocation for both activities to achieve performance for both functions as requirements. 
	\item \textit{Transmission efficiency} ($\kappa$): This is an internal parameter which depends on the hardware configuration of the JRC system. 
	\item \textit{Channel condition} ($P_0$): This is an external parameter which depends on the communication environment conditions, e.g., channel gain and noise. 
\end{itemize}

\subsubsection{Radar Performance}

To evaluate the performance of a radar system, we only can rely on the received signals reflected from the objects because these signals can provide useful information such as distance, directions and velocities. In general, the more power the system receives from reflected signals, the higher the accuracy information the system can obtain from the targets, and thus the greater performance the system can achieve. From~(\ref{eq_Ra_ra}), there are some important factors which have significant impacts on the radar performance, and they can be divided into two catalogs, i.e., internal and external factors. External factors are the communications environment and target locations, while internal factors are related to hardware configurations of radar system such as antenna gains and transmit power. In practice, we are unable to control external factors, but we can control some internal factors to improve the radar performance. For example, we can increase the transmit power at the transmitter and/or transmit signals at a low frequency to increase the received signal power. 

\paragraph{Radar rate estimation} For JRC systems, estimating radar signal rates is a challenging task, and this is also an important metric to evaluate the system performance. The estimation rate can be determined by the minimum number of bits that need to be used to encode the Kalman residual~\cite{chiriyath2017radar}. In general, this estimation is a statistical deviation from the radar prediction of a target parameter, for a given channel
degradation. Given a radar channel with the transmitted information $X$ and the addition of some noise $N$, the estimation rate can thus be calculated as follows~\cite{cover1999elements}:
\begin{equation}
R_{\text{e}} = \frac{I(X; X+N)}{T_\text{p}},
\end{equation}
where $I(x;y)$ is the radar estimation information function and $T_\text{p}$ is the pulse repetition interval of the radar system which can be calculated by $T_\text{p} = \frac{T_{\text{pulse}}}{\delta}$. Here, $T_{\text{pulse}}$ is the radar pulse duration and $\delta$ is the radar duty factor.

Then, if the radar estimation error follows the Gaussian distribution with variance $\langle ||r_{\tau,\text{e}}||^2 \rangle = \sigma^2_{\tau,\text{e}}$, we can express $R_{\text{e}}$ in the following way~\cite{cover1999elements}:
\begin{equation}
R_{\text{e}} \leq \frac{1}{2 T_\text{p}} \log_2 \Bigg( 1 + \frac{\sigma^2_{\tau,\text{p}}}{\sigma^2_{\tau,\text{e}}}	\Bigg),
\end{equation}
where $\sigma^2_{\tau,\text{p}}$ is the variance of a process noise. A process noise in a radar system can be expressed as the information extracted from the target based on prior observations.

\subsection{ Potential Application Scenarios and Implementation Challenges of JRC Systems}
\label{apps_JRC}

In this section, we are going to study the applications of JRC systems in practice. In general, we can divide applications into  military and civilian uses. We summarize some of the important applications in Table~\ref{tab_applications} that are described in the following.

\subsubsection{Military Applications}


\paragraph{Shipborne JRC systems} The main aim of using radar systems onboard is to detect enemies on the sky and on the ground/sea at long distances. In the past, radar and data communication systems (e.g., voice and text) are usually separately operated. However, due to the development of digital technologies, more and more applications of using JRC have been introduced recently to facilitate both functions. For example, when a battleship performs a radar scanning activity, it can include some information to transmit data to its allies on the sky or on the sea. In this case, the battleship can not only detect enemies but also carry out strategic communications to ensure shipborne electronic warfare with its allies and command post.

\paragraph{Airborne JRC systems} Similar to shipborne JRC systems, both radar and data communications functions are expected to be implemented on modern airborne systems to enhance electronic warfare by communicating and detecting objects at long distances simultaneously. However, there are several fundamental differences between airborne and shipborne JRC systems. First, while airborne systems are usually moving very fast (up to few thousands kmph, e.g., Lockheed SR-71 Blackbird~\cite{Blackbird}), shipborne systems' movements are pretty slow (less than 100 kmph). In addition, while shipborne JRC systems are usually used to detect and communicate with targets moving above the horizon, the airborne JRC systems are often used to detect and communicate with targets moving below the horizon as seen by the radar that is also known as look-down/shoot-down ability to combat aircrafts. Specifically, targets of airborne systems are usually below the radar, and thus the radar has to ``look down'' to search for the target, it will cause many difficulties in detecting the target. Thus, to address this problem, look-down/shoot-down radars~\cite{Roulston2008post,Abdallah2017Study} have been developed recently with outstanding features of detecting and tracking air targets moving below the horizon as seen by the radar. However, integrating communications systems with such radar system still needs further investigations.

\paragraph{Ground-based JRC systems} In a similar way, the command post and many mobile military vehicles such as tanks, reconnaissance vehicle and light utility vehicle on the group also can perform both radar detection and data communications by using JRC systems to improve electronic warfare. However, different from applications of JRC in airborne and shipborne systems which mainly focus on communicating and detecting targets at long distances without many obstacles, ground-based JRC systems mainly focus on communicating and detecting objects at short distances, e.g., battlefields, with many obstacles in surrounding environments, e.g., vehicles, trees, and buildings. As a result, designing ground-based JRC systems needs to take these factors into considerations. For example, low-frequency signals are usually used in airborne JRC systems due to long-range communications and detections requirements, while high-frequency signals are often used in ground-based JRC systems because of short-range communications and detections demands.

\begin{table*}[!h]
	\centering
	\caption{Applications of JRC Systems} 
	\centering 
	\begin{tabular}{|c|c|c|c|c|}
		\hline
		\textbf{Applications} & Coverage & Frequency & Implementation Challenges & Commercialized products \\
		& (up to)  & GHz & &   \\
		\hline
		\hline
		Shipborne & Long & 2-12 & Long distances and  & Raymarine (\url{www.raymarine.com.au}) \\
		& 100-150 km & \cite{radartutorial} & fast movement & Garmin (\url{www.garmin.com}) \\
		\hline
		Airborne & Very long & 8-12 & Very long distances and & Aeroexpo (\url{www.aeroexpo.online}), \\
		& 100-300 km & \cite{radartutorial} & very fast movement & Leonardocompany (\url{www.leonardocompany.com}) \\
		\hline
		Ground-based & Very long & 0.3-2 & Very long distances & Lockheedmartin (\url{www.lockheedmartin.com}) \\
		& 100-300 km & \cite{radartutorial} & & Raytheon (\url{www.raytheonmissilesanddefense.com}) \\
		\hline
		Autonomous & Short & 24 or 75-79 & Fast movement and & Nxp (\url{www.nxp.com}) \\	
		vehicles & 100-200 m & \cite{radartutorial}~\cite{infineon} & high density environment & Infineon (\url{www.infineon.com}) \\
		\hline
		Indoor localization and & Short & 2-12 or 24 & High density environment & Sensingproducts (\url{www.sensingproducts.com}) \\	
		activity recognition & 10-50 m & \cite{radartutorial} & with many reflecting objects & Parametric (\url{www.parametric.ch}) \\	
		\hline		
		UAV communication & Medium & 8-12 or 24-40 & Fast movement and & Echodyne (\url{www.echodyne.com})   \\	
		and radar sensing & 2-3 km & \cite{radartutorial}~\cite{Remy2012The} & high density environment & Orbisat (\url{www.orbisat.com.br})  \\	
		\hline 
	\end{tabular}
	\label{tab_applications}
\end{table*}

\subsubsection{Civilian Applications}

\paragraph{Autonomous vehicular systems} Over the last five years, we have experienced a huge demand on autonomous vehicular systems, especially self-driving cars. However, there is a tremendous barrier that is hindering the development of autonomous vehicular systems, that is safe for both people in the car and others in traffic. Current safety systems, e.g., based on sensor systems and cameras, do not guarantee an extra safety for autonomous vehicular systems because many unexpected events on the road are out of control by these systems, e.g., moving objects from blinded zones and impacts by weather as well as other environmental impacts. As a result, automotive radar has recently considered to be an enabling technology for future autonomous vehicles with a significant improvement for road safety~\cite{HouraniMillimeter2018}. In practice, automotive radar systems, e.g., NXP (\url{www.nxp.com}), Rohde\&Schwarz (\url{www.rohde-schwarz.com}) and Infineon (\url{www.infineon.com}), working at 77/79 GHz are able to detect and recognize objects at a range of up to 250 meters, which enables the driver assistance capabilities required to obtain a five-star rating from Euro NCAP (European New Car Assessment Program)~\cite{infineon}. These radar systems are currently used at the same frequency as the vehicle networks. As a result, many applications of JRC can be implemented in order to simultaneously enhance communication efficiency and road safety for autonomous vehicles.

\paragraph{Wi-Fi communications integrated with Indoor localization and activity recognition}

Wi-Fi systems are typically used to help people connect to the Internet through using access points for short-range communications such as indoor environments. However, some recent research works have found that the Wi-Fi systems can be very useful for indoor localization and activity recognition~\cite{wang2019joint,he2015wi,liu2019wireless}. For example, the authors in~\cite{he2015wi} review emerging technologies used in Wi-Fi fingerprint localization with focus on advanced localization techniques and efficient system deployment. Furthermore, in~\cite{liu2019wireless}, many new advanced techniques for Wi-Fi sensing reviewed, which enables a wide range of human activity recognition such as gesture recognition, vital signs monitoring and occupancy monitoring. The main idea of these techniques is extracting information from Wi-Fi signals backscattered/reflected from the target objects (e.g., human, animal, and devices), thereby identifying location and activities of the target objects. As a result, Wi-Fi systems are also considered to be JRC systems in which Wi-Fi signals can be used for both communications and sensing/detection at the same time~\cite{Tan2014Areal,Wenda2020Passive,Kevin2011Through}.

\paragraph{Joint UAV communications and radar functions} Similar to application scenarios in airborne systems, JRC is also useful to implement in Unmanned Aerial Vehicles (UAVs) to help them simultaneously communicate and identify targets. However, there is a fundamental difference between JRC functions used in military airborne systems and in civilian UAVs. Specifically, civilian UAVs are usually used for short-range communications with the main aims of collecting and providing sensing data, while communications in military airborne systems are mostly for tactical communications and transmitted at very long distances. In addition, unlike military airborne systems, radar functions in UAVs are mainly used for collision avoidance, e.g., to avoid crashes with high building, trees, and other flying objects. Therefore, JRC used in UAVs is very similar to those in autonomous vehicles systems. Some examples of commercialized collision avoidance radars used in UAVs such as MR72~\cite{MR72} and NRA15~\cite{NRA15} which are currently working at 77GHz and 24GHz, respectively, and thus JRC function can be integrated effectively to improve communications for such systems.

\section{Spectrum Sharing}
\label{sec:radar-commu-mul}




JRC systems such as DFRC perform both radar and communication functions by using a common hardware device. Therefore, these functions need to share system resources such as spectrum and energy. In particular, sharing the frequency spectrum is very important since the frequency spectrum is becoming increasingly congested due to the plethora of connected devices and services. To enable the frequency spectrum sharing between the radar and communication functions while guaranteeing the requirement performance of each function, several resource sharing approaches have been proposed. In general, the approaches can be divided into four categories~\cite{ma2019joint}: communication signal-based approaches, radar signal-based approaches, time division approaches, and spatial beamforming approaches. 
\begin{itemize}
\item \textit{Communications signal-based approaches:} These approaches use standard communication signals such as OFDM for the radar probing. In particular, a JRC system transmits OFDM signals including data bits to a remote communication receiver. The OFDM signals that are reflected from radar targets can be used by the radar subsystem to obtain the targets' parameters. However, issues of these approaches are the randomness of the data bits and the high peak-to-average power ratio (PAPR) of the communication signals. 
\item \textit{Radar signal-based approaches:} These approaches use conventional radar signals such as frequency-modulated continuous wave to transfer communication symbols. One major issue of these approaches is that embedding the communication symbols in the radar signals can compromise the radar performance. Thus, advanced waveform designs need to be investigated. 
\item \textit{Time-division approaches:} These approaches perform time allocation to the radar and communication functions separately. The key issue of these approaches is how to optimize the trade-off between radar and communication performance. 
\item \textit{Spatial beamforming approaches:} These approaches design beamforming for the communication signals, and then the radar signal is projected into the null space of its channel to the communication receiver. 
\end{itemize}

\subsection{Communication Signal-Based Approaches}

\begin{figure}[t!]
\centering
\includegraphics[width=\linewidth]{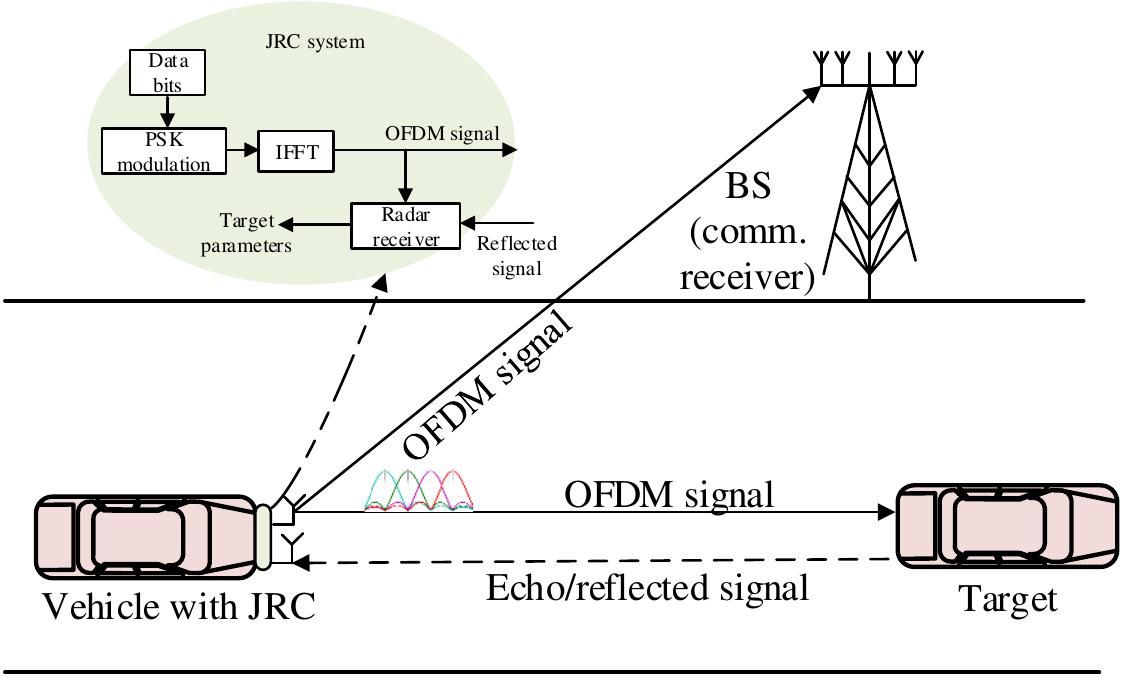}
 \caption{\small JRC system based on OFDM signal.}
 \label{comm_waveform_based}
\end{figure}

Two communication signals that are commonly used for the radar probing are the spread spectrum and OFDM as shown in Fig.~\ref{comm_waveform_based}. 

\subsubsection{Spread Spectrum}
\label{spread_waveform}
In digital communication systems, e.g., the code-division multiple access (CDMA), each communication signal with a bandwidth can be transmitted with a larger spectral band by using the spread coding technique. The spread coding technique allows communication signals to be modulated with pseudorandom sequences. Moreover, the pseudorandom sequences have good autocorrelation properties that facilitate radar target detection. Thus, the communication signal modulated with the spread coding can be used for the JRC systems as proposed in~\cite{sturm2011waveform} and \cite{mizutani2001inter}. The general idea of such an approach is as follows. First, data bits are mapped into data symbols, e.g., by using the PSK modulation. Then, each data symbol is modulated, i.e., multiplied, with a code sequence, e.g., an m-sequence. The code sequences are assumed to be known at receivers, e.g., by using synchronization schemes. Thus, the radar receiver and the communication receiver can, respectively, estimate target parameters and detect data symbols by using the matched filter based on correlation algorithms. The simulation results in~\cite{sturm2011waveform} show that the m-sequences with higher spreading factors, i.e., longer sequence lengths, are able to estimate the targets with higher ranges. However, the maximum velocity that the proposed scheme can estimate is very limited, e.g., $6.25$ m/s. This limits the application of the proposed scheme to practical real-time applications such as autonomous vehicles. Moreover, high speed analog-to-digital converters are required for the wideband spread-spectrum waveforms that increases cost and complexity. 
\subsubsection{OFDM Waveform}
\label{OFDM_waveform}
The OFDM allows multiple orthogonal subcarrier signals with partially overlapping spectra to carry data in parallel. The OFDM thus improves spectral efficiency significantly. The OFDM has several other advantages such as robustness against multipath fading, easy synchronization and equalization, and high flexibility. These advantages enable the OFDM to be effectively used for the target detection of the radar function. 

The pioneering work that uses the OFDM for JRC is~\cite{sturm2009ofdm}. The system model is a monostatic system, e.g., an autonomous vehicle (AV) as shown in Fig.~\ref{comm_waveform_based}, that is equipped with one transmitter and one radar receiver, i.e., DFRC. The transmitter first modulates data bits to OFDM signals by using a conventional OFDM modulation (see Section~\ref{OFDM_basic}).~Accordingly, the data bits are mapped into data symbols, e.g., by using BPSK, and then an inverse fast Fourier transform (IFFT) algorithm is applied to transfer the data symbols to OFDM signals. At the same time, the transmitter also shares the OFDM signals with the radar processing of the radar receiver. The OFDM signals are transmitted to a distant communication receiver. Some OFDM signals reflected from targets are received by the radar receiver. The radar processing calculates the range of the target by simply correlating the transmitted signal with the reflected signal. Note that this process generates a range profile. The simulation results in~\cite{sturm2009ofdm} show that the proposed scheme can accurately calculate the ranges of two close targets, i.e., with a spacing between them being $1.9$ m.

However, the scheme proposed in~\cite{sturm2009ofdm} has a drawback that the correlation function of the time domain OFDM signal depends on the data bits. Thus, the  range profile may have high sidelobes, and this makes the radar receiver difficult to distinguish between the peak and the sidelobes that consequently reduces the detection accuracy. To circumvent the drawback, the authors in~\cite{sit2011ofdm} propose to remove the data bits from the received signal before estimating the target parameters. In particular, the radar processing at the radar receiver applies a fast Fourier transform (FFT) algorithm to transform the reflected signal in the time domain to received modulation symbols in the frequency domain. The received modulation symbols include the transmitted modulation symbols, velocity and range information of the target. The transmitted modulation symbols are removed from the received modulation symbols by using the element-wise division. The IFFT algorithm is then applied to the received modulation symbols to calculate the velocity and the range of the target. The simulation results in~\cite{sit2011ofdm} show that the peak-to-sidelobe ratio of the range profile obtained by the proposed scheme is much higher than that of the range profile obtained by the baseline scheme from~\cite{sturm2009ofdm}. This facilitates the detection process and significantly improves the accuracy of radar target parameter estimation. For example, the proposed scheme can estimate the target velocity of up to $252$ m/s~\cite{sit2011doppler}.

Different from~\cite{sit2011ofdm}, the authors in \cite{liang2019design} propose to combine the OFDM technique with the \textit{P4 code}~\cite{han2019artificial} to address the randomness of data bits. The P4 codes are basically similar to the phase values generated by the phase-shift keying (PSK) modulation. First, a sequence of P4 codes, i.e., phases, is generated according to the data bit rate. By cyclically shifting the positions of the P4 codes in the sequence, new sequences of P4 codes are generated that constitute a complementary set. Before the random bits are modulated by the OFDM technique, they are mapped into one of the P4 code sequences in the complementary set. The complementary set has one important feature that reduces the sidelobes of autocorrelation functions implemented at the radar receiver. This enables the radar receiver to accurately determine the range of the target with high speed. Indeed, the simulation results show that with the proposed scheme, the JRC system is able to clearly detect targets with a velocity up to $300$ m/s.

Unlike~\cite{liang2019design}, the authors in~\cite{tian2017radar} propose to combine the OFDM with the m-sequence~\cite{tedesso2018code} instead the P4 code. The m-sequence is also known as a maximum-length sequence that includes bits generated using maximal linear feedback shift registers. The m-sequence and its cyclic shifted versions have
an ideal periodic autocorrelation function. Thus, the m-sequence can be used to design the radar and communication signals to enhance the resolution range and velocity estimation of the radar. In particular, before the random bits are modulated with the OFDM, they are mapped into a time shift value that is used to generate the corresponding m-sequence. At the radar receiver, the cross-correlation and discrete Fourier transform (DFT) are used to estimate the range and the velocity of the targets. The simulation results in~\cite{tian2017radar} show that by using a m-sequence with the size of $127$, the proposed scheme is able to detect two close targets, i.e., the distance between them is $0.3$ m, with a range of up to $12$ km. Moreover, the data transmission rate can achieve up to $8.96$ Mbps. 

Apart from the P4 code and the m-sequence, the Golay code~\cite{davis1999peak} has recently been combined with the OFDM as proposed in \cite{li2019waveform}. The Golay code, also known as the Golay complementary sequence, is a type of linear error-correcting code used in digital communications. Thus, the Golay code can not only eliminate the data dependency but also can improve the error-correction capability of the joint radar-communication system. The modulation of the Golay code is implemented similarly to that of the m-sequence as presented in~\cite{tian2017radar}. The simulation results in \cite{li2019waveform} show that the BER obtained by the proposed scheme is much lower than that obtained by the original OFDM scheme, e.g.,~\cite{sturm2009ofdm}. Moreover, the proposed scheme significantly decreases the side lobes of the ambiguity functions that results in improved radar performance.

 Most of the aforementioned approaches assume that the phase shifts on different OFDM subcarriers are the same. However, when a large number of OFDM subcarriers are used, i.e., the wideband OFDM, the received signals on different subcarriers are incoherent~\cite{podkurkov2019efficient}, meaning that the phase shifts on different subcarriers may be different. In this case, the traditional detection algorithms such as the correlation algorithms as proposed in~\cite{tian2017radar} may not accurately estimate the phase shifts of the received signal. Since the target velocity is estimated according to the phase shifts, the inaccurate estimation of the phase shifts reduces the accurate estimation of target velocity. For this, the authors in~\cite{zhang2017efficient} propose to transform the OFDM wideband system into an approximately equivalent narrowband system by using the linear interpolation
method and the cubic spline interpolation method~\cite{oppenheim1999discrete}. Then, the traditional detection algorithms such as correlation algorithms can be applied to estimate the radar target parameters, i.e., the range, velocity, azimuth and elevation angles. By using Monte Carlo simulations, the results show that the proposed scheme outperforms the traditional detection approaches without using the interpolation method in terms of a lower total root mean square estimation error of the estimated parameters.

Note that in the aforementioned OFDM waveform-based approaches, the IFFT algorithm is typically used to transform data symbols on subcarriers in the frequency domain to samples that constitute OFDM symbols in the time domain.~Due to the central limit theorem, some output samples have very large magnitudes. This results in the PAPR problem~\cite{paredes2015problem} in the OFDM approaches. The PAPR in the time domain is defined as the ratio of the maximum instantaneous power to the average power over output samples. The high PAPR forces the transmit circuit operating in the saturation region, and the signal distortion, i.e., in-band distortion, will arise. This results in reducing the Signal-to-Noise Ratio (SNR) at the communication receiver and the detection range for the radar function. To reduce the PAPR, clipping-based active constellation extension (ACE) technique~\cite{bae2009adaptive} or the tone reservation (TR)~\cite{krongold2004active} can be used.

To improve the spectrum efficiency and data rate while reducing the PAPR, the authors in~\cite{lv2018joint} propose to combine the OFDM with the orthogonal chirp division multiplexing (OCDM) signals~\cite{lv2018novel}. The OCDM signal consists of a number of chirp waveforms that are mutually orthogonal with each other in the chirp domain. Some reserved chirp waveforms are reserved for generating peak canceling signals, and the other chirp waveforms are used for embedding the communication data. In particular, at the transmitter, the data symbols, e.g., QAM symbols, are first modulated with the OCDM signals generated using the inverse discrete Fresnel transform (IDFnT) algorithm~\cite{ouyang2015discrete}. Then, the OFDM technique is applied to the modulated symbols to generate OFDM signals. At the communication and radar receivers, the discrete Fresnel transform (DFnT) and FFT algorithms are used to detect the data symbols and the targets. The simulation results in~\cite{lv2018joint} show that the communication rate obtained by the OCDM-OFDM-based waveform scheme is $4N^2$, while those obtained by both the OCDM-based waveform scheme and the OFDM-based waveform scheme are $4N$. Here, $N$ is the number of chirps. Moreover, the ambiguity function obtained by the OCDM-OFDM-based waveform has a sharp shape and low sidelobes that improves the radar performance.

The aforementioned approaches discuss how to address two major issues, i.e., the data bit randomness and the high PAPR, of using the OFDM for the JRC. Since both the functions share the OFDM symbols, choosing OFDM modulation parameters such as the subcarrier spacing and length of guard interval, i.e., cyclic prefix (CP), has a considerable effect on the performance of both the radar and communication functions. The authors in~\cite{braun2009parametrization} analyzed and presented conditions that some important OFDM parameters should satisfy to guarantee the performance of both the functions. The conditions generally depend on the characteristics of radar and communication channels. In particular, the CP length of the OFDM symbols needs to be larger than the maximum excess delay to prevent the inter-symbol interference. Second, the subcarrier spacing needs to be smaller than the coherence bandwidth, i.e., the frequency span over which the channel is assumed to be constant.

\subsubsection{Orthogonal Time Frequency Space (OTFS) Waveform}
\label{OFDM_waveform}

\begin{figure}[t!]
\centering
\includegraphics[width=\linewidth]{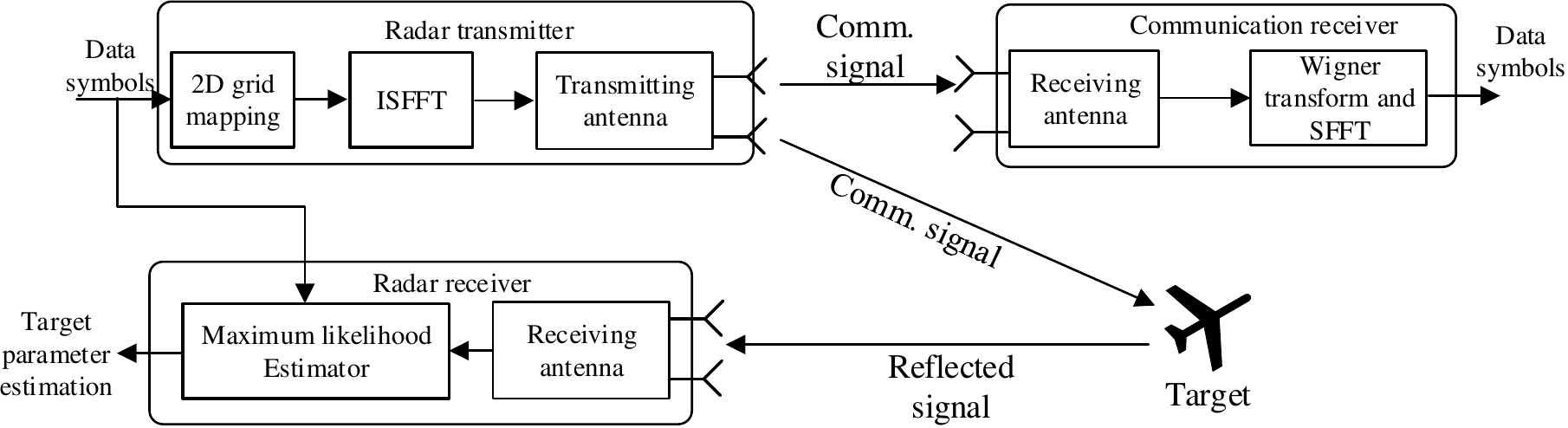}
 \caption{\small OTFS waveform modulation for JRC systems.}
 \label{OTFS_based}
\end{figure}

In this section, we discuss an emerging modulation technique called \textit{orthogonal time frequency space (OTFS)}~\cite{gaudio2019effectiveness} that can be used for the JRC system. The OTFS is considered to be a generalization of the OFDM and the  CDMA, i.e.,  the spread-spectrum technique. The OTFS technique enables data symbols to experience a near-constant channel gain even for the channels with high Doppler frequencies, massive MIMO, or at high frequencies such as mmWave. Therefore, the OTFS has recently been proposed for the JRC system as in~\cite{gaudio2019performance}. The system model is shown in Fig.~\ref{OTFS_based} that includes a communication transmitter, a radar receiver collocated with the communication transmitter, and a remote communication receiver. At the communication transmitter, the data symbols, e.g., QAM symbols, are first arranged on a 2D grid. Then, the inverse symplectic finite Fourier transform (ISFFT) is applied to represent the data symbols in the time-frequency domain. This process is also called \textit{Heisenberg transform}~\cite{hadani2017orthogonal}.  At the communication receiver, the Wigner transform and the SFFT are applied to the received signal to detect the data symbols. The radar receiver estimates the target parameters by using a maximum-likelihood algorithm applied on the reference signal, i.e., the transmitted symbols, and the received signal. The simulation results show that the radar performance obtained by the OTFS scheme is similar to that obtained by the OFDM scheme using the same bandwidth and time resources. However, the communication rate obtained by the OTFS is much higher than that obtained by the OFDM scheme. The reason is that the OTFS has a higher multiplexing gain and does not use an overhead from the CP sequence, which is used in the OFDM scheme. The future works need to evaluate the OTFS scheme in dynamic mobile environments with high Doppler frequency.

\subsection{Radar Signal-Based Approaches}
This section discusses spectrum sharing approaches in which communication symbols are embedded into the emission of the radar signals. The traditional radar systems typically use two signals, i.e., frequency-hopping and chirp or sweep signal, with constant-modulus waveforms to avoid the signal distortion and to improve the energy efficiency~\cite{hassanien2018dual}. The two signals have recently been proposed for the JRC system.

\subsubsection{Frequency-hopping signal}

\begin{figure}[t!]
\centering
\includegraphics[width=\linewidth]{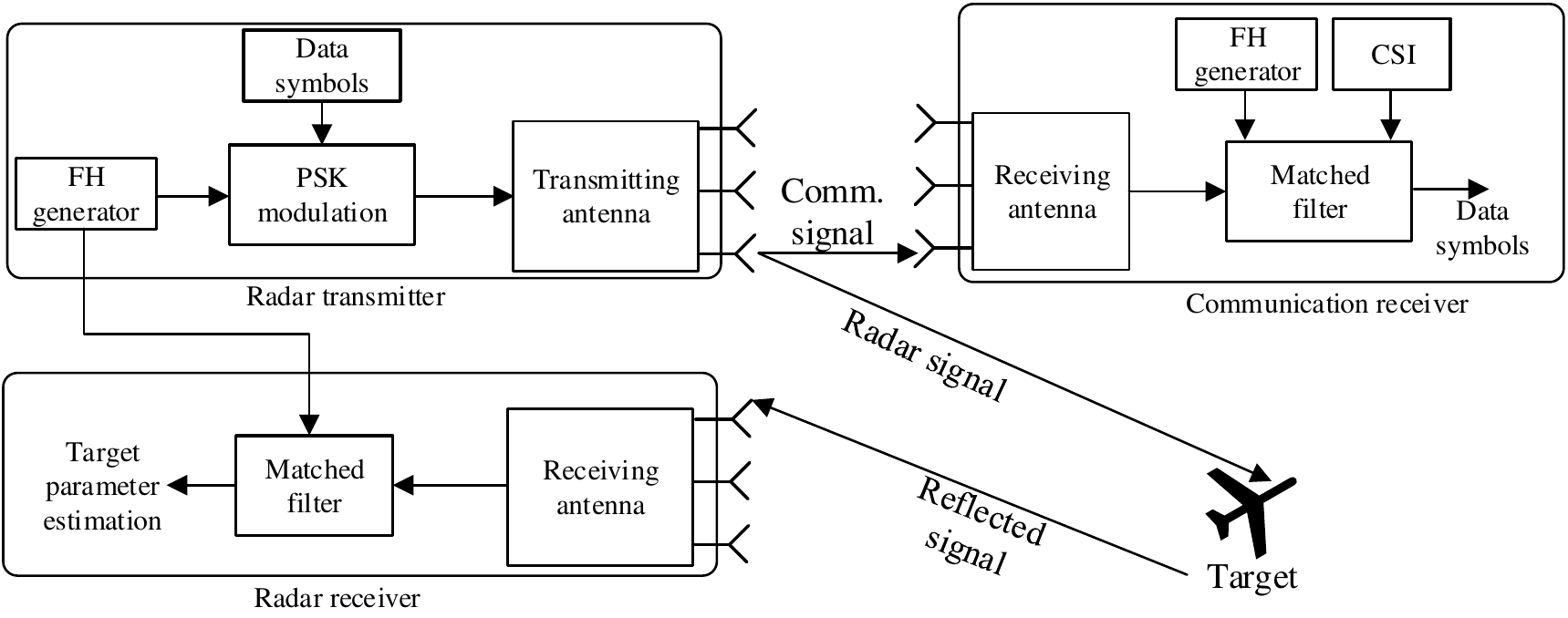}
 \caption{\small FH waveform modulation for JRC systems.}
 \label{frequency_hopping_based}
\end{figure}
 Frequency-hopping (FH) technique is a method of transmitting radio signals by rapidly changing the frequency among many distinct frequencies. The FH signals have the constant-modulus feature and are easily generated. Thus, they are commonly used for radar systems in military areas. Moreover, the FH waveform signals are resistant to interference and eavesdropping. Thus, FH signals can be used to embed data symbols in the JRC systems to enhance the data transmission security as proposed in \cite{hassanien2017dual}. The system model is a DFRC system equipped with a common dual-function transmit platform, i.e., a MIMO radar system, as shown in Fig.~\ref{frequency_hopping_based}. The system first generates a set of $M$ orthogonal FH waveforms by using the code optimization algorithm~\cite{costas1984study} that guarantees a good ambiguity function for the radar detection. Then, the communication symbols that are modulated by PSK are embedded into the FH waveforms. In particular, $M$ orthogonal FH waveforms are transmitted in each radar pulse. The phase of each FH waveform is modulated with $Q$ FH codes. Here, each FH code represents a communication symbol, meaning that $MQ$ communication symbols are embedded during the radar pulse. At the communication receiver, the matched-filtering algorithm is adopted to estimate the embedded phases and the communication symbols.~The simulation results in \cite{hassanien2017dual} show that the symbol error ratio (SER) can reach up to $10^{-6}$ when the BPSK is used for the symbol modulation. However, how the proposed scheme alters or compromises the radar performance is not shown. Moreover, the proposed scheme requires an accurate channel state information (CSI) estimation for estimating the embedded phases and communication symbols that is very challenging to obtain in practice. The achievable SER of the communication function may be higher with inaccurate CSI.

To address the challenge, the authors in~\cite{wang2019co} proposed to use the FFT algorithm instead of the matched-filtering algorithm at the communication receiver. In particular, the received signal at the communication receiver is partitioned into $Q$ continuous non-overlapped sub-pulses. Then, the FFT algorithm is implemented with the $Q$ sub-pulses to determine their dominant frequency components. Based on these frequency components, the FH codes are estimated, and the embedded data symbols are detected. The simulation results show that the ambiguity function has a sharp shape with lower sidelobe levels, improving the radar performance. Furthermore, the BER of the communication symbol detection obtained by the proposed scheme is very low, i.e., up to $10^{-6}$, given the SNR of $-9$ dB.~Especially, compared with \cite{hassanien2017dual}, the proposed scheme does not require the CSI estimation to detect the data symbols that significantly reduces the complexity in designing the communication receiver.


Recently, the Costas hopping waveform~\cite{golomb1984constructions} has been used as FH waveform as proposed in~\cite{nusenu2018dual}. The Costas hopping waveforms can have nearly ideal range-Doppler ambiguity properties and exhibit a thumbtack-shaped ambiguity function. Moreover, the Costas waveforms are simple to generate and immune to interference that helps the radar receiver to accurately estimate the target velocity. The system model is a collocated MIMO radar system with $M$ transmit antenna elements. The system first generates the Costas waveforms using the construction algorithm~\cite{golomb1984constructions}. Then, the phase modulation approach as proposed in \cite{hassanien2017dual} is adopted to embed the information symbols in every Costas waveform. The frequency diverse array (FDA) technology is used to transmit the embedded Costas waveforms. Specifically, each waveform emitted from an individual antenna element is orthogonal with a frequency increment. This means that the FDA allows the system to use more transmit and receive degrees-of-freedom than the conventional MIMO radar in~\cite{hassanien2017dual} and~\cite{wang2019co}. This makes the system easy to distinguish the targets even if they have the same angle but different ranges. The simulation results show that the proposed scheme outperforms the FH waveform-based approach~\cite{hassanien2017dual} in terms of SER at the communication receiver and of SINR at the radar receiver. This indicates that the proposed scheme has better robustness against the interference and noise. However, the proposed scheme requires the phase synchronization between the transmit platform and the communication receiver that may be challenging to implement.

To address the phase synchronization challenge, the authors in \cite{hassanien2015dual} developed the \textit{phase-rotational invariance approach} for embedding the communication symbols in the radar emission. The main idea can be described as follows. Assume that the transmitter side aims to embed a sequence of $Q$ bits to the communication receiver into each radar pulse. The transmitter side generates $2^Q$ pairs of beamforming vectors. The $2^Q$ pairs of beamforming vectors generate $2^Q$ different phase rotation values. Then, during the radar pulse, the sequence of $Q$ bits is mapped into a pair of beamforming vectors, i.e., a phase rotation value. In other words, a pair of beamforming vectors are embedded into the radar pulse. To avoid the interference, a pair of orthogonal waveforms are associated with the pair of beamforming vectors. As such, the same pair of waveforms is used during all pulses, while the pair of beamforming vectors changes from pulse to pulse based on which bit sequence is transmitted, i.e., which phase rotation value is selected. At the communication receiver, the phase rotation value and the corresponding bit sequence are estimated by taking the difference in phase between the two beamforming vectors. The simulation results show that the proposed scheme outperforms the information embedding schemes based on sidelobe diversity~\cite{euziere2014dual}. Especially,
the proposed scheme does not require phase synchronization
as it does not need the phase estimation of the received
signal. 

In fact, the data rate obtained by the scheme in~\cite{hassanien2015dual} can be significantly improved if more pairs of orthogonal waveforms are used as proposed in~\cite{hassanien2016non}. As such, instead of transmitting one-bit sequence during the radar pulse, multiple bit sequences can be transmitted on different pairs of orthogonal waveforms. The simulation results show that the proposed scheme can embed up to $15$ bits per pulse, while the baseline scheme from~\cite{hassanien2015dual} can embed only $8$ bits per pulse given the same BER and SNR. This means that the proposed scheme improves the data rate significantly.



\subsubsection{Chirp signal}
\label{chirp_waveform}
Apart from the FH signals, the chirp signal is commonly applied to radar systems. The chirp signal, also known as sweep signal, is a signal in which the frequency increases or decreases with the \textit{chirp rate}~\cite{wang2018joint}. Before transmitting the chirps, the radar transmitter performs a so-called \textit{chirp modulation}, LFM, or FMCW (see Section~\ref{FMCW_tutorial}). The echo signal reflected from the target is received by the radar receiver. Then, the radar receiver estimates the range, velocity, and direction of the target based on the differences in phase and frequency between the transmitted signal and the echo signal, e.g., through matched filters. With low sidelobe levels, the LFM can detect two small targets that are located at a long range with a very small separation between them. Moreover, the LFM-based radar system is highly resistant to interference, e.g., jamming and eavesdropping. In addition, the LFM-based radar system can simplify hardware components due to the constant modulus feature of the chirp waveform. Recently, the chirp waveform has been used to convey data bits~\cite{sahin2017novel}. For example, a positive chirp rate is to transmit bit ``1'', and a negative value is to transmit bit ``0''.  This means that the chirp waveform can be used for the JRC systems. 

However, generating the sweep signal typically requires a large range of frequencies that results in low spectrum efficiency. To improve the spectrum efficiency, the LFM modulation can be combined with the OFDM as proposed in~\cite{wang2018poster}. In particular, data symbols are first modulated by a typical OFDM transmitter to generate OFDM signals. The OFDM signal is then multiplied with the LFM waveform. At the radar receiver, the radar processing is implemented by mixing, i.e., multiplying, the reflected signal with the conjugate LFM waveform generated from a local oscillator. This process is called dechirping. Then, the 2D-FFT algorithm is applied to the baseband signal after the dechirping to estimate the range and the velocity of the target. Since the baseband signal is the same as that in conventional OFDM systems, the baseband processing at the OFDM receivers can be used to demodulate the data symbols. The proposed scheme is able to detect one target with a distance up to $60$ m and a velocity of $3$ m/s. However, parameters to generate the LFM waveform at the transmitter need to be known at the receiver, and this requires some synchronization schemes. Moreover, the OFDM is a non-constant envelope modulation technique with high PAPR that may result in the serious distortion of transmitted signals in the nonlinear region of radar amplifier at the receiver.

To address the shortcomings of the scheme proposed in~\cite{wang2018poster}, minimum shift keying (MSK) is proposed to combine with the LFM modulation as proposed in~\cite{dou2017radar}. Such a combination scheme is namely MSK-LFM. MSK is known as a continuous phase modulation (CPM) scheme in which the data symbols are modulated with signals that have continuous phases. The integration of the MSK signal with the chirp waveform is implemented similarly to the integration of the OFDM signal with the chirp waveform as presented in~\cite{wang2018poster}. Since the MSK signal has the constant envelope feature, the MSK-LFM signal can avoid the distortion caused by the nonlinearity of the radar amplifier. However, as analyzed in~\cite{chen2017energy}, the spectrum of the MSK-LFM signal is a function of data bits. Thus, in the cases that all the data bits included in the LFM pulse, are ``0'' or ``1'', the spectrum of the MSK-LFM signal exceeds the original bandwidth of radar system, i.e., the LFM signal bandwidth. This results in increasing the energy leakage and degrading both the detection and communication performances.

To strict the spectrum within the original bandwidth of the radar system, a simple solution is to place data bits in the middle of the LFM pulse, and the edge of the signal has no data bits. However, this leads to discontinuous phases between consecutive LFM pulses, and thus a large spectrum extension may occur. The authors in \cite{zhipeng2015communication} propose a modified three-phase integrated waveform algorithm. The idea is to add a sequence of bits ``1'' at the beginning of the LFM pulse and a sequence of bits ``0'' at the end of the LFM pulse. This is to avoid the above two cases, i.e., all the bits included in the LFM signal are ``1'' or ``0''. The simulation results show that the spectrum of the MSK-LFM signal obtained by the proposed scheme is always within that of the LFM signal. However, the data rate obtained by the proposed scheme is very limited due to a number of redundant bits. To reduce the redundant bits, the approaches based on the partial response of the CPM~\cite{zhang2017modified} and rate-shift algorithm~\cite{ni2019high} can be used. In particular, the rate-shift algorithm is implemented based on the time-varying property of the upper bound on  available transmission rate of the MSK-LFM signal. Then, the data rate is adjusted such that it is always approximately close to the maximum available rate at any time. The simulation results in~\cite{ni2019high} show that the rate-shift approach can improve the data throughput up to $60$\% compared with the constant rate MSK-LFM signal given the same BER.

The radar signal-based approaches as proposed in~\cite{dou2017radar}, \cite{zhipeng2015communication}, and \cite{ni2019high} are considered in single-user scenarios in which the JRC transmitter transmits communication symbols to a single communication receiver. It is worth noting that the frequencies, also known as subcarriers, included in the LFM radar signals are orthogonal with each other. Therefore, the aforementioned approaches, e.g.,~\cite{dou2017radar}, can be applied to multi-user transmissions in which each user is assigned to one subcarrier in the LFM radar signal. Such an approach can be found in~\cite{li2013integrated} and \cite{zhang2017waveform}. In particular, the authors in~\cite{li2013integrated} considered a scenario including one JRC transmitter and multiple communication users. First, each communication bit to be sent to a user is modulated with a discontinuous phase modulation, i.e., BPSK. Then, the data symbol is embedded into/multiplied with a subcarrier of the LFM radar signal. This means that one data symbol is transmitted on a subcarrier in one LFM radar pulse. Thus, if the LFM radar signal has $N$ subcarriers, there are only $N$ symbols, i.e., $N$ bits, transmitted during the LFM pulse. The proposed scheme consequently has a low capacity that cannot meet the need for high-speed transmission in practical systems.    

To achieve high-speed transmission, the authors in \cite{zhang2017waveform} propose to modulate the communication bits sent to the users by using the continuous phase modulation, i.e., the CPM, instead of the discontinuous phase modulation. First, the bit sequence to be sent to a user is converted into a bipolar amplitude modulated sequence. Then, the CPM is applied to convert the bipolar amplitude modulated sequence into phase symbols. The symbols to be sent to the user are embedded into one subcarrier of the LFM signal. At each communication receiver, the low-pass filtering, CPM demodulation and decoding are applied to detect the communication symbols. Since a large number of communication symbols are transmitted in one LFM pulse, the proposed scheme can significantly improve the data transmission and spectrum efficiency. In particular, the spectrum efficiency achieved by the proposed scheme is almost $D\log_2D$ times higher than that achieved by the baseline scheme~\cite{li2013integrated}. Here, $D$ is the number of communication symbols transmitted in one pulse on each subcarrier. Moreover, the proposed scheme can achieve the BER close to that obtained by the baseline scheme without adjacent channel interference. In addition, the distance and velocity ambiguity functions are almost the same with the LFM waveform, meaning that the proposed scheme can well accomplish the detection. These benefits are of great significance for the development of intelligent transportation systems. 

The BER obtained by the scheme proposed in~\cite{zhang2017waveform} can be improved when it is combined with the low-density parity-check (LDPC) code~\cite{moon2005error} as proposed in \cite{li2019integrated}. Accordingly, the LDPC codes are inserted in the symbol sequence before this symbol sequence is embedded into the LFM radar signal. At the communication receiver, the BCJR algorithm~\cite{bahl1974optimal} is used together with the traditional CPM demodulation, e.g., as used in \cite{zhang2017waveform}, to decode and detect the communication symbols. The simulation results show that compared with the baseline scheme in~\cite{zhang2017waveform}, the proposed scheme can improve the BER around $1.8$ dB. However, the proposed scheme occurs a high latency due to the introduction of the LDPC coding and BCJR algorithm. 

\subsection{Time-Division Approaches}
\begin{figure}[t!]
\centering
\includegraphics[width=6.8 cm, height=3.6 cm]{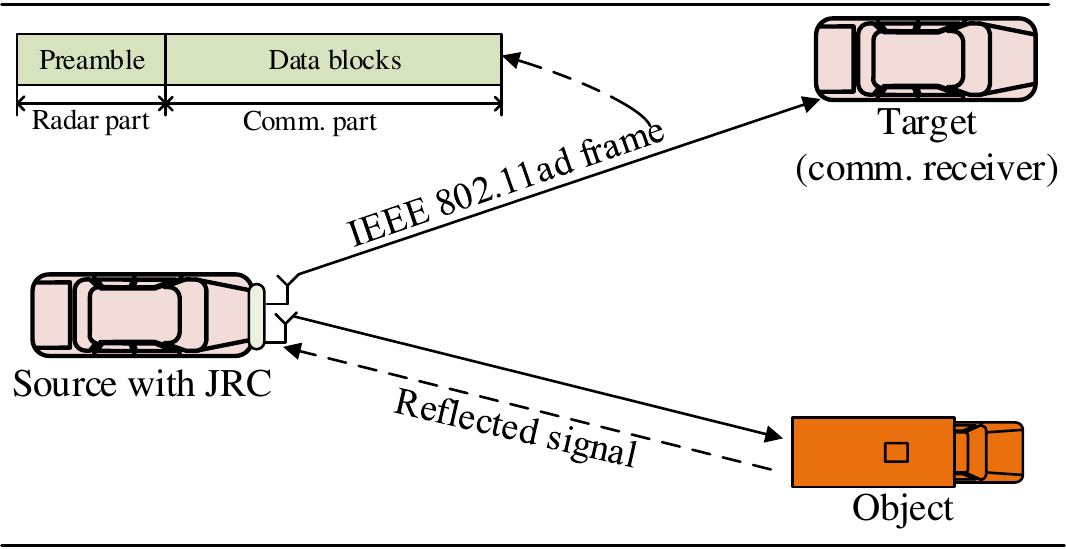}
 \caption{\small A JRC system based on IEEE 802.11ad frame in which the preamble is used for object parameter estimation and the data blocks are used for data communication.}
 \label{Time_division}
\end{figure}
Time division approaches are simple methods that allow the radar and communication functions to coexist and share the same waveform or the same frequency band. A straightforward time division approach is to allocate time slots to the radar function and the communication function in a fixed manner. Such an approach is found in \cite{kumari2015investigating}. The system model is an autonomous vehicle system including a source vehicle, a target vehicle, and surrounding objects, e.g., other vehicles, as shown in Fig.~\ref{Time_division}. The source vehicle is equipped with the DFRC in which the radar function is to sense the surrounding environment to detect the objects and the communication function is to exchange information such as velocity, braking, and entertainment content, with the target vehicle.~To provide both high data rate for the communication and high accuracy and resolution for the radar, the IEEE 802.11ad standard, i.e., a wireless LAN (WLAN) specification operating at the millimeter wave (mmWave) band, is used for the dual radar-communication system. It is worth noting that the mmWave band is able to provide a vast amount of bandwidth, e.g., up to $4$ GHz~\cite{kishida201579}, that significantly improves the performance of the radar function and communication function in the DFRC systems. The reasons are that the performance of the communication function, e.g., the data transmission rate, and the performance of the radar function, e.g., the range resolution, directly depend on the amount of allocated bandwidth. Consequently, mmWave JRC is suitable to be used for autonomous vehicles (AVs) since the mmWave JRC is able to provide accurate distance measurements of nearby obstacles, i.e., other cars and people, and to minimize the sizes of hardware of the AVs. Moreover, the high communication performance of mmWave JRC allows the AVs to communicate with each other with low latency that effectively supports the warning systems of the AVs.~The frame of the IEEE 802.11ad consists of preamble and data blocks. The source vehicle reserves the preamble block in the frame for the radar, i.e., to detect objects and to estimate their ranges and velocities, and uses data blocks for the data transmission. In particular, the autocorrelation algorithms are applied to the preamble for target detection, coarse and fine range estimation, and velocity estimation. The simulation results show that the proposed scheme can achieve a radar detection rate up to $99$\% given low SNR, i.e., above $-2$ dB. However, the proposed scheme does not achieve the desired velocity accuracy, i.e., of $0.1$ m/s, due to the short preamble duration~\cite{kumari2017ieee}. Moreover, the mmWave band has short wavelengths, thus the radar and communication signals suffer from high propagation loss and have shorter range due to the high rain attenuation and atmosphere. In other words, the DFRCs have limited range of both data transmission and target detection.

To improve the radar performance, one potential solution is to increase the preamble duration frame. However, this significantly degrades the communication performance. To optimize the trade-off, two approaches are proposed in \cite{kumari2017performance} and \cite{kumari2019adaptive}. The authors in \cite{kumari2017performance} introduced the concept of ``fraction of data symbols''.~The fraction of data symbols is the ratio of the number of data symbols possibly included in the frame to the number of data symbols in the standard frame as used in~\cite{kumari2015investigating}. The problem is to determine the fraction of data symbols to minimize the mean-squared error (MSE) bounds for the range estimation, velocity estimation, and data symbol estimation. The optimization problem is proved to be convex, that can be solved, e.g., by using subgradient projection method. The simulation results show that the proposed scheme can improve the range minimum mean square error (MMSE) by $3.3$ cm$^2$ compared with the baseline scheme~\cite{kumari2015investigating}. 

Unlike~\cite{kumari2017performance}, the authors in \cite{kumari2019adaptive} proposed to use sparse sensing techniques to optimize the trade-off
between the communication performance and radar performance. The idea is to add virtual preambles in a coherent processing interval (CPI). Here, the CPI consists of some frames in which the relative acceleration and velocity of the objects and the target with respect to the source vehicle are small enough and can be assumed to be constant. The virtual preambles located in the CPI in a sub-Nyquist fashion~\cite{kumari2018virtual} that maximizes the velocity estimation accuracy and minimizes the communication rate distortion. The simulation results show that the velocity error and the rate-distortion achieved by the virtual pulse scheme are much lower than those obtained by the baseline scheme in~\cite{kumari2015investigating}.

Another time-division approach can be found in~\cite{chiriyath2017radar}. Different from~\cite{kumari2017performance}, cycle
times are used instead of the standard frames. Then, time
portions in each cycle time are allocated to the radar and
communication so as to maximize the radar estimate rate and communication rate of the radar-communication system. Here, the estimation rate is a
metric similar to the communications rate that provides
a measure of the information about the target gained
from radar illumination. The estimation rate is determined based on the mutual information (MI) achieved by the radar receiver. One advantage of the proposed scheme is that the time portion allocated to the radar function in the current cycle time can vary depending on the radar information measured in the previous cycle time. For example, if little information is gained through the radar function in the previous cycle time, then the time portion for the radar function should decrease. 

The simulation results for the schemes proposed in \cite{kumari2017performance}, \cite{kumari2015investigating}, and \cite{chiriyath2017radar} show the benefits of using the time frames or cycle times to improve both the data communication rate and sensing accuracy. The time-division approaches are thus applicable to autonomous vehicle systems. However, the vehicles in the systems can access the same channel at the same time that causes an access collision. To avoid the collision, the authors in \cite{petrov2019unified} introduced a mechanism called Radar-Aware Carrier-Sense Multiple Access (RA-CSMA). This mechanism enables each vehicle in the collision domain, i.e., the area including vehicles potentially affecting each others' transmissions, to sense the channel status and reserve the channel for the duration of the frame. The proposed mechanism outperforms the random channel access scheme in terms of the spectral efficiency and the achievable communications range. The performances in terms of spectral efficiency and SINR obtained by the proposed mechanism are also close to those obtained by the idealistic channel access scheme, i.e., the perfect TDMA access with no inter-vehicle interference and ideal synchronization. However, the RA-CSMA mechanisms may repeat several requests before accessing the channel that results in increased latency significantly.

\subsection{Spatial Beamforming}
This section discusses spectrum sharing approaches based on beamforming design. Such an approach is typically used in coexistent radar and communication systems~\cite{ahmedoptimized}. The key idea is to project the radar signal into the null space of its channel to the communications receiver~\cite{khawar2014mimo}.
\begin{figure}[t!]
\centering
\includegraphics[width=7.5 cm, height=3.6 cm]{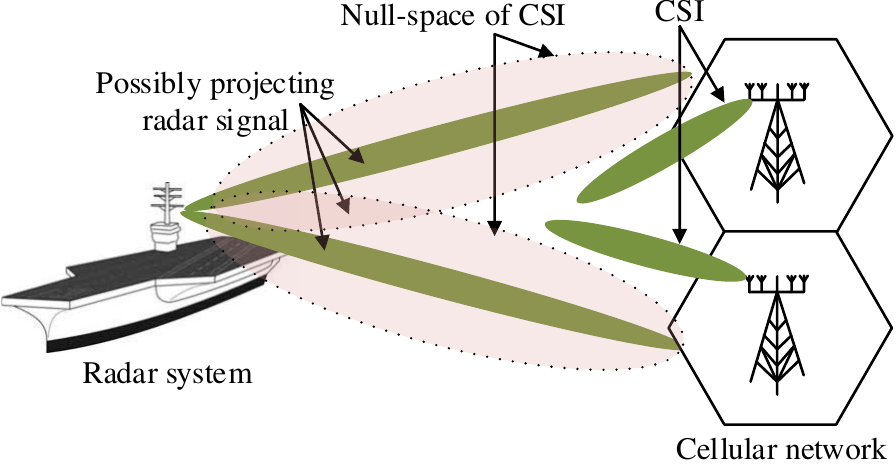}
 \caption{\small Spatial beamforming for JRC systems.}
 \label{spatial_beamforming_allocation}
\end{figure}

In particular, the authors in \cite{shajaiah2014impact} proposed a channel-selection algorithm for a radar-communication coexistence system in which a military radar system and LTE base stations (BSs) share the $3.5-3.6$ GHz bands. First, the CSI matrix of each BS interference channel, i.e., the channel shared between the BS and the radar system, is estimated by using the blind null space learning algorithm~\cite{noam2013blind}.~The problem is to determine the CSI matrix that minimizes the difference between the original radar signal and the radar signal projected onto the null space of the CSI matrix. Then, the singular value decomposition (SVD) method is adopted to determine the null space of each CSI matrix of the BS, i.e., namely projection matrix of the BS.~The radar signals are then projected on the best CSI matrix. The simulation results show that the performance, e.g., in terms of estimation accuracy, obtained by the proposed scheme is close to that obtained by the original radar signal scheme. This means that the proposed scheme is able to minimize the degradation in the radar performance. However, how the proposed scheme affects the performance of the cellular BSs is not shown. 

\begin{figure}[h!]
\centering
\includegraphics[width=6.8 cm, height=3.4 cm]{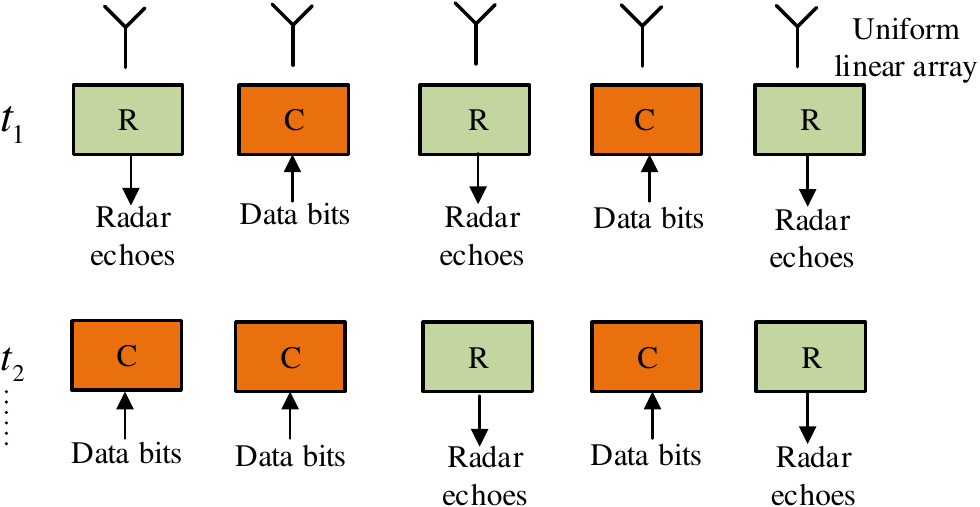}
 \caption{\small Spectrum-sharing based on antenna allocation. ``C'' and ``R'' stands for radar function and communication function, respectively. $t_i$ refers to each data transmitting time.}
 \label{antenna_allocation}
\end{figure}

Different spatial beamforming approaches can be implemented by allocating antenna elements to the radar function and communication function. For example, the authors in~\cite{ma2018novel} address the antenna allocation problem for the DFRC systems, and the authors in~\cite{wang2018sparse} and \cite{wang2018dual} consider the problem of sparse transmit array design for the DFRC systems. Here, we discuss the scheme proposed in~\cite{ma2018novel} in detail to understand how the antenna elements are assigned to the radar and communication. The system model includes a communication transmitter and a radar receiver that share the same uniform linear array (ULA) as shown in Fig.~\ref{antenna_allocation}. Each antenna element of the ULA can be connected with the communication transmitter or the radar receiver. The communication transmitter is to transmit communication signals, and the radar receiver is to receive radar echoes. At each communication transmitting time, antenna elements of the ULA are dynamically selected to transmit data bits according to the data bit stream, and the rest is used by the radar receiver. In particular, the data bit stream is divided into blocks, and each block consists of constellation bits and spatial bits. The constellation bits are modulated by traditional modulation techniques, e.g, BPSK. The spatial bits are to determine the combination of transmit antenna elements of the ULA. This is similar to the Global System for Mobile (GSM) that can increase the channel capacity and improve the spectral efficiency. Compared with the traditional MIMO system, the proposed scheme can improve the channel capacity by the number of spatial bits~\cite{wang2012generalised}. To evaluate the radar performance, the CRB of the radar resolution is used. The simulation results show that given the small number of radar receiving antennas, the CRB of the proposed scheme is much lower than that of the baseline scheme, i.e., in which the antennas are spatially partitioned into the two subsystems. This implies the efficiency of the proposed scheme. 

\subsection{Summary and Lessons Learned}

 In this section, we have discussed four spectrum sharing approaches for the JRC systems. The approaches and their references are summarized in Table~\ref{spectrum_sharing}. We observe that the approaches based on communication signals, i.e., OFDM, and radar signals, i.e., LFM, receive more attentions than the other ones. From this section, the lessons learned are as follows:
 \begin{itemize}
 \item In JRC systems, designing spectrum sharing is important to combine the radar signal and the communication signal while guaranteeing the performance requirements of both the radar function (system) and communication function (system). 
 \item For spectrum sharing in JRC systems, we can leverage any communication signals to implement the radar function in JRC systems. One common principle is that the radar receiver needs to know the communication symbols or communication signals. Then, the radar receiver can use signal detection algorithms, e.g., cross-correlation algorithms, to extract necessary information, e.g., velocity and range, about the targets from the received communication signal, i.e., the echo signal, and the known communication signal. To achieve a high spectral efficiency, the approaches reviewed in this section use the OFDM-based communication signal for the radar function. As such, communication signal-based approaches are simple. 
 \item Similarly, we can leverage a radar signal to perform the communication function in JRC systems. However, the radar signal-based approaches are more complicated than the communication signal-based approaches since the radar signal-based approaches require modulation/demodulation to embed/extract the communication data into/from the radar signals.
 \item Spectrum sharing enables the JRC systems to improve the spectrum efficiency as well as to reduce the size and cost of the hardware. However, this affects the performance of each function. For example, we can leverage the OFDM communication signals for the radar target detection based on the correlation function as discussed in Section~\ref{OFDM_waveform}. However, the correlation function of the time domain OFDM signal depends on the data bits, and the range profile may have high sidelobes that drastically reduce the detection accuracy of the radar function. Another example is that we can leverage the radar signals such as LFM radar signals to carry communication data symbols as discussed in Section~\ref{chirp_waveform}. However, the small number of waveforms used in each LFM radar signal results in the low communication data rate. Therefore, choosing which spectrum sharing approach, i.e., based on communication signal and based on radar signal, depends on the major objectives, i.e., improving communication performance or radar performance, of the JRC systems.
\item In fact, each spectrum sharing approach proposed for JRC systems has its own advantages and disadvantages. For example, the OFDM-based approaches can improve spectrum efficiency, but they incur high PAPR and require relatively costly hardware. Also, the LFM waveform-based approaches have continuous constant modulus waveform that can be generated and detected by using simplified hardware, but they have lower peak output power that limits the target detection range. Thus, understanding the advantages and disadvantages of each approach is necessary to properly design spectrum sharing schemes for the JRC systems. In fact, in addition to the spectrum sharing, the radar function and communication function share system energy supply. Thus, it is important to consider power allocation in the JRC systems that are discussed in the next section.
 \end{itemize}

\begin{table*}
\caption{Spectrum sharing approaches for JRC systems.}
\label{spectrum_sharing}
\begin{centering}
\begin{tabular}{|>{\arraybackslash}m{1.7cm}|>{\arraybackslash}m{0.4cm}|>{\arraybackslash}m{5cm}|>{\arraybackslash}m{2cm}|>{\arraybackslash}m{2.5cm}|>{\arraybackslash}m{2.5cm}|}
\hline 
 \cellcolor{mygray}  \textbf{Approaches} &  \cellcolor{mygray} \textbf{Ref.}&  \cellcolor{mygray} \textbf{Key ideas}&  \cellcolor{mygray} \textbf{Techniques} & \cellcolor{mygray} \textbf{Performance improvement} &  \cellcolor{mygray} \textbf{Shortcomings}  \tabularnewline
\hline 
\hline 
&\cite{sturm2011waveform}& Data symbols are modulated with m-sequences, and the matched filters are used at the radar and communication receivers &Spread coding and collection algorithms &Target detection with high range &Low velocity targets, high cost and complexity of receiver design \tabularnewline \cline{2-6}

 &\cite{sturm2009ofdm}& Data symbols are modulated OFDM technique by using the IFFT. The OFDM demodulation is used at the communication receiver to detect data symbols, and the correction algorithm is used at the radar receiver to detect targets.&FFT algorithm and correlation algorithm &High range resolution and spectrum efficiency&High sidelobes\tabularnewline \cline{2-6}
Communication signals&\cite{sit2011ofdm}& Same as \cite{sturm2009ofdm}, but the element-wise division and the IFFT algorithm are used at the radar receiver to estimate the target parameters.&Element-wise division, FFT/IFFT algorithms  &Sidelobe reduction and velocity performance improvement&High PAPR\tabularnewline \cline{2-6}
& \cite{liang2019design} \cite{tian2017radar} \cite{davis1999peak}& P4 code, m-sequence, or Golay code are combined with OFDM to modulate the data symbols. &  Maximal linear feedback
shift registers, cross-correlation algorithm, and DFT algorithm&High range resolution and high velocity estimation accuracy&High PAPR\tabularnewline \cline{2-6}
&\cite{lv2018joint}&OCDM is combined with OFDM to modulate data bits.&IDFnT/DFnT/FFT algorithms  &Sharp shape and low sidelobes& High modulation and demodulation complexity\tabularnewline \cline{2-6}
\hline 
& \cite{hassanien2017dual}& Data symbols are embedded into FH waveforms. Then, the matched-filtering algorithm is adopted to estimate the embedded data symbols. & Code optimization algorithm and matched filtering algorithm &Low SER, and interference resistance &Accurate CSI estimation requirement.\tabularnewline \cline{2-6}
&\cite{wang2019co}& Same as \cite{hassanien2017dual}, but the FFT algorithm is used instead of the matched-filtering algorithm.& Code optimization algorithm and FFT algorithm&Sharp shape and low sidelobes, and low BER&CSI estimation is not required. \tabularnewline \cline{2-6}
Radar signals&\cite{nusenu2018dual}&Data symbols are embedded into Costas waveforms before they are transmitted using the FDA technology.& Construction algorithm, and frequency diverse array technology &Low SER, high SINR, and robustness against interference&Phase synchronization requirement \tabularnewline \cline{2-6}

&\cite{wang2018poster}&LFM is combined with OFDM. The 2D-FFT algorithm is used at the radar receiver, and the conventional OFDM demodulation is used at the communication receiver. &2D-FFT algorithm&High-range resolution, long-range detection and high spectrum efficiency&High PAPR \tabularnewline \cline{2-6}
&\cite{li2013integrated}&Data symbols of different users are modulated with subcarriers in the LFM signal. &  BPSK modulation technique&High range resolution, and application capability in multi-user scenarios. &Low communication throughput\tabularnewline \cline{2-6}
&\cite{li2019integrated}&LFM is combined with LDPC technique. BCJR algorithm is used at the communication receiver to detect data symbols. & CPM demodulation and decoding, LDPC technique and BCJR algorithm&High-range resolution, and low BER&High modulation and demodulation latency\tabularnewline \cline{2-6}
\hline 
Time division/& \cite{kumari2015investigating}&The JRC transmitter uses the IEEE 802.11ad standard in which the preamble is used for radar function and the data blocks are used for communication function. & Autocorrelation algorithms &High radar detection rate and high data rate&Low-velocity estimation accuracy\tabularnewline \cline{2-6}
spatial beamforming & \cite{kumari2017performance}&Same as \cite{kumari2015investigating}, but the fraction of data symbols is determined to minimize the MSE bounds for the range estimation, velocity estimation, and data symbol estimation.& Standard algorithms for convex optimization problem &Minimization of the MSE bounds
for the range estimation, velocity estimation, and data symbol
estimation &Low spectral efficiency\tabularnewline \cline{2-6}
& \cite{shajaiah2014impact}&Radar system uses the blind null space learning algorithm to determine the null space of the CSI matrix of the cellular system. Then, the radar signals are projected on the null space of the CSI matrix. &Channel selection algorithm, blind null space learning algorithm, and singular value decomposition method &Minimization off interference caused by the radar system to the cellular system &CSI estimation requirement\tabularnewline \cline{2-6}
\hline 


\end{tabular}
\par\end{centering}
\end{table*}

\section{Power Allocation}
\label{power_all_sec}

Power allocation in JRC systems is mainly different from that in existing communication systems in two aspects: 1) the transmitted signals for radar and communications in DFRC systems are drawn from the same transmit power budget while in existing communication systems each transmitter has its own transmit power budget and 2) transmit power is allocated to optimize the joint performance of radar and communication in JRC systems while in existing communication systems the transmit power is allocated to optimize communication performance. 
The studies of power allocation in JRC systems can be classified into two groups based on the considered system models, i.e., power allocation over different antennas/sub-carriers in multi-antenna/multi-carrier systems. 
This section reviews the power allocation designs in JRC systems.

\subsection{Multi-antenna JRC Systems}
 
\subsubsection{Power Allocation in DFRC Systems}

\begin{figure} 
 \centering  
 \includegraphics[width=0.4\textwidth]{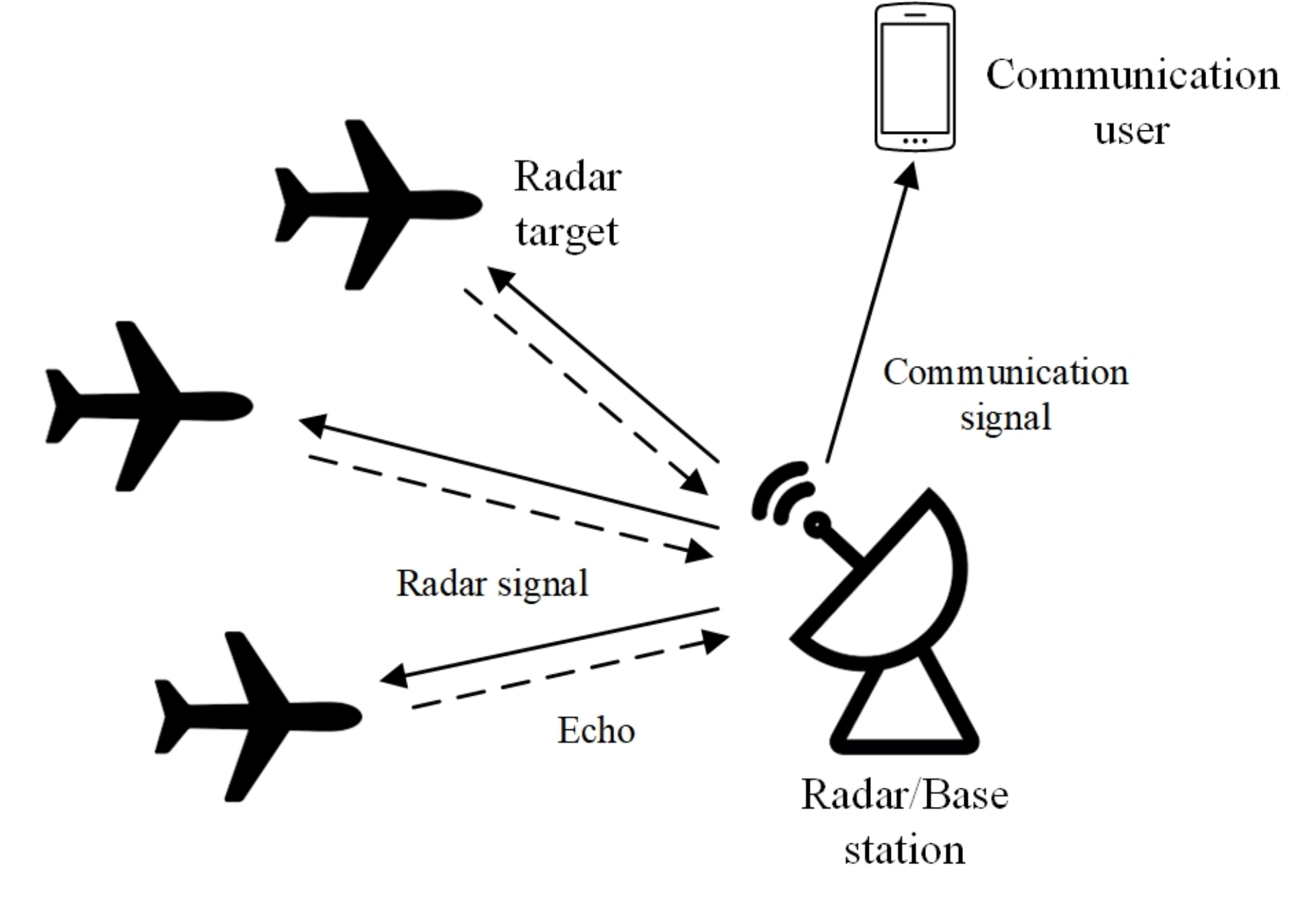}  
 \caption{System model of DFRC in \cite{F.2019Liu}.} \label{DF} 
 \end{figure}

Reference \cite{F.2019Liu}  considers a DFRC system as shown in Fig.~\ref{DF}, where a multiple-antenna BS transmits data to a downlink user while detecting multiple radar targets simultaneously over the same mmWave channel.
With the objective to minimize the sum of the communication and radar beamforming errors, the authors formulate a weighted beamformer design problem subject to the non-convex constant-modulus constraint and transmit power budget. Moreover, to obtain the solution efficiently, the authors decompose the formulated problem into three sub-problems and introduce a {\em triple alternating minimization} algorithm to solve the sub-problems with low complexity. The numerical results demonstrated that the introduced algorithm is able to achieve a near-optimal solution. The performance gap between the optimal solution and the proposed solution decreases with the decrease of SNR. However, a limitation of the proposed beamforming design is that it only works for the case when the number of data streams is larger than the number of targets.

Different from \cite{F.2019Liu} which only studies power allocation for communication phase, reference \cite{X.2020Yuan}  considers additional power allocation for training symbols during the channel estimation phase.
To minimize the channel estimation error, the authors first derive the optimal power allocation between communication and training symbols in closed-form. 
Then, based on the optimal power allocation, the authors propose three waveform designs to maximize communication MI only, radar MI only, and weighted sum of communication and radar MI. 
It is shown that the optimal power allocation scheme can significantly increase the MI of communications while causing a trivial impact on the MI of radar sensing.
Moreover, when the number of communication symbols is larger than that of the training symbols, the MI of radar sensing with the proposed power allocation exceeds the case without the power allocation.  
The simulation results demonstrate that weighted communication and radar waveform design is less affected by the channel correlation and its performance gain over the  waveform designs to maximize communication MI only
increases with the channel correlation. However, the proposed designs only work for the ideal scenario without
cochannel interference and the CSI perfectly known by the dual-function transmitter.

References \cite{F.2018Liu} and \cite{ChengchengMar2020Xu}  extend the system model in \cite{F.2019Liu} to the scenario with multiple downlink users. %
The authors in \cite{F.2018Liu} consider both separated and shared antenna allocation strategies that separate and share the transmit antennas, respectively, for mmWave radar detection and communications. With the aim to match the desired beam pattern for detecting radar targets while meeting the SINR requirements for the downlink users, beamformers have been designed based on semi-definite relaxation (SDR) under both antenna allocation strategies. Moreover, by including the SINR requirements as a penalty, the authors simplify the beamformer design problems to manifold optimizations~\cite{A.2009Absil}. Subsequently, the authors introduce a Riemannian conjugate gradient (RCG) algorithm \cite{P.2006Petersen} to solve the  optimizations with low complexity.
It is demonstrated through numerical results that the shared antenna allocation strategy significantly outperforms the separated counterpart in terms of the tradeoff between the beampattern quality and the downlink SINR. The  performance gain of the shared strategy over separated counterpart is up to 8 dB in terms  peak-sidelobe-ratio (PSLR). Moreover, the performance of the simplified manifold optimizations is comparable to that of original problems. Nevertheless, the beamformer design
is based on the assumption that the transmitted signals for
communication and radar sensing are independent. The impact
of channel correlation is not taken into account.

Reference \cite{ChengchengMar2020Xu} aims to maximize the weighted sum rate of a DFRC system with rate-splitting multiple access (RSMA) while maintaining a desirable radar beampattern under the average transmit power constraint of each antenna of the dual-function transmitter. 
As the formulated maximization problem is non-convex, the authors propose a solution based on the {\em alternating direction method of multipliers}  (ADMM)~\cite{S.2011Boyd} to find local optima with low complexity. The simulation results indicate that the RSMA-based DFRC system achieves a better tradeoff between the weighted sum rate and mean-squared error compared to the conventional space-division multiple access (SDMA)~\cite{orn1995Ottersten}. The reason is the generation of the common stream in the RSMA renders reduced interference at the transmit beampattern angles compared to SDMA.  In terms of weighted sum rate, the RSMA-based DFRC system could outperform the SDMA counterpart by up to 48$\%$.

Reference \cite{A.2018Hassanien} considers full-duplex (FD) communications in the DFRC system with multiple communication users and a single target. 
The objective is to embed the downlink transmission into the emission of radar signals while separating the uplink communication signals from the radar target returns (i.e., echo) and clutter.  To this end, the authors design a joint transmit and receive beamformer for the multi-antenna BS under the transmit power budget. The proposed design is shown to distinguish communication signals at the radar receiver 
from signal-dependent radar target returns even when they share the same spatial angle with the radar signals. 
Compared to the case with no transmit processing gain, the performance gain of the proposed transmit beamforming  is shown to be up to 8 dB in terms of achieved SINR. However,
this work assumes perfect timing synchronization between
the MIMO radar platform and the communication users. The
robustness of the beamformer design to synchronization errors
is unknown.

References \cite{P.2018Kumari} and \cite{Liuarxiv2020F} study the applications of DFRC in mmWave vehicular systems. 
Specifically, reference~\cite{P.2018Kumari} considers a mmWave vehicular DFRC system where a source vehicle uses IEEE 802.11ad waveforms to communicate with a recipient vehicle while detecting another target vehicle based on the received echo.  
To facilitate target detection, the authors introduce an analog beamforming algorithm that adopts random subsets of transmit antennas to concurrently generate coherent beams towards the recipient vehicle at an angle of departure and perturb the sidelobes of the beams for radar target detection in the angular field. 
It reveals that there exists a tradeoff between the communication rate and radar recovery rate which can be balanced by adjusting the subset size of the transmit antennas. 
The cause of the tradeoff is that a smaller subset size of the transmit antennas facilitates sidelobe perturbation while reducing communication SNR. 
By optimizing the subset size, the introduced beamforming design is shown to enable multiple-target detection with high accuracy at the cost of a minimal decrease in communication performance.
For a target distance of 30 meters, the communication rate declines less than $5\%$ while maintaining the target detection probability to be one.~However, the considered system assumes constant velocity for the target vehicle which considerably simplifies mobility estimation.
 Moreover, this work assumes that directional LoS communication links always exists, which restricts the proposed designs in more realistic NLoS  scenarios. 
The mmWave signals attenuate markedly when penetrating obstacles. For vehicular communications, the existence of LoS link may be severely hindered by mobile vehicles. Therefore, it is imperative to design solutions for mmWave vehicular networks.

To generalize the target mobility model, reference \cite{Liuarxiv2020F} considers time-varying velocity of the targets. Moreover, the system model in~\cite{P.2018Kumari} is extended in \cite{Liuarxiv2020F} to the scenarios of multi-vehicle communications and multi-vehicle  detection. The authors investigate the problem that optimally allocates transmit power among multiple beams to to minimize the tracking
errors for multiple vehicles at a roadside unit under the sum
rate requirement for the downlink transmission to multiple
vehicles and transmit power constraint. The authors prove the convexity of the minimization problem  and design an optimal power allocation scheme based on the posterior Crame\'r-Rao bound (PCRB) \cite{pcrb}. Compared with the conventional water-filling scheme, the proposed power allocation scheme is shown to achieve a better tradeoff between radar sensing and communication performance. 
Moreover, the proposed scheme achieves high estimation accuracy, e.g., root-mean-squared-error (RMSE) of the target angle can be reduced to less than 0.01$^\circ$. Nonetheless, the proposed scheme is tailored to minimize the sum-PCRB which causes SNR loss of the vehicles with high channel gains.
Moreover, similar to \cite{P.2018Kumari}, the limitation of this work is that the case with NLoS is not taken into account.


Reference \cite{A.2018HassanienSPAWC} considers  uplink communication in a half-duplex DFRC system. Specifically, the DFRC system operates as a radar during transmit mode and simultaneously receives and radar target returns and uplink transmission from a communication user in communication mode. The design objective is to separate the uplink transmission and radar target returns from the aggregated signals so as to minimize cross-interference.
To this end, the authors employ the minimum variance distortionless response (MVDR) principle~\cite{Wolfel2005M} to deeply null the radar target returns located outside the main radar beam.
It is shown through numerical evaluation that the introduced receive beamforming based on MVDR effectively mitigates the cross-interference between the reflected radar signal and the communication signal even if they are received from the same angle. The difference between the achieved SINR in the cases with and without the use of MVDR is up to 12 dB.~A shortcoming of this work is that the beamformer design is based on the  assumption that the cross-correlation of the waveforms can be ignored.
However, it is not practical to realize perfectly orthogonal waveforms. It is important to take into account the cross-correlation of the waveforms in designing the beamformers.
 
  \begin{figure} 
  \centering 
  \includegraphics[width=\linewidth]{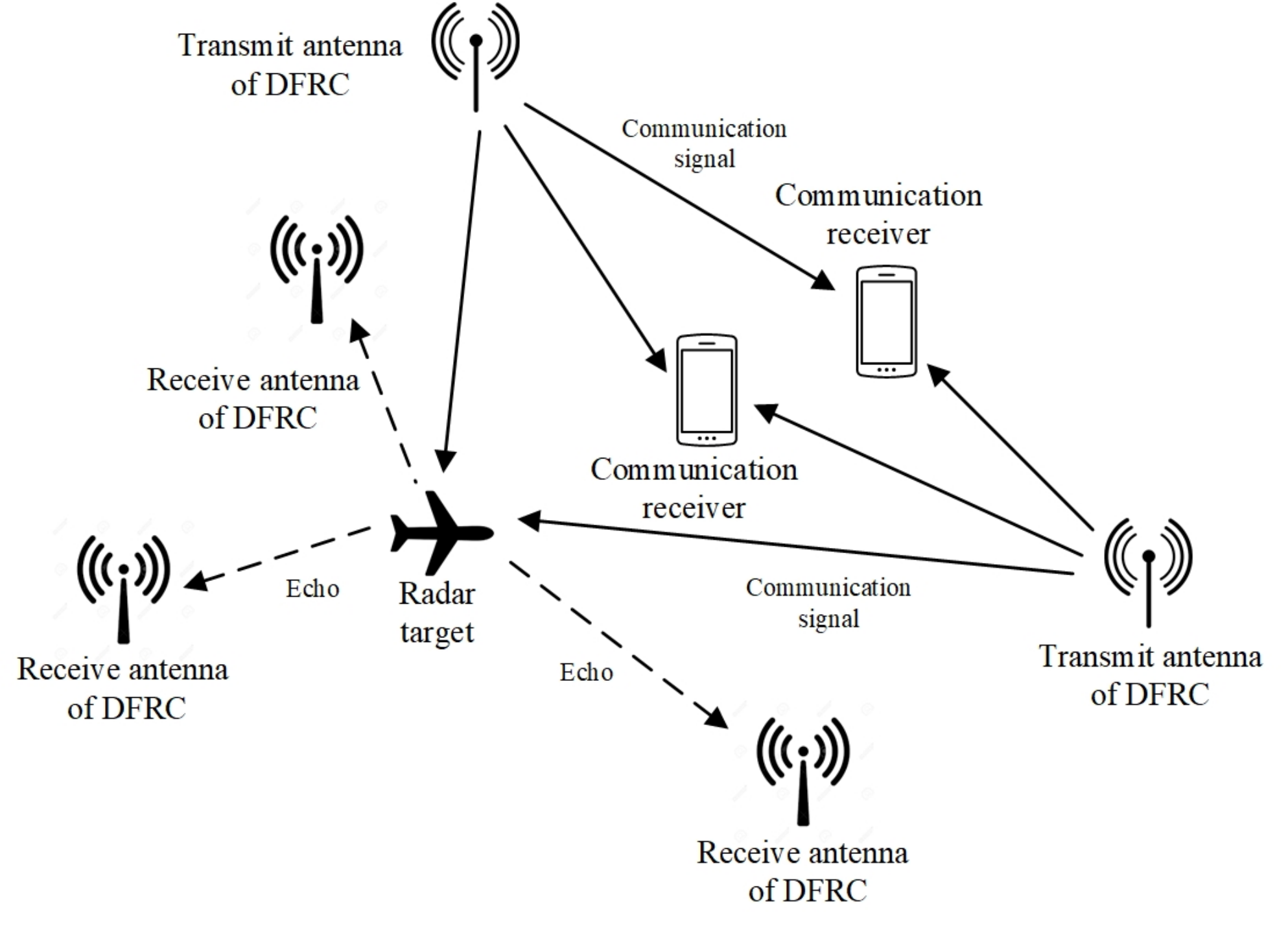}  
  \caption{System model of distributed CRC in ~\cite{A.2019Ahmed}.}  \label{DCRC} 
  \end{figure}

Different from above-reviewed works which consider co-located antennas,
the authors in~\cite{A.2019Ahmed} consider a general distributed DFRC system which concurrently serves multiple signal-antenna transmitters and detects multiple targets with spatially separated transmit and receive antennas, as shown in Fig. \ref{DCRC}. The design goal is to achieve localization accuracy and communication rate in terms of Cramer-Rao bound (CRB) and Shannon's capacity, respectively. For this, the authors formulate a problem that determines the power allocation
for distributed DFRC transmitters to minimize the mean-squared localization error, subject to the target communication rate requirements
of multiple receivers. The authors demonstrate the convexity
of the formulated problem and adopt a standard water-filling
algorithm to obtain the optimal power allocation solution. It is shown by the simulations that the proposed power allocation method allows the DFRC system to achieve much lower localization error compared to both radar-only and communication-only systems as well as comparable communication rate to the communication-only system. However, the power allocation problem is based on the assumptions that 1) the signals from the target can be ignored at the communication receiver, and 2) estimates of the target position and RCS for the next cycle are available.
Both assumptions are hard to realize in practice, which largely limits the applicability of the proposed solution.

   \begin{figure}
   \centering  
   \includegraphics[width=\linewidth]{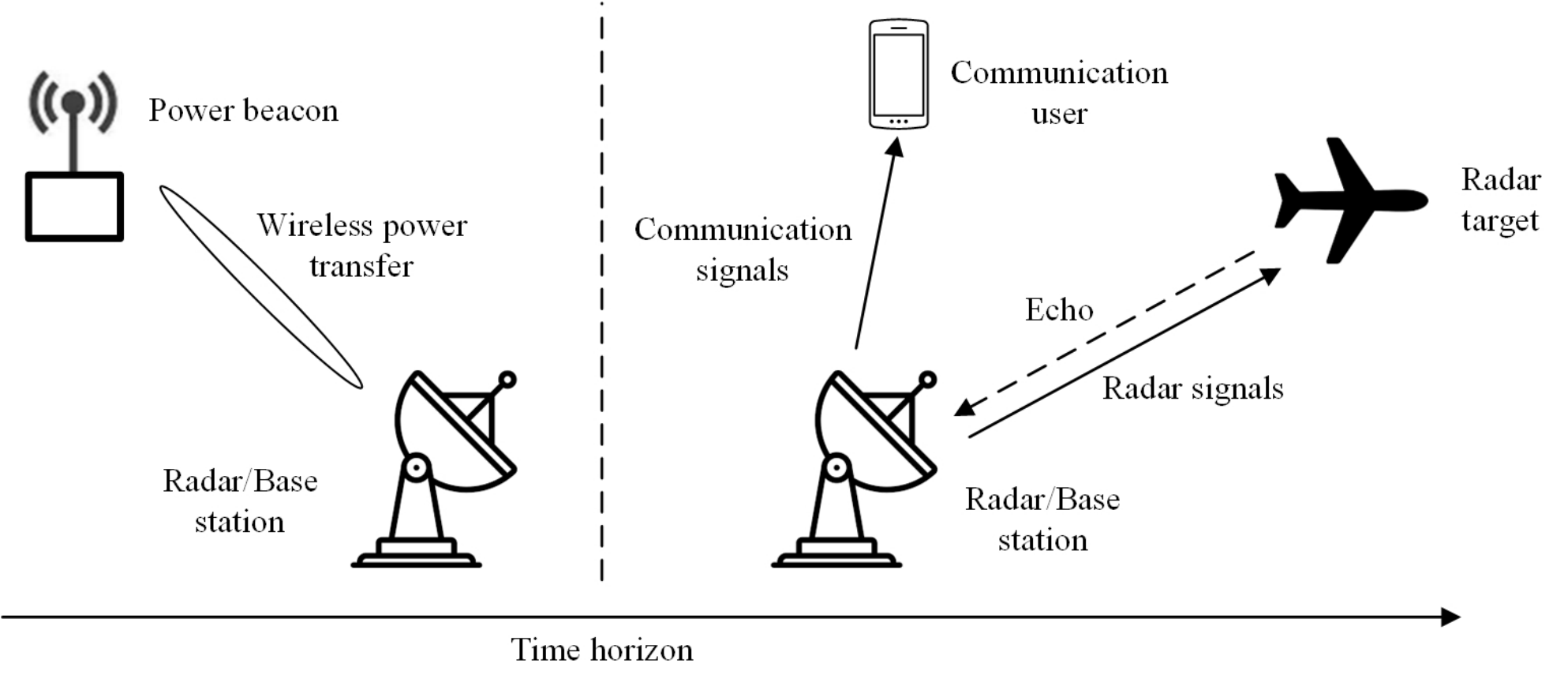}  
   \caption{System model of wireless-powered dual function radar/communications.}   \label{WPDF}
  \end{figure}

The above-reviewed works all consider DFRC systems with internal power supply. Differently, reference  \cite{ZhouY2018} considers a DFRC transmitter powered by RF energy harvesting~\cite{X.SecondQuarter2015Lu}.  
Specifically, the DFRC transmitter first harvests RF energy from a wireless power beacon and then functions as a JRC base station, as shown in Fig~\ref{WPDF}. The objective is to minimize the transmit power of the power beacon under the constraints of radar and communication performance. To address this issue, the authors propose a semi-definite relaxation and auxiliary variable method to jointly optimize the energy beamforming vector, energy transfer time and the transmit power of the DFRC transmitter. It is analytically proven that the optimal solution always exists and is rank one. 
Compared with an equal power allocation scheme that assigns the same transmit power to each sub-channel, the proposed scheme is shown to consume less than half energy to achieve the same data information rate and MI.~However, the authors do not consider the circuit power consumption of the DFRC transmitter which overestimates the achieved performance. 

Different from the above-reviewed references which do not take into account communication security, reference \cite{C2017Shi} considers the scenario where the radar receiver could be malicious attackers to intercept the information transmitted simultaneously with the radar waveform. To maintain a low probability of interception (LPI) \cite{A.G.2004Stove} at the radar receiver, the authors formulate an LPI-based power allocation problem to minimize the transmit power for communication signal and radar waveform  subject to the probability of false alarm and probability of detection requirements at the radar receiver. Moreover, a bisection search-based approach is employed to attain the optimal solution of the formulated problem.~The simulations show that the probability of detection is affected by the transmit power of the dual-function transmitter, probability of false alarm, and SNR of signal paths, and the communication rate. However, a drawback of the proposed design is that it is based on the assumption that the exact information (i.e., position and speed) of the target is known. Such information greatly affects the probabilities of target detection and false alarm at the radar receiver.

  \begin{figure}
   \centering  
   \includegraphics[width=0.8\linewidth]{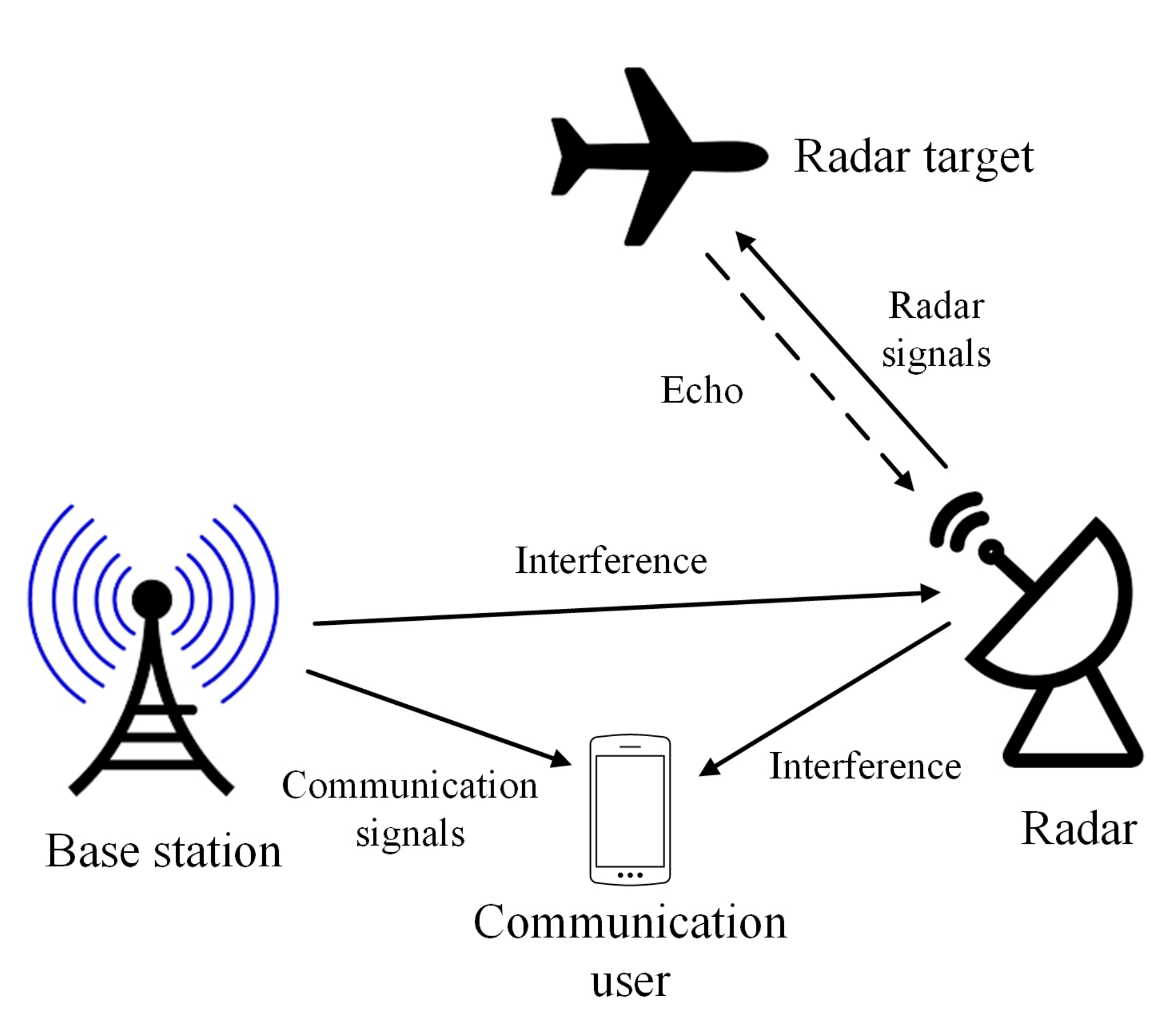}
  \caption{System model of CRC.}   
   \label{CRC}
 \end{figure} 
    
\subsubsection{Power Allocation in CRC Systems}    
    
Another group of works focus on power allocation issues in CRC systems where the communication subsystem operates on the same frequency band with the radar subsystem, that significantly improves the spectrum efficiency.  
Reference \cite{B.2017Li} 
considers a CRC system with a single communication user and radar, as shown in Fig.~\ref{CRC}. With the objective to maximize the radar SINR with respect to a single target subject to the downlink communication rate and power constraints, the authors in \cite{B.2017Li} formulate a problem that determines the transmit covariance matrix for communication, the transmit precoder for radar, and the radar subsampling to maximize the radar SINR with respect to the downlink communication rate and power constraints. The problem is then solved by a sequential convex programming technique. The simulation results show that the JRC system with the proposed design outperforms traditional radars that do not coordinate communication systems in RCS estimation accuracy and saves up to $60\%$ of data samples required for the radar estimation. Moreover, compared with null space projection precoding which emits the radar waveforms towards the orthogonal directions of the communication receiver to avoid interference, the proposed precoding scheme achieves 77.6 dB higher radar SINR.~However, the  considered system relies on a control center that has instant CSI of all the channels to implement the proposed design. The real-world operation of such a system is greatly affected by CSI estimation errors, synchronization errors, communication overhead and delay. 



References~\cite{J.2018LiuGlobalSIP}, \cite{F.WCLLiu2017} and~\cite{F.Liu2017} extend the system model in  \cite{B.2017Li} to the scenarios with multiple downlink users.
The objective of \cite{J.2018LiuGlobalSIP} is to maximize the weighted MI rate of radar and communication users by configuring the transmit covariance matrices under the overall transmit power budgets. As the configuration problem involves non-convex optimization, the authors propose an alternating optimization-based iterative approach based on the principle of Gauss-Seidel iteration~\cite{D.2006P}.
The proposed approach is proven to achieve at least a local optimum solution. The numerical simulations demonstrate the local convergence of the proposed approach. However, the performance gap between the global optimal solution and the resulting solution is unknown. 
Moreover, transmit covariance matrices are devised based on the assumption that 
perfect CSI is known to all radar and communication transmitter.

Differently, reference \cite{F.WCLLiu2017} focuses on the design of transmit beamforming taken into account the impact of CSI inaccuracy. In particular, the authors formulate a non-convex problem to maximize the detection probability of the radar constrained by the communication rate and transmit power constraints with CSI quantization errors. 
Due to the non-convexity of the problem, the authors instead optimize its 
upper bound and norm bound in the case with imperfect CSI. 
The authors transform the proposed problem into a semi-definite program and then obtain the solution based on the standard semi-definite relaxation techniques.
Compared with zero-forcing and MMSE beamforming methods, even the upper bound minimization based on the proposed method is shown to achieve
higher average detection probability. However, the detection performance in this work can only be deemed as the upper bound as the radar waveforms are assumed to be orthogonal to the communication signals which cannot be perfectly achieved.  
 


Reference \cite{F.Liu2017} extends the system models in \cite{B.2017Li} and \cite{F.WCLLiu2017} to the cases with multiple radar target detection under both perfect and imperfect CSI.  
Considering the case with perfect CSI, the authors first introduce two transmit beamforming designs based on convex optimizations. 
One beamforming design is based on the BS transmit power minimization subject to the SINR requirements at downlink users and interference level to the radar while the other is based on interference minimization at the radar under the maximum transmit power and minimal SINR constraints of the users. To solve the formulated problem with perfect CSI, the authors devise a gradient projection method based on the Armijo rule \cite{S1999Wright}.  Moreover, in the case with imperfect CSI, the authors develop a worst-case robust beamforming design based on the S-procedure \cite{S.2004Boyd}. 
The simulation results show that, compared with the SDR-based beamformers, proposed beamformers are shown to consume less than half of the energy when QPSK modulation is adopted and up to 4 dB radar SNR gain in the case with perfect CSI. Moreover, in the case with imperfect CSI, the proposed robust beamformer exhibits higher tolerance for CSI errors and saves up to 1 dB average transmit power compared with the SDR-based counterpart.

  \begin{table*}   \scriptsize
  \caption {Beamforming Designs for DFRC Systems}
  \begin{center}
  \begin{tabular}{ p{12mm} p{15mm} p{45mm} p{20mm} p{10mm}  p{15mm}  p{17mm} }  
   \hline
   Reference &  System & Objective & Implementation & CSI & Frequency  & Communication $\&$ radar channels \\
       \hline  
       \cite{F.2019Liu}  & MT-SU-DFRC  &  Minimization of the weighted sum of the communication and radar beamforming
       errors subject to the constant-modulus and power constraints of  the beamformers  &   Transmitter  &  Perfect   & mmWave & MIMO $\&$ MIMO  \\
   \hline
  \cite{X.2020Yuan} & MT-SU-DFRC  &  Maximization of MI of communications subject to the total power constraint & Transmitter  & Perfect & Microwave &  MIMO $\&$ MIMO \\
   \hline
  \cite{ChengchengMar2020Xu}   & ST-MU-DFRC &  Maximization of weight sum rate subject to the radar beampattern and transmit power constraints  &  Transmitter & Perfect & Microwave &   MISO $\&$ MIMO  \\
   \hline
    \cite{F.2018Liu}  &  MT-MU-DFRC &  Minimization of detection error subject to the SINR requirements for the downlink users & Transmitter  &  Perfect  &  Microwave  &  MISO $\&$ MIMO \\
    \hline 
   \cite{A.2018Hassanien} &  ST-MU-DFRC &  Separating the
   uplink transmission from the radar target returns  & Transmitter $\&$ receiver  & Perfect   & Microwave & MISO $\&$ MIMO  \\
     \hline  
  \cite{P.2018Kumari} & MT-SU-DFRC  & Balance the tradeoff between radar detection and communication rate  performance  & Transmitter $\&$ receiver &  Perfect   & mmWave &  MIMO $\&$ MIMO  \\
     \hline
    \cite{Liuarxiv2020F} &  MT-MU-DFRC &  Balance the tradeoff between radar detection and communication rate  performance &  Transmitter $\&$ receiver  &  Perfect   &  mmWave &  MIMO $\&$ MIMO   \\
     \hline
       \cite{A.2018HassanienSPAWC} & MT-SU-DRFC  & Uplink communication signals separation from the accumulated received signals & Receiver  & Perfect  & Microwave & MIMO $\&$ MIMO \\
     \hline 
        \cite{ZhouY2018}  & ST-SU-DFRC & Minimization of the transmit power of the power beacon  & Power beacon & Perfect   & Microwave & SISO $\&$ SISO \\
     \hline  
      \cite{A.2019Ahmed} & MT-MU-DFRC &   Mean squared localization error minimization subject to the communication rate constraints of multiple receivers & Transmitter  &  Perfect  & Microwave & MISO $\&$ MIMO  \\
     \hline   
        \cite{C2017Shi}   & ST-SU-DFRC  &  Minimization of the total transmit power subject to the information rate and covertness requirements & Transmitter & Perfect & Microwave &  SISO $\&$   SISO  \\
              \hline  
   \end{tabular}
   \end{center} \label{Compare} 
   \end{table*}

      \begin{table*}   \scriptsize
      \caption {Beamforming Designs for CRC Systems}
     \begin{center}
     \begin{tabular}{  l p{12mm} p{45mm} p{20mm} p{16mm}  p{15mm}  p{17mm} }  
     \hline
     Reference &  System & Objective & Implementation & CSI & Frequency  & Communication $\&$ radar channels \\
      \hline 
      \cite{B.2017Li} & ST-SU-CRC  &    Radar SINR maximization subject to the communication rate and power constraints &  Transmitter &  Perfect & Microwave & MIMO \& MIMO \\
         \hline
      \cite{J.2018LiuGlobalSIP}   & ST-MU-CRC  &  Weighted system MI rate maximization under the transmit power budgets & Transmitter & Perfect & Microwave & MISO $\&$ MIMO \\
            \hline       
       \cite{F.WCLLiu2017}  &  ST-MU-CRC   &  Radar detection probability maximization  subject to the communication rate and transmit power constraints     & Transmitter &  Perfect $\&$ Imperfect  &  Microwave & MISO $\&$ MIMO \\
    \hline        
        \cite{F.Liu2017}  & ST-MU-CRC   & 1) Communication transmit power minimization subject to the SINR requirements at downlink users and interference level at radar receiver; 2) Interference minimization at the radar receiver subject to the transmit power budget and minimal SINR constraints of the communication users  & Transmitter  &    Imperfect  & Microwave &   MISO $\&$ MIMO \\
         \hline   
   \end{tabular}
   \end{center} \label{Compare2} 
   \end{table*}

 \subsection{Multicarrier Communications}

Multi-carrier communication relies on orthogonal subcarrier waveforms with a lower data rate to suppress inter-symbol interference (ISI). Compared to single carrier communication, ISI can be effectively mitigated at the cost of reduced spectral efficiency~\cite{Chen2004S}. Due to the practicality, frequency diversity, waveform diversity and ease of implementation  \cite{S.2011Sen},  multicarrier waveforms have attracted increasing attention in communication-only and radar-only systems and have been adopted in many systems, such as LTE, LTE-Advanced and WiMAX systems~\cite{K.2008Fazel}. However, there has been a very limited amount of research on the power allocation designs in multicarrier JRC systems.

References \cite{M.2016Bica}, \cite{T.2019Tian}, and \cite{F2019WangTSP} consider power allocation problems in multi-carrier CRC systems.
Focusing on non-overlapping subcarrier allocation,  
the authors in \cite{M.2016Bica} design a power allocation scheme for each subcarrier to maximize the MI rate of radar under communication rate and transmit power constraints.  
Based on the Lagrange multipliers, the authors derive an optimal power allocation solution in closed-form. 
An interesting finding is that a higher maximizing MI rate does not ensure optimal radar performance. Moreover, it is shown that communication signals scattered off the target largely affect the radar performance, especially when the radar target returns are weak. The proposed scheme incurs heavy communication overhead as it assumes that the transmitted communication signal is perfectly known by the radar receiver. Thus, signaling between the communication transmitter and radar receiver is inevitable.

References~\cite{T.2019Tian} and~\cite{F2019WangTSP} consider both overlapping and non-overlapping subcarrier allocation 
between radar and communications, respectively. 
In~\cite{T.2019Tian}, the authors first introduce a power allocation solution for independent radar MI and communication rate maximization problems based on the non-overlapping subcarrier allocation scheme. The optimal solution is derived in closed-from by using the Karush-Kuhn-Tucker (KKT) condition. Then, the authors formulate a joint radar MI and communication rate maximization problem based on the  overlapping subcarrier allocation scheme and propose a sequential optimization algorithm to obtain the optimal transmit powers for communication and radar iteratively.  It is validated through simulations that the   overlapping subcarrier allocation scheme brings significant gains in radar MI and communication rate compared to the orthogonal counterpart at the cost of higher complexity. However, this work assumes that in the case of overlapping  subcarrier allocation, the number of shared subcarriers is fixed. An interesting extension is to optimize the number of  subcarriers to be shared.

Similar to \cite{T.2019Tian}, reference~\cite{F2019WangTSP} also considers both overlapping and non-overlapping subcarrier allocation schemes, however, targets on 
 communication throughput maximization.
Specifically, the authors formulate communication throughput maximization problems under the SINR target for the radar and transmit power constraints. 
Under the non-overlapping subcarrier allocation scheme, the authors show the non-convexity of the formulated problem and propose a sequential convex programming method based on alternating direction~\cite{Fukushima1992M}. Moreover, under the orthogonal subcarrier allocation scheme, the formulated problem involves mixed-integer nonlinear optimization, the optimum and suboptimum of which can be solved by a branch and bound algorithm and a penalized sequential convex programming approach, respectively. The proposed methods are shown to offer superior performance compared with the penalized
sequential convex programming method \cite{P-SCP} and a branch-and-bound method \cite{S.2011Boyd} under both overlapping and non-overlapping subcarrier allocation schemes, especially when the radar SINR is low or when the communication transmit power is high. 
Nevertheless, this work only studies the direct interference across the communication and
radar subsystems. The impact of secondary interferences such as clutter is also worth investigating.

Different from the system models in  \cite{M.2016Bica,T.2019Tian,F2019WangTSP},  another group of works~\cite{Bica2019M} and \cite{Ahmed2019} consider DFRC systems where the radar and communication subsystems are allocated with non-overlapping subcarriers. 
In~\cite{Bica2019M}, the radar subsystem aims to maximize the MI between the radar target returns  and the impulse response of a target while the communication subsystem aims to maximize that between the transmitted and the received communication signals at the communication receiver.
Through joint optimizing power allocation, the authors design a selfish radar scheme that prioritizes radar performance and a cooperative scheme that maximizes the weighted MI rate. Both schemes yield optimal solutions in Lagrangian forms based on the KKT conditions.
The simulation results suggest that the selfish radar scheme and the cooperative scheme render better radar performance and communication performance, respectively. However, spectrum sharing between the subsystems is not investigated in this work.

Reference~\cite{Ahmed2019} extends the system model of \cite{Bica2019M} to the case with multiple communication receivers. The authors target maximizing the MI between the target response and the transmit waveform for the radar subsystem. 
Power and subcarrier allocation schemes have been developed for both the radar-centric design and cooperative design which does not and does guarantee target communication performance, respectively.
The simulations result show that, compared with the radar-centric design, the cooperative design can considerably improve  the communication MI and the radar MI at the cost of a small decrease in the radar MI, e.g., a 37.7$\%$ increase of communication MI at the cost of a 5$\%$ decrease in radar MI. However, similar to \cite{Bica2019M}, the frequency reuse between communication and radar subsystems is not investigated.

The goal of \cite{C.2019Shi} is to minimize the total transmit power of the system through joint optimization of subcarrier and power allocation under the radar MI and communication rate requirements. 
This design goal results in a mixed-integer nonlinear program shown to be non-convex. The authors propose a three-step approach which sequentially solves the subproblems of subcarrier assignment, power allocation for radar and power allocation for communication based on the waterfilling operation and bisection search. Compared with the existing subcarrier and power allocation approaches in \cite{M.May2019Bica} and \cite{C.May2019Kai}, the proposed approach is shown to be superior in saving the total transmit power. 
The performance gain of the proposed approach increases with higher MI requirements for radar and/or communication. 
However, this work does not consider the impact of cross-interference between the radar and communication subsystems by assuming the use of orthogonal channels. Moreover, the performance gain between the proposed approach and the optimal solution remains unknown.

   \begin{table*}  
   \caption {Multicarrier Power Allocation Designs}
  \begin{center} \label{MC_PA}
  \begin{tabular}{  p{12mm} p{20mm} p{20mm} p{20mm}  p{65mm}       }  
  \hline
  Reference &  System model  &  Subcarrier \hspace{2mm}  allocation     & Communication channels & Design objective       \\
  \hline
  \cite{M.2016Bica} &  ST-MU-CRC  & Orthogonal  & SISO   & Radar MI rate maximization under communication rate and transmit power constraints           \\
  \hline  
   \cite{T.2019Tian} & ST-SU-CRC  &  Orthogonal and non-orthogonal  & SISO& Radar MI rate maximization under communication rate and transmit power constraints        \\
    \hline 
  \cite{F2019WangTSP}  & ST-SU-CRC & Orthogonal and non-orthogonal & SISO& Communication throughput maximization subject to the SINR target for the radar and transmit power constraints.         \\
  \hline 
  \cite{Bica2019M} & ST-SU-DFRC  &   Orthogonal &   SISO &  The MI maximization between the radar target returns and the impulse response and that between the transmitted and the received communication signals   \\
 \hline 
  \cite{Ahmed2019} & ST-MU-DFRC &  Orthogonal  & SISO &  MI maximization for radar subject to communication and transmit power constraints    \\
 \hline
  \cite{C.2019Shi} & ST-SU-DFRC &  Orthogonal   &  SISO & Transmit power allocation subject to the radar MI target and the communication rate target. 
    \\
 \hline
  \end{tabular}
  \end{center}  
  \end{table*}


\subsection{Summary and Lessons Learned}
\label{sum_lesson_power}
  
 In this section, we have presented power allocation approaches for the JRC systems. The approaches and their references are summarized in Tables~\ref{Compare}, \ref{Compare2}, and \ref{MC_PA}. In particular,
in Subsection IV-A, we have discussed power allocation approaches for multiple antenna JRC systems with both DFRC and CRC configurations. We summarize the key lessons learned regarding power allocation in Subsection IV-A as follows.

\begin{itemize}
 \item  Rate distortion-based minimum mean-squared error is a commonly used metric to characterize the tradeoff between communication throughput and estimation accuracy of radar's parameters.   
 \item Power allocation in RSMA systems allows achieving both higher spectrum and energy-efficiency than SDMA and NOMA with both perfect CSI and imperfect CSI at the sender-side~\cite{ChengchengMar2020Xu}. The reason is that rate splitting has the potential to decode partial interference through the use of common and private streams.  
 \item  The ratio of mean path loss of communication to that of radar plays a critical role in the range of optimal MI that can be achieved through power allocation. A larger ratio usually indicates a greater range of achievable MI.
 \item DFRC systems with distributed antennas outperform their counterparts with collocated antennas by further exploiting spatial diversity. 
 \item In a real-world system, the environmental clutter can be estimated in the absence of radar targets. The negative effects of clutter can be mitigated by precoding techniques (e.g.,~\cite{S1999Wright}). 
 \end{itemize}

In Subsection~IV-B, we have reviewed power allocation approaches for multi-carrier JRC systems. The key lessons learned are summarized as follows. 
 
 \begin{itemize}
 \item The use of broadband OFDM carriers in JRC systems can increase the frequency diversity of both subsystems by mitigating the fading effects and resolving the multipath reflections.
 \item Typically, multi-carrier waveforms cannot achieve a constant envelope as each sub-carrier has a time-variant envelope. This causes the PAPR problem which requires a linear operation region for power amplifiers. As the power amplifiers in practical systems have a limited linear region, it is difficult to meet the requirements of JRC systems that demand for a high value of PAPR.   
 Therefore, the PAPR problem is an important problem to be dealt with in the design of power allocation schemes for multi-carrier JRC systems that employ orthogonal carriers.   
\item In multi-carrier JRC systems, a greater maximized MI usually does not indicate an optimal communication and/or detection performance. The reason is that ensuing the maximized communication MI through optimal power allocation incurs a loss in the radar MI~\cite{F2019WangTSP}.
 \item OFDM carriers can be contaminated by secondary interference, i.e., clutter.
 The power allocation problems for multi-carrier systems can be remarkably complicated as the clutter is signal-dependent. Doppler-robust power allocation should be investigated in multi-carrier JRC systems  in the presence of clutter.
  \end{itemize}

 \section{Interference Management}
 \label{interference_man}
 Due to spectrum sharing,  the radar and communication subsystems are mutually impaired by the interference imposed by each other.
 Interference management performs a pivotal role in the performance of both communication and radar subsystems. Different from interference management in existing wireless communication systems which mainly focus on mitigating the interference among communication transmissions,  JRC systems target mitigating the cross-interference for both radar and communication subsystems.
  This section reviews interference cancellation approaches for JRC systems. 
 
 \subsection{CRC systems}
 \label{sec:crc_IM}
 
 From the perspective of the availability of system information, the designs of CRC systems can be classified into three types: {\em non-cooperative}, {\em cooperative} and {\em co-designed}. With non-cooperative and cooperative types of systems, interference management schemes are devised with and without the information exchange between the subsystems. Moreover, the co-designed  systems jointly configure the subsystems, e.g., in waveforms and transmit power.
  From the perspective of the design objective, the research of interference management can be sorted into two categories: radar interference cancellation at the communication receiver and communication interference cancellation at the radar receiver

 At a radar receiver, other than the radar target returns, there are unwanted signals reflected from other objects, such as buildings and trees. These unwanted signals are referred to as \textit{clutters}, which are deemed as interference  to be removed at the radar receiver. 
 References \cite{B.2016LiTSP} and ~\cite{B.ICASSPLi2015} target on effective interference power (EIP) minimization problems at the radar in CRC systems. 
  Subject to the constraints of preserving a communication capacity target under the transmit power constraint, the authors in \cite{B.2016LiTSP} first propose a cooperative design that adopts a fixed radar sampling scheme and a Lagrangian
  dual decomposition-based algorithm to optimize the precoding  matrix  of the communication subsystem.  Then, the authors introduce a joint design of radar sampling and communication precoding matrix based on alternating optimization. 
   The simulation results indicate that the second design obtains a significantly lower EIP and recovery error than those of the first one especially when the number of radar transmit and receive antennas is large.~However, the proposed design assumes perfect orthogonality of transmit waveforms and requires perfect synchronization between the radar and communication subsystems in terms of sampling times which are only ideal to be realized in real-world systems.

 Compared with \cite{B.2016LiTSP},   reference~\cite{B.ICASSPLi2015}   
   additionally considers non-cooperative and codesigned CRC system. Under the constraints of transmit power budget and target communication capacity, the authors  investigate the designs of the radar sampling and communication transmit covariance matrices to minimize EIP.
  In cooperative and non-cooperative CRC systems, the design problem is shown to be convex and solved by the interior point method. In co-designed CRC systems, the authors demonstrate the non-convexity of the problem and propose an alternating algorithm that searches only for the optimum sampling among matrices which are row-permutation and column-permutation of the original sampling matrix. 
  Compared to the cooperative and non-cooperative design, the co-design is shown to result in lower EIP by at least 20$\%$. Moreover, the performance gain of the co-design in terms of recovery error increases with the radar sampling rate. Nevertheless, the performance gap between the proposed low-complex  solution for the co-design and the optimal solution is not investigated.

  Unlike the works in \cite{B.2016LiTSP} and \cite{B.ICASSPLi2015} that focus on minimizing EIP at radar, references  \cite{L.2017ZhengJSSP}, \cite{Y.2019LiarXiv}, and \cite{B2019Jain} target on interference removal at communication receiver.
 The authors in \cite{L.2017ZhengJSSP} design two algorithms for joint interference removal and data demodulation in a non-cooperative CRC system composed of one communication subsystem and multiple radar subsystems. The first one is based on the on-grid
    compressed sensing technique which exploits the  sparse representation of radar signals and the sparsity
     of the demodulation error. 
    The second algorithm is a compressed sensing-based technique that forces an atomic norm constraint. As the second algorithm involves high computation complexity, the authors also implement a fast method for the second algorithm based on the non-convex factorization \cite{S.2003Burer}. It is shown that both algorithms outperform the original demodulation and the second algorithm achieves a lower SER. A limitation of this work is that the radar target returns in the direction of the communication receiver are neglected.

    Reference \cite{Y.2019LiarXiv} extends \cite{L.2017ZhengJSSP} by explicitly accounting for both radar target returns and secondary interference at the communication receiver.
  The authors introduce two algorithms to facilitate reliable communication data demodulation. The first one utilizes a sparse representation of the interference and estimates the radar signals through a convex optimization with relaxations. The second algorithm 
   performs radar signals estimation and the communication demodulation error estimation via two-stage processing. In particular, a local optimum is first obtained by an alternating optimization approach. Then, 
   the global optimum  can be inferred in a higher-dimensional space based on a signed shift truncation from the local optimum. 
   Simulation results reveal that, with QPSK, the symbol error rate and computation time obtained by the two-stage processing algorithm are more than one order of magnitude and two orders of magnitude lower than those obtained by the convex relation algorithm.
   However, it is noted that \cite{Y.2019LiarXiv} considers only a single radar, while \cite{L.2017ZhengJSSP} considers multiple radars. Besides, this work only considers the case when communication and radar signals fully overlap the same bandwidth. Partial bandwidth reuse between radar and communication subsystems is worth to be explored for potential performance enhancement.

   \begin{figure*}[]
	\centering
	\begin{minipage}[t]{0.4\textwidth}
	 \includegraphics[width=1.1\textwidth]{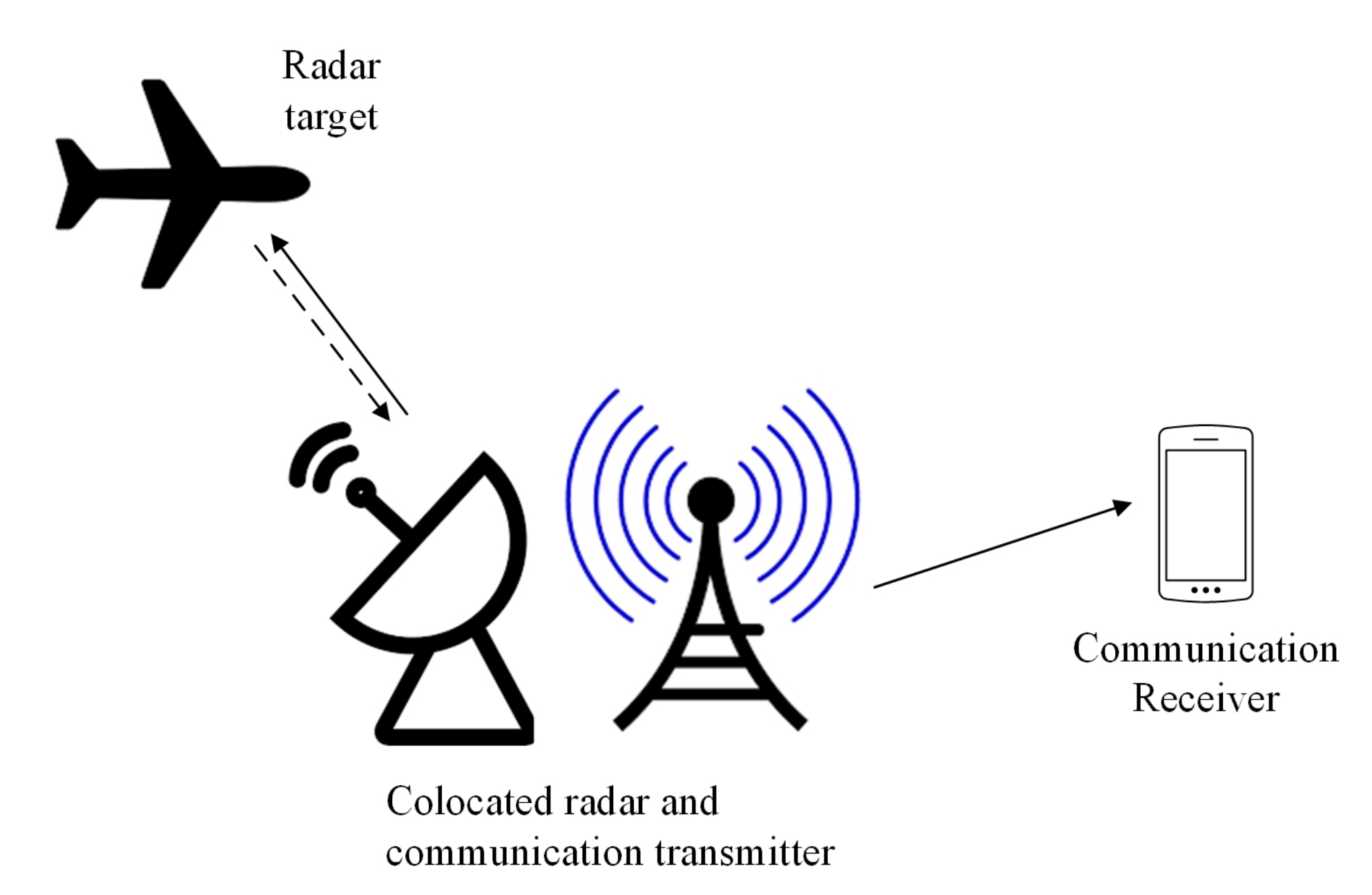}
	\subcaption{Colocated radar and communication transmitter}
	\label{subfig:reward}
	\end{minipage}
	\begin{minipage}[t]{0.4\textwidth}
		\includegraphics[width=1.1\textwidth]{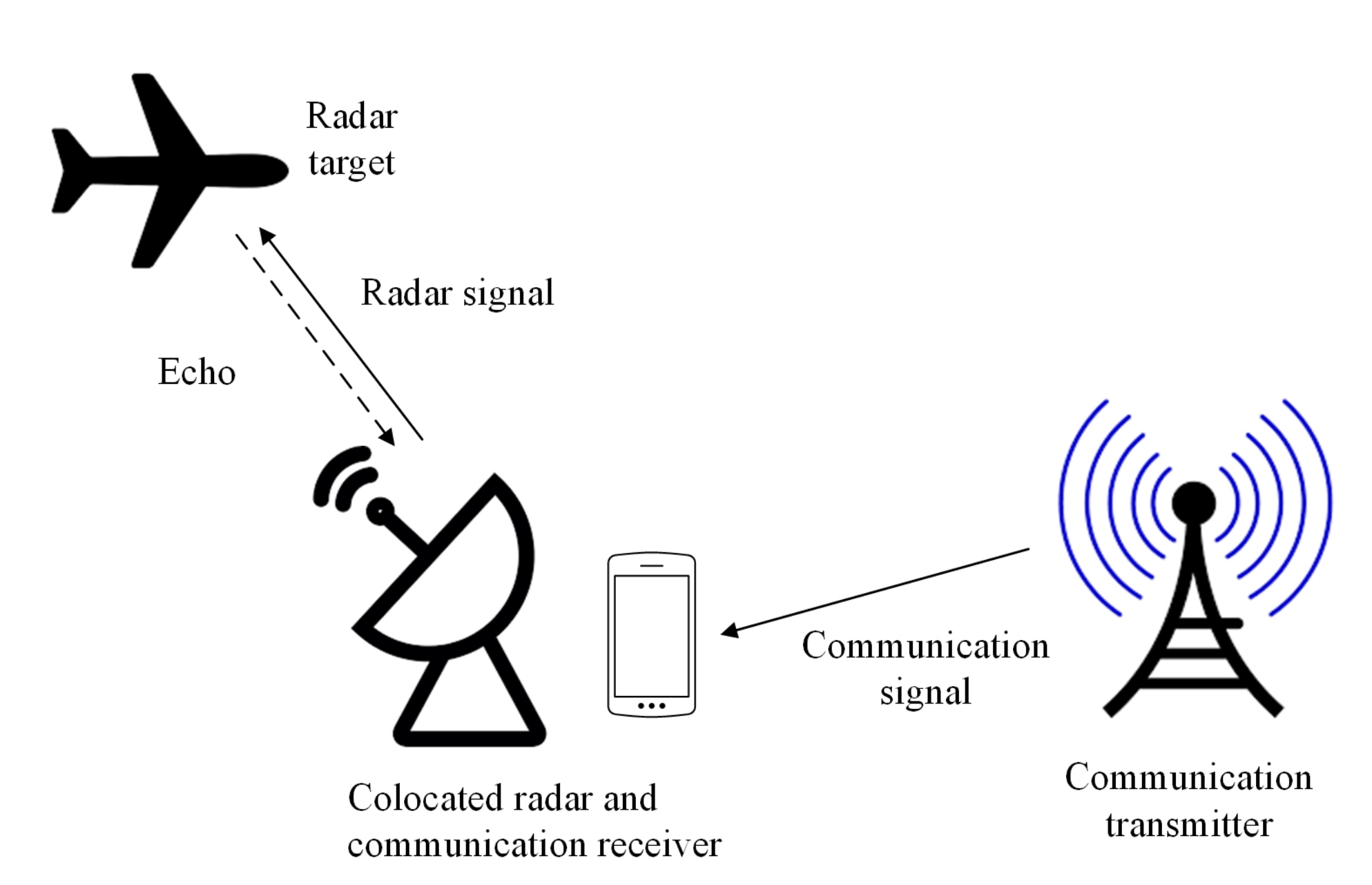}
		\subcaption{Colocated radar and communication receiver }
		\label{subfig:miss-detection-vs-speed}
	\end{minipage}
	\caption{System model of mmWave CRC in \cite{B2019Jain}.}
   \label{fig:colocated}
\end{figure*}

Different from the above works which consider microwave communication, reference \cite{B2019Jain} considers mmWave CRC systems with colocated radar and communication transmitter/receivers as shown in Fig~\ref{fig:colocated}. Assuming perfect CSI between  radar and communication, the authors aim to mitigate the cross-interference between radar and communication. In particular, the authors develop a two-stage beamformer which first subtracts the mmWave signals imposed onto the radar and then mitigates the radar's signals at the communication receiver. The gap between the system performance with and without the proposed designs is shown to widen when the transmit power of the communication transmitter is greater than that of the radar. The SIR gain at radar receiver is up to 50 dB.~However,  this work only studies a special scenario where LoS communication links exist. In mmWave systems, blockages significantly affect the propagation behavior of mmWave signals. Considering the cases with non-LoS communication links is essential in the beamformer design. Besides, this work considers a fixed transmit power for both communication and radar which largely simplifies the beamformer design. The transmit power and the beamformer can be jointly optimized for better performance.

  
     \begin{table*}  
     \caption {Summarizations of self-interference cancellation approaches for CRC systems}
    \begin{center} \label{IC_CRC}
    \begin{tabular}{  p{12mm} p{20mm} p{25mm} p{20mm} p{12mm}  p{50mm}       }  
    \hline
    Reference &  System  Model & Design type  & Communication channels & Radar channels & Design objective     \\
    \hline
    \cite{B.2016LiTSP}  & ST-SU-CRC & Cooperative $\&$ co-designed & MIMO & MIMO &  Radar EIP minimization subject to the communication capacity and transmit power constraints \\ 
            \hline   
     \cite{B.ICASSPLi2015}  & ST-SU-CRC & Non-cooperative $\&$ cooperative   & MIMO & MIMO & EIP minimization at the radar receiver  \\
       \hline
        \cite{L.2017ZhengJSSP} & ST-SU-CRC & Non-cooperative & SISO & SISO &  Radar signals mitigation at the communication receiver  \\
                   \hline
        \cite{Y.2019LiarXiv} & MT-SU-CRC & Non-cooperative & SISO& SISO& Radar signals and clutters mitigation at the communication receiver \\
           \hline
       \cite{B2019Jain}    & mmWave ST-SU-CRC & Cooperative & MIMO & MIMO & Cross-interference mitigation at the radar and communication receivers  \\
            \hline  
      \hline   
    \end{tabular}
    \end{center}  
    \end{table*}

 \subsection{DFRC}
  \label{sec:DRRC_IM}

 
  In CRC systems  the cross-interference is independent and the cancellation is performed at each individual radar and/or receiver. Differently, in DFRC systems, the communication signals are correlated with the radar target returns as they come from the same source. 
 Cancellation of communication signals in the radar target returns involve reconstructing the received signals in the original form which is more challenging than interference cancellation in communication systems as the subtraction of reconstructed signals imprecisely can cause more residues resulting in severe degradation of the radar's dynamic range, measured by the radar's capability to handle a range of signal strengths.

 References \cite{L.2011SitEuropean} and \cite{L.2011SitRADAR} aim to improve the radar dynamic range in   OFDM  DFRC systems.   The authors in~\cite{L.2011SitEuropean} propose two interference cancellation methods, namely, serial cancellation and selective cancellation. The former first identifies the strongest interferers with an amplitude above a threshold and then starts decoding from the interferer with the highest power to the one with the lowest power.  
 Differently, the latter method only  reconstructs the strongest identified interferer. It is shown that the
  selective cancellation outperforms the serial cancellation in terms of processing time and radar dynamic range. 
  The authors in \cite{L.2011SitRADAR} devise an interference cancellation scheme by utilizing the available communication signal  extracted from the regularly spaced pilot symbols. Specifically, a simple frequency offset estimator is implemented to extract   
  the frequency offset information from the estimated CSI to reduce the frequency offset errors in the reconstructed signals.
  It is shown that the radar performance can be improved up to $34$ dB by canceling the LOS path of the interferer with the devised scheme. Nevertheless, neither \cite{L.2011SitEuropean} nor \cite{L.2011SitRADAR} accounts for the impact of secondary interference. Moreover, both of the works consider perfect frequency offset, and thus the resulted radar dynamic ranges can only be deemed as the performance upper-bounds. 
  
  To address the impact of imperfect frequency offset, reference~\cite{Y.2018Zeng} considers estimation errors in frequency offset. 
  The objective is to 
  reconstruct the interfering signals at the radar subsystem in a FD IEEE 802.11-based OFDM DFRC so that the impact of the erroneous frequency offset is minimal. A self-interference cancellation 
  approach for the radar receiver based on a combined approach with 2D fast Fourier transform and one-dimensional search. The numerical results reveal that larger  analog to digital converter dynamic range benefits the self-interference  cancellation performance and up to $100$ dB can be canceled without degrading the detection accuracy. Though erroneous frequency offset is taken into account, the impact of filtering is not studied. In practice, the transmit signals need to be filtered to avoid out-of-band emission, which inevitably introduces delay and inter-symbol interference which are both ignored.

\subsection{Summary and Lessons Learned}
 In this section, we have reviewed the interference management approaches for CRC systems (in Subsection~\ref{sec:crc_IM}) and DFRC systems (in Subsection~\ref{sec:DRRC_IM}).  The approaches and their references are summarized in Table~\ref{IC_CRC}. The key lessons learned from Subsection~\ref{sec:crc_IM} are summarized as follows. 
\begin{itemize}
\item To eliminate radar interference, an effective technique that has recently been proposed is to project the radar waveform onto the null space of the channels between the communication and radar subsystems. Nevertheless, with imperfect CSI, the null space projection technique may be misaligned with the targets that impair the detection performance. Thus, to use such a technique, an accurate CSI is required. 
\item With the knowledge of radar waveforms, a communication system can suppress the radar interference through direct subtraction techniques. Nevertheless, if the waveform information is intercepted by a jammer during information sharing, jamming signals can be designed to harm radar detection. Hence, safeguarding the information sharing between communication and radar subsystem is as important as interference mitigation designs. 
\end{itemize} 

The key lessons learned from Subsection~V-B are summarized as follows. 
 
\begin{itemize}
 \item In DFRC systems, the communication performance is rather affected by the cumulative cochannel interference in the entire system, while radar performance is mainly limited by the strongest interferer~\cite{L.2012SitMicrowave}.
 \item Owning to the mutual interference, the DFRC transceivers must be equipped with sufficient dynamic range and analog-to-digital resolution to ensure the digitized signals remain undistorted.
  \item Despite wider antenna directionality that empowers radar detection over a greater field, the overall detectable range may decrease due to increased interference level. Hence, antenna directionality should be properly configured considering interference mitigation.
 \end{itemize}

\section{Security}
\label{sec:security}

In this section, we review security approaches that applied to safeguard the communication secrecy and covertness of JRC systems. In JRC systems, the most common attacks faced by the communication subsystem includes:
\begin{itemize}

\item  {Jamming attack}: A jamming attack is meant to disrupt legitimate communications with artificial noise. Specifically, a jammer transmits random RF signals over the same frequency band of  a legitimate transmission  to decrease the receive SINR at the target receiver so that the legitimate transmission is difficult to be decoded. 

\item  {Eavesdropping attack}: Theft of information during the transmission.
In an eavesdropping attack, the eavesdropper passively overhear the broadcast channels and attempt to decode the received signal to extract private and confidential information, such as identification numbers, business secret, and sensitive data.

 
\end{itemize}

 
Under jamming and eavesdropping attacks, the common performance metrics of interest to measure communication performance are communication rate and secrecy rate defined as follows.

\begin{itemize}

\item  {\em Communication rate}: The achievable rate at the communication receiver.

\item  {\em Secrecy rate}: The difference between the achievable rates at the communication receiver and the eavesdropper. 




\end{itemize}

In the following, we review the existing literature related to security designs according to the roles of attackers in the JRC systems.

\subsection{Literature Review}

 References~\cite{A.Deligiannis2018} and  \cite{N.2019Su}  consider DFRC systems where the radar targets perform eavesdropping attacks to the dual-function transmitter.  Specifically, in a  DFRC with a single target and a single communication receiver, the authors in  \cite{A.Deligiannis2018} formulate optimization problems to   
 1) maximize the secrecy rate of communication; 2) maximize the radar SINR for target detection; and 3) minimize transit power of the dual-function transmitter while preventing the eavesdropper to decode the communication signals. As the original problems are non-convex due to the non-convexity of the secrecy rate expression, the authors transform them into convex problems by approximating the secrecy rate function based on the Taylor expansion.
 To minimize the signal leakage to the eavesdropper, the authors propose to let the dual-function transmitter to  beamform additional pseudorandom distortion signals other than the communication signals to disturb the eavesdropping.
 In particular, under the assumption that the eavesdropper's location and CSI are both unknown, the authors devise transmit covariance matrices of the distortion and communication signals to optimize the three considered objectives.  
 Compared with the isotropic noise generation scheme \cite{S.2008Goel} which projects noise uniformly towards the orthogonal directions of communication channels, the proposed beamforming designs are shown to achieve a higher secrecy rate by up to more than 10$\%$. However, the performance gap between the origin non-convex problems and the approximated problem are not investigated. Besides, the radar is assumed to know the perfect location of the targets and thus channel state information which is too  ideal to realize. 
 
 To relax the hard assumptions of the target in \cite{A.Deligiannis2018}, reference \cite{N.2019Su} considers that the location of the target is  estimated with errors.
 Moreover, the authors extend the system model in \cite{A.Deligiannis2018} to the scenarios with multiple eavesdropping targets and communication receivers.  
 To guarantee the sum of secrecy rate of the system, the authors consider artificial noise in the transmit beamformer design to minimize the SINR at the eavesdropping targets subject to the SINR requirements of communication receivers. Under both perfect and imperfect CSI of the eavesdroppers, the authors use fractional programming approaches along with semi-definite relaxation to solve the beamformer design problems. The simulation results demonstrate that under both cases with and without perfect CSI, the proposed approaches coverage to the optimum quickly. The SINR at the eavesdropper in the case with inaccurate location estimation is up to two orders of magnitude lower than that in the case with accurate location estimation. However, it is difficult to extend the proposed beamformer with the target's location uncertainty to the scenario of multiple targets as the complexity expands exponentially with the number of targets.

\begin{figure} 
  \centering
  \includegraphics[width=0.8\linewidth]{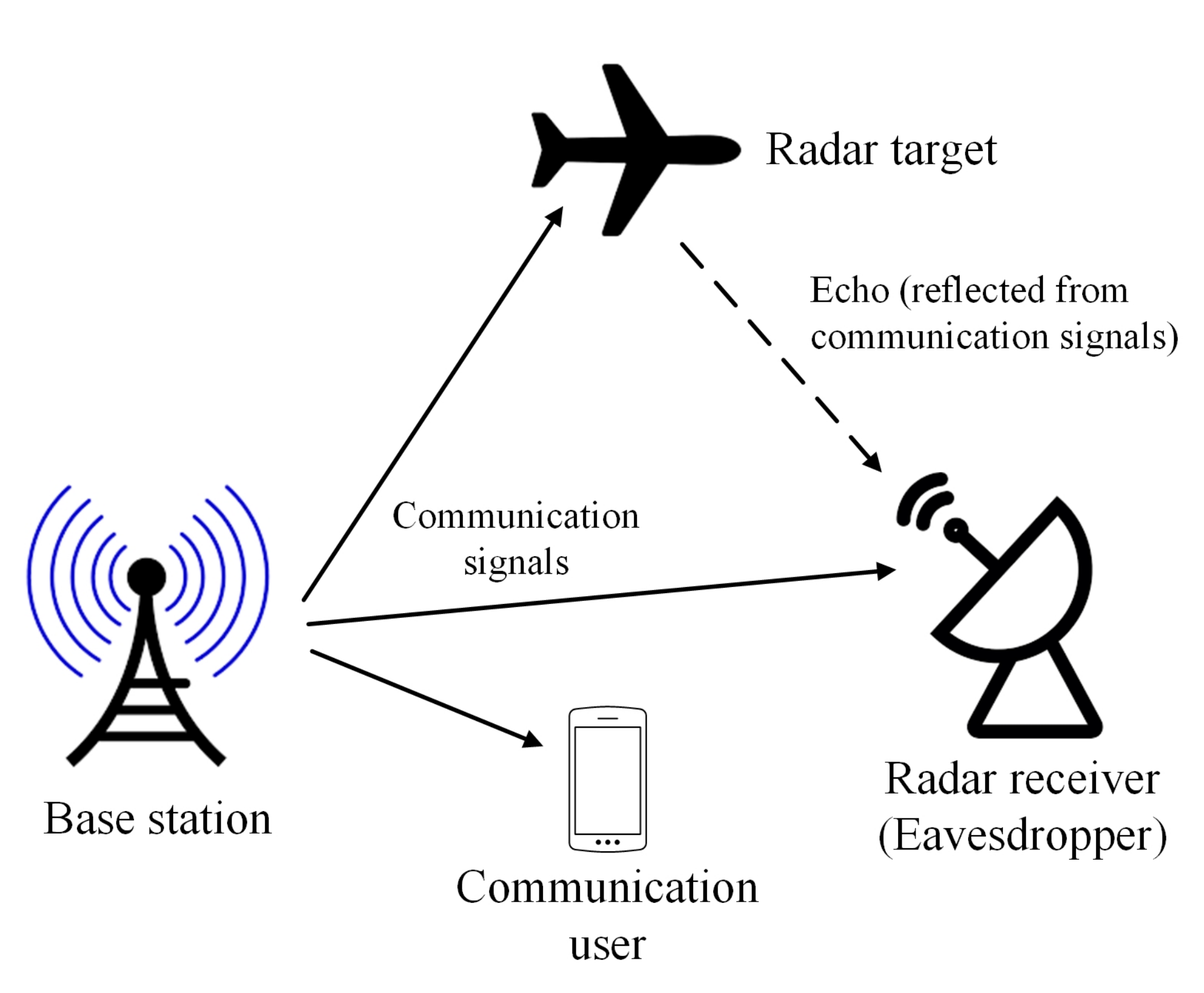}  
  \caption{System model of DFRC with eavesdropping  radar receiver \cite{K2018Chalise}.}    \label{Eav}  
  \end{figure}

 Different from \cite{A.Deligiannis2018} and  \cite{N.2019Su} which consider the radar target as the eavesdropper, reference  \cite{K2018Chalise}  
 considers the potential risk of the radar receiver as the attacker.
 In particular, the authors consider a CRC system, as shown in Fig. \ref{Eav}, in which the radar receiver detects the target's reflected signals originated from the communication transmitter while acting as the eavesdropper to overhear the signals directly come from the communication transmitter. The authors aim to maximize the radar SINR while maintaining required secrecy rate in the presence of eavesdropping. Under the cases where the radar and communication subsystems use orthogonal and non-orthogonal channels, the considered maximization problems are shown to be non-convex.   
 To cope with the non-convexity, the authors introduce an iterative algorithm that employs semidefinite programming and a semi-analytical approach for the former case and an alternating optimization approach based on semi-definite relaxation for the second case to find sub-optimal solutions.
  The numerical results illustrate that in the case with orthogonal channels, the proposed iterative approach brings about considerable performance gains over the pure semidefinite programming-based approach~\cite{Q.2010Luo} especially when the secrecy rate requirement is high. Furthermore, despite the interference induced in the case with non-orthogonal channel allocation, the performance of the case with  non-orthogonal channels exceeds that with orthogonal channels under the joint optimization of radar waveform and transmit covariance matrix. The performance gain of orthogonal channel allocation increases the secrecy rate target. However, this work assumes a clutter-free environment. Therefore, the achieved radar and communication performance can only be deemed as the upper bounds.
  
  Different from references \cite{A.Deligiannis2018}, \cite{N.2019Su} and \cite{K2018Chalise} that all consider passive eavesdropping attacks, reference  
  \cite{A.2018Garnaev} studies a DFRC system faced by the active jamming attack from a jammer. 
  The objective of the dual-function transmitter is to maximize a weighted payoff function of communication rate and radar SINR. On the other hand, the jammer aims to reduce the payoff function of the DFRC system through jamming. Under the assumption that the DFRC system knows a
  priori distribution of the jamming attack, the authors formulate a Bayesian game between the DFRC system and the jammer the strategies of which are the transmit power and jamming power, respectively. The authors prove the uniqueness of Bayesian Nash equilibrium of the formulated game and derive a water-filling equation to find the Nash equilibrium strategy. The simulations reveal that an increase in a priori probability of the jamming attack causes the DFRC system to adopt a more power-consuming strategy.~However, the performance gain between the proposed equilibrium solution and the optimal solution is unknown. Besides, this work assumes that the jamming signal does not reach the targets. In practice, the radar target returns from jamming signal also cause non-neglectable impact on radar performance.

   \begin{table*}  
       \caption {Security Approaches for JRC Systems}
      \begin{center}
      \begin{tabular}{ |  l | p{20mm} | p{40mm} | p{15mm} | p{18mm} |  p{20mm} |  p{17mm} }  
      \hline
      Reference &  System  & Objective &   CSI & Attacker  & Communication $\&$ Radar channels \\
       \hline 
    \cite{A.Deligiannis2018} & ST-SU-DFRC & Maximization of secrecy rate,  maximization of radar SINR, minimization of transmit power  & Perfect &   Eavesdropping target  &  MIMO $\&$ MIMO  \\
                   \hline 
      \cite{N.2019Su}   &  MT-MU-DFRC & Minimization of the
     SINR at the eavesdropping target  & Perfect $\&$ Imperfect   & Eavesdropping target &  MISO $\&$  MIMO \\
          \hline   
    \cite{K2018Chalise}  &  ST-SU-CRC &  Maximization of SINR  at communication receiver subject to the information secrecy rate requirement &  Perfect  & Eavesdropping radar receiver  &   MIMO $\&$ MIMO   \\
    \hline 
     \cite{A.2018Garnaev}   &  ST-SU-DFRC & Maximization of a weighted  communication throughput and radar SINR   &    Perfect & Jammer & SISO $\&$ SISO   \\
     \hline
    \end{tabular}
    \end{center} \label{Compare3} 
    \end{table*}

\subsection{Summary and Lessons Learned}

 In this section, we have reviewed security approaches for JRC systems. The approaches and their references are summarized in Table~\ref{Compare3}. As shown in the table, the two most common attacks considered in the JRC systems are jamming and eavesdropping. In addition, compared with the jamming attack, the eavesdropping has received more attentions. The key lessons learned are summarized as follows. 

\begin{itemize}
\item DFRC systems allow data symbols to be embedded in radar signals. Thus, radar targets can eavesdrop the information from the legitimate radar signals. Therefore, encryption methods/algorithms should be adopted to enhance the security of the information before transmitted. 
\item Adopting the encryption algorithms can complicate the hardware of the radar transmitter and the communication receiver. Thus, similar to cellular systems, physical layer security with power allocation algorithms can be used for the JRC systems to minimize the decodability of the legitimate signal at the eavesdropper. Note that since the JRC system consists of the radar function and the communication function, the performance of both the radar function and communication function needs to be guaranteed when the power allocation algorithms are applied. 
\item A solution to minimize the signal leakage to the eavesdropper is to let the transmitter beamform additional pseudorandom distortion signals other than the communication signals to disturb the eavesdropping as proposed in~\cite{N.2019Su}. This not only decreases the decodability of the legitimate signal at the eavesdropper but also benefits target detection. However, this algorithm requires the information about the target's location, and thus it is applicable in scenarios with single target.
\item The eavesdropper can perform the jamming and eavesdropping attacks simultaneously, and combating both the two attacks may be challenging for the JRC system. A potential approach that has recently been proposed in~\cite{A.2018Garnaev} is to use the game theory to model and analyze interactions between the active eavesdropper and the JRC system. Through the interactions, the JRC system can effectively learn or predict the attack strategy of the active eavesdropper, then having defending and reaction strategies based on the equilibrium analysis. Thus, more game theory approaches should be investigated for the JRC systems.
\end{itemize}

\section{Challenges and Future Research Directions}
\label{sec:challenges_open_issues}
Apart from the aforementioned issues, there are still challenges and
new research directions in deploying the JRC systems to be discussed as follows~\cite{sturm2009ofdm}.

\
\subsection{Synchronization Requirements for JRC Based on LFM and LFM-OFDM Waveforms}
\label{syn_requirements}

\subsubsection{JRC Based on LFM Waveform}
As mentioned in Section~\ref{chirp_waveform}, the LFM waveform has the constant modulus feature that enables the JRC systems to achieve a high range resolution. However, the current LFM-based  JRC approaches mostly assume that the chirp parameters used at the transmitter and the communication receiver are the same. Due to a potential mismatch in oscillators at the transmitter and the receiver, it is challenging to guarantee that the LFM waveforms generated at the transmitter and the receivers are exactly the same. The chirp
mismatch may potentially degrade communication performance. Accurate clock synchronization algorithms at the communication receiver are thus required to compensate for the chirp
mismatch.  

\subsubsection{JRC Based on LFM-OFDM Waveform}
In fact, LFM can be combined with OFDM as a promising solution to address the inefficient spectrum utilization of the LFM and the low range resolution of the OFDM. However, there are two challenges when implementing the proposed waveform. 
\begin{itemize}
\item \textit{Dechirp timing:} To estimate the target parameters and detect the data symbols, the radar and communication receivers need to determine the starting time of dechirping and the starting time of each OFDM symbol. This requires a very accurate time synchronization algorithm at the receivers. 
\item  \textit{Multipath effect:} In conventional OFDM
systems, the received signal at the receiver has a transmission delay. The transmission delay is translated to a frequency shift on the OFDM subcarriers when the receiver demodulates the received signal, i.e., via the FFT algorithm. Then, simple algorithms such as autocorrelation can be used to estimate the frequency shift. However, when the LFM is combined with the OFDM, the subcarriers included in the received signal often have different frequency shifts which may result in the inter-carrier interference and the non-orthogonality between OFDM subcarriers. Advanced frequency synchronization algorithms such as ESPRIT algorithm \cite{salberg2005doppler} and~\cite{nguyen2012time} may be used to estimate and completely compensate for the frequency shifts. 
\end{itemize}

\subsection{Massive Access and Resource Management}
JRC is expected to be implemented in a dynamic wireless environment with high density of base stations and mobile devices. This raises the massive access and resource management issues in the JRC systems. 
\subsubsection{Massive Access Management}
In the existing JRC approaches as discussed in Section~\ref{sec:radar-commu-mul}, the radar receiver detects targets using echoes from them. In practice, the radar receiver also receives communication signals transmitted from a massive number of mobile users. Thus, a problem for the radar receiver is to distinguish between the
echoes from targets and communication signals from the mobile users
in the presence of noise and interference. Given the independent statistical characteristics of
the two kinds of signals, machine learning (ML) can be used for detecting the radar echoes and the communication signals from the mobile users. For this, it is important to construct a dataset that includes the radar echoes, communication signals from the mobile users, and the corresponding labels. Then, a neural network is trained based on this dataset for the signal classification. However, constructing such a dataset may be difficult.

\subsubsection{Location-Dependent Resource Management}

Apart from the communication signals from a massive number of mobile users, the spatial locations of the radar and communication components of a JRC system play an important role in the JRC performance.  However, most of the existing literature reviewed in Section III and Section IV does not model the locations of system components explicitly.  The main challenge is to understand the impact of mobility (e.g., in radar targets and communication receivers) which incurs temporal corrections in the system performance. Any resource allocation scheme that fails to adapt to the dynamic location variations would result in a deficient system. 
Therefore, designing location-dependent resource allocation taking into account mobility has great potential to improve efficiency in utilizing system resources (e.g., frequency, time, and energy).

\subsubsection{Resource Management with Large-Scale JRC Systems}
In addition to the mobility, the performance of a JRC system is significantly affected by the spatial distribution of coexisting radar, communication, and JRC systems operating on the same frequency band.  Moreover, characterizing the impact of a large-scale system based on their spatial distribution is the key to understanding of JRC performance and the design of resource management in a real-world implementation. Stochastic geometry~\cite{ElSawy2016H}, a powerful tool to model and analyze the randomness in the spatial distribution of large-scale systems, can be exploited for the analytical study of JRC systems.

\subsubsection{Resource Management for Autonomous Vehicles (AVs) Equipped with JRC}
 JRC such as DFRC has been a promising technology for autonomous vehicles, namely DFRC-equipped AVs. However, one main issue is how to effectively schedule the radar function and the communication function of the DFRC. To address the issue, the authors in \cite{kumari2015investigating} and \cite{kumari2017performance} adopt time schedule schemes in which the portions in a time frame are allocated to the radar and communication functions in a static fashion. In practice, the surrounding environment of the vehicle is uncertain and dynamic, and thus adaptive algorithms for the radar and communication mode selection need to be investigated to maximize spectrum efficiency. For example, when the weather is in bad condition, e.g., heavy rain, the vehicle can select the radar mode more frequently to improve the
radar performance to detect unexpected events, i.e., the nearby objects, on the road. On
the contrary, when the weather and the communication channel
are in good conditions, the vehicle can select the communication
mode more frequently to transmit its data. This is due to the fact that other types of sensors, e.g., video and LIDAR, can work more effectively when the weather is good. The problem for the AV is to determine optimal decisions, i.e., radar mode or communication mode, to maximize the radar performance and the communication performance. This is challenging since  the environment states, e.g., weather and road states as well as
the communication channel state are dynamic and uncertain. To solve the problem, learning algorithms such as reinforcement learning (RL) or deep reinforcement
learning (DRL) can be developed to allow the AV to quickly obtain the
optimal policy without requiring any prior information about the
environment.

\subsubsection{Resource Management with Incentive Mechanisms}
It is worth noting that the DFRC-equipped AVs actually act as IoT devices that sense surrounding environments, e.g., traffic conditions, and then transmit sensing data to aggregation units, e.g., road-side units, for further processing. The data can be image or video files that have a large size. Thus, the DFRC-equipped AVs can require a huge amount of spectrum from the service providers (SPs) to simultaneously perform the radar function and the communication function. To motivate the service providers and the DFRC-equipped AVs to participate in the
spectrum allocation market, the problem is to enhance the utility of both the service providers and the DFRC-equipped AVs. In such a multi-buyer multi-seller market, Stackelberg game and matching theory can be used as effective solutions to maximize the the utility of all the stakeholders.



\subsection{Security Issues in JRC Systems}

Apart from the massive access and resource management, security issues need to be further investigated in JRC systems. In particular, currently, there are two common power allocation approaches for the security issues in JRC systems as discussed in Section IV-C. However, the approaches are proposed for the separate security issues. In particular, the approaches proposed in \cite{N.2019Su} and \cite{K2018Chalise} are for the eavesdropping attack, and the approach proposed in \cite{A.2018Garnaev} is for the jamming attack. In fact, the attacks can be equipped with a FD technology that enables them to launch the eavesdropping and jamming attacks simultaneously. The attacks are named \textit{FD active eavesdroppers} that may be more challenging to be prevented. Due to the uncertainty about the jamming pattern as well as the location of the FD active eavesdropper, learning algorithms such as RL and DRL can be effectively used that enable the JRC systems to find optimal defense strategy.

\subsection{Potential Integration of JRC and other Emerging Systems}
Future works need to also investigate on the integration of JRC and emerging technologies such as Intelligent Reflecting Surface (IRS) and edge computing.
\subsubsection{Integration of JRC and IRS}

IRS~\cite{wu2019intelligent} has been introduced to improve the communication performance by using a number of low-cost passive radio-reflecting elements. The element can reflect RF signals with an adjustable phase shift so that three-dimensional passive beamforming is established without an active RF transmission device, requiring negligible energy consumption. Thus, IRS can combine with JRC systems to improve the data throughput and spectrum efficiency. In such a system, a problem is to determine the optimal phase shifts of IRSs, the beamformer for the communication signal, which performs over the forward channel, and the beamformer for the radar signal, which performs over the cascaded (forward and reverse) channel, to maximize the performance of the communication and radar functions. However, to meet time requirements of the radar and communication functions, the optimization problem needs to be quickly solved, that is challenging.

\subsubsection{JRC and Edge Computing}
\label{JRC_edge}

JRC systems can utilize and access edge computing~\cite{shi2016edge} facilities in the next-generation wireless networks 5G and beyond. In particular, an AV equipped with JRC can offload its computing to edge devices (e.g., the road side units) to analyze video captured by the cameras for safety purposes. However, as the AV transfers its video data to the edge devices, it may consume a large amount of bandwidth that affects the communication and radar functions of JRC. Thus, one problem is to determine the optimal amounts of bandwidth assigned to the video data transmission, communication function, and radar function, so as to maximize the performance of the radar and communication functions while guaranteeing the QoS requirement, e.g., latency, of video data analytics. However, solving the problem may be challenging due to the high mobility of the AV.

\subsection{Summary}
To enable JRC to be widely implemented in the applications, there still exist issues needed to be considered such as synchronization, dechirp timing, multipath channel effect, and security (as discussed in Subsection~\ref{syn_requirements}). This is generally challenging since the radar signal and communication signal are different in waveform. Moreover, the radar function and the communication function have different performance requirements. Nevertheless, JRC is a promising technology for several military and civilian applications since it is able to perform both the communication function and radar function simultaneously that improve the efficiency of resources, i.e., spectrum and energy, and minimize the system cost. In addition, there will be several open research directions related to JRC. For example, JRC can be combined with edge computing systems as discussed in Subsection~\ref{JRC_edge}. However, in such a scenario, the performance of communication function, radar function, and computing service needs to be jointly considered.


\section{Conclusions}
\label{sec:conclusions}

This paper has presented a comprehensive survey on resource management issues for JRC. First, we have presented the fundamental concepts related to JRC and important performance metrics used in the JRC systems, followed by the discussions of applications of JRC. Then, we have provided detailed reviews, analyses, and comparisons of resource management approaches in the JRC systems. The approaches include spectrum sharing with waveform design, power allocation, and interference cancellation. In addition, we have discussed the security issues and countermeasures in the JRC systems. Finally, we have outlined important challenges as well as future research directions related to the JRC systems.

\bibliographystyle{IEEEtran}
\bibliography{radar_communication}{}

\begin{thebibliography}{}

\end{thebibliography}


\begin{thebibliography}{100}
\providecommand{\url}[1]{#1}
\csname url@samestyle\endcsname
\providecommand{\newblock}{\relax}
\providecommand{\bibinfo}[2]{#2}
\providecommand{\BIBentrySTDinterwordspacing}{\spaceskip=0pt\relax}
\providecommand{\BIBentryALTinterwordstretchfactor}{4}
\providecommand{\BIBentryALTinterwordspacing}{\spaceskip=\fontdimen2\font plus
\BIBentryALTinterwordstretchfactor\fontdimen3\font minus
  \fontdimen4\font\relax}
\providecommand{\BIBforeignlanguage}[2]{{%
\expandafter\ifx\csname l@#1\endcsname\relax
\typeout{** WARNING: IEEEtran.bst: No hyphenation pattern has been}%
\typeout{** loaded for the language `#1'. Using the pattern for}%
\typeout{** the default language instead.}%
\else
\language=\csname l@#1\endcsname
\fi
#2}}
\providecommand{\BIBdecl}{\relax}
\BIBdecl

\bibitem{liu2020joint}
F.~Liu, C.~Masouros, A.~Petropulu, H.~Griffiths, and L.~Hanzo, ``Joint radar
  and communication design: Applications, state-of-the-art, and the road
  ahead,'' \emph{IEEE Transactions on Communications}, to appear.

\bibitem{Sean_5G_auction}
\BIBentryALTinterwordspacing
S.~Kinney. (2018, Aug.) Update on global 5g spectrum auctions. [Online].
  Available: \url{https://www.rcrwireless.com/20180821/5g/5g-spectrum-auctions}
\BIBentrySTDinterwordspacing

\bibitem{auction_result}
\BIBentryALTinterwordspacing
(2018, Apr.) Results of auction. Ofcom. [Online]. Available:
  \url{https://www.ofcom.org.uk/__data/assets/pdf_file/0018/112932/Regulation-111-Final-outcome-of-award.pdf}
\BIBentrySTDinterwordspacing

\bibitem{IoT_growth}
\BIBentryALTinterwordspacing
(2020, May) Number of active iot devices expected to reach 24.1 billion in
  2030. Help Net Security. [Online]. Available:
  \url{https://www.helpnetsecurity.com/2020/05/22/active-iot-devices/}
\BIBentrySTDinterwordspacing

\bibitem{griffiths2014radar}
H.~Griffiths, L.~Cohen, S.~Watts, E.~Mokole, C.~Baker, M.~Wicks, and S.~Blunt,
  ``Radar spectrum engineering and management: Technical and regulatory
  issues,'' \emph{Proceedings of the IEEE}, vol. 103, no.~1, pp. 85--102, 2014.

\bibitem{feng2020joint}
Z.~Feng, Z.~Fang, Z.~Wei, X.~Chen, Z.~Quan, and D.~Ji, ``Joint radar and
  communication: A survey,'' \emph{China Communications}, vol.~17, no.~1, pp.
  1--27, 2020.

\bibitem{chiriyath2017radar}
A.~R. Chiriyath, B.~Paul, and D.~W. Bliss, ``Radar-communications convergence:
  Coexistence, cooperation, and co-design,'' \emph{IEEE Transactions on
  Cognitive Communications and Networking}, vol.~3, no.~1, pp. 1--12, 2017.

\bibitem{daniels2017forward}
R.~C. Daniels, E.~R. Yeh, and R.~W. Heath, ``Forward collision vehicular radar
  with ieee 802.11: Feasibility demonstration through measurements,''
  \emph{IEEE Transactions on Vehicular Technology}, vol.~67, no.~2, pp.
  1404--1416, 2017.

\bibitem{logistic}
\BIBentryALTinterwordspacing
K.~Ghaffarzadeh and N.~Jiao. Mobile robots, autonomous vehicles, and drones in
  logistics, warehousing, and delivery 2020-2040. [Online]. Available:
  \url{https://www.idtechex.com/en/research-report/mobile-robots-autonomous-vehicles-and-drones-in-logistics\\
  -warehousing-and-delivery-2020-2040/706}
\BIBentrySTDinterwordspacing

\bibitem{khawar2014mathematical}
A.~Khawar, A.~Abdelhadi, and T.~C. Clancy, ``A mathematical analysis of
  cellular interference on the performance of s-band military radar systems,''
  in \emph{Wireless Telecommunications Symposium}, Washington, DC, Apr. 2014,
  pp. 1--8.

\bibitem{Labib2017}
M.~{Labib}, V.~{Marojevic}, A.~F. {Martone}, J.~H. {Reed}, and A.~I.
  {Zaghloui}, ``Coexistence between communications and radar systems: A
  survey,'' \emph{URSI Radio Science Bulletin}, vol. 2017, no. 362, pp. 74--82,
  2017.

\bibitem{ma2019joint}
D.~Ma, N.~Shlezinger, T.~Huang, Y.~Liu, and Y.~C. Eldar, ``Joint
  radar-communications strategies for autonomous vehicles,'' \emph{arXiv
  preprint arXiv:1909.01729}, 2019.

\bibitem{gameiro2018research}
A.~Gameiro, D.~Castanheira, J.~Sanson, and P.~P. Monteiro, ``Research
  challenges, trends and applications for future joint radar communications
  systems,'' \emph{Wireless Personal Communications}, vol. 100, no.~1, pp.
  81--96, 2018.

\bibitem{SkolnikRADAR2001}
I.~S. Merrill \emph{et~al.}, ``Introduction to radar systems,'' \emph{Mc
  Grow-Hill}, vol.~7, no.~10, 2001.

\bibitem{Radar_Range_Resolution}
\BIBentryALTinterwordspacing
(2020) Range resolution. radartutorial. [Online]. Available:
  \url{https://www.radartutorial.eu/01.basics/Range\%20Resolution.en.html}
\BIBentrySTDinterwordspacing

\bibitem{Campbell2012Surface}
C.~Campbell, \emph{Surface acoustic wave devices and their signal processing
  applications}.\hskip 1em plus 0.5em minus 0.4em\relax Elsevier, 2012.

\bibitem{Weiss2019mmwave}
L.~G. Weiss, ``Wavelets and wideband correlation processing,'' \emph{IEEE
  signal processing magazine}, vol.~11, no.~1, pp. 13--32, 2019.

\bibitem{HieuiRDRCIEEE}
N.~C.~L. Nguyen Quang~Hieu, Dinh Thai~Hoang and D.~Niyato, ``irdrc-an
  intelligent real-time dual-functional radar-communication system for
  automotive vehicles,'' \emph{IEEE Wireless Communications Letters}, to
  appear.

\bibitem{Ma2020Joint}
D.~Ma, N.~Shlezinger, T.~Huang, Y.~Liu, and Y.~C. Eldar, ``Joint
  radar-communication strategies for autonomous vehicles: Combining two key
  automotive technologies,'' \emph{IEEE Signal Processing Magazine}, vol.~37,
  no.~4, pp. 85--97, Jun. 2020.

\bibitem{Roberton2003Integrated}
M.~Roberton and E.~Brown, ``Integrated radar and communications based on
  chirped spread-spectrum techniques,'' in \emph{IEEE International Microwave
  Symposium Digest}, vol.~1, 2003, pp. 611--614.

\bibitem{Sahin2017Anovel}
C.~Sahin, J.~Jakabosky, P.~M. McCormick, J.~G. Metcalf, and S.~D. Blunt, ``A
  novel approach for embedding communication symbols into physical radar
  waveforms,'' in \emph{2017 IEEE Radar Conference (RadarConf)}.\hskip 1em plus
  0.5em minus 0.4em\relax IEEE, 2017, pp. 1498--1503.

\bibitem{Saddik2007Ultra}
G.~N. Saddik, R.~S. Singh, and E.~R. Brown, ``Ultra-wideband multifunctional
  communications/radar system,'' \emph{IEEE Transactions on Microwave Theory
  and Techniques}, vol.~55, no.~7, pp. 1431--1437, 2007.

\bibitem{sturm2011waveform}
C.~Sturm and W.~Wiesbeck, ``Waveform design and signal processing aspects for
  fusion of wireless communications and radar sensing,'' \emph{Proceedings of
  the IEEE}, vol.~99, no.~7, pp. 1236--1259, 2011.

\bibitem{wang2018joint}
C.-H. Wang and O.~Altintas, ``A joint radar and communication system based on
  commercially available fmcw radar,'' in \emph{2018 IEEE Vehicular Networking
  Conference (VNC)}, Taipei, Taiwan, Dec. 2018, pp. 1--2.

\bibitem{WinklerRange2007}
V.~Winkler, ``Range doppler detection for automotive fmcw radars,'' in
  \emph{2007 European Radar Conference}.\hskip 1em plus 0.5em minus 0.4em\relax
  Munich, Germany: IEEE, Oct. 2007, pp. 166--169.

\bibitem{Huang2012Decentralized}
H.~Huang and V.~K. Lau, ``Decentralized delay optimal control for interference
  networks with limited renewable energy storage,'' \emph{IEEE Transactions on
  Signal Processing}, vol.~60, no.~5, pp. 2552--2561, 2012.

\bibitem{cover1999elements}
T.~M. Cover, \emph{Elements of Information Theory}, 2nd~ed.\hskip 1em plus
  0.5em minus 0.4em\relax NJ, USA: John Wiley \& Sons, 2006.

\bibitem{Blackbird}
\BIBentryALTinterwordspacing
The fastest plane in the world - sr-71 blackbird. [Online]. Available:
  \url{https://migflug.com/jetflights/remarkable-airplanes-of-the-world-part-1-the-fastest/}
\BIBentrySTDinterwordspacing

\bibitem{Roulston2008post}
J.~Roulston, ``The post-war development of fighter radar in europe-a british
  perspective,'' in \emph{2008 IEEE International Radar Conference}.\hskip 1em
  plus 0.5em minus 0.4em\relax IEEE, 2008, pp. 1--9.

\bibitem{Abdallah2017Study}
B.~Abdallah, H.~Yousif, I.~Ali, and M.~Asmail, ``Study and analysis of radar
  system,'' Ph.D. dissertation, Sudan University of Science and Technology,
  2017.

\bibitem{radartutorial}
\BIBentryALTinterwordspacing
Radar tutorial. available online. [Online]. Available:
  \url{https://www.radartutorial.eu/07.waves/Waves\%20and\%20Frequency\%20Ranges.en.html}
\BIBentrySTDinterwordspacing

\bibitem{infineon}
\BIBentryALTinterwordspacing
Radar sensors for automotive. [Online]. Available:
  \url{https://www.infineon.com/cms/en/product/sensor/radar-image-sensors/radar-sensors/radar-sensors-for-automotive/}
\BIBentrySTDinterwordspacing

\bibitem{Remy2012The}
M.~A. Remy, K.~A. de~Macedo, and J.~R. Moreira, ``The first uav-based p-and
  x-band interferometric sar system,'' in \emph{IEEE International Geoscience
  and Remote Sensing Symposium}, 2012, pp. 5041--5044.

\bibitem{HouraniMillimeter2018}
A.~Al-Hourani, R.~J. Evans, P.~M. Farrell, B.~Moran, M.~Martorella,
  S.~Kandeepan, S.~Skafidas, and U.~Parampalli, ``Millimeter-wave integrated
  radar systems and techniques,'' in \emph{Academic Press Library in Signal
  Processing, Volume 7}.\hskip 1em plus 0.5em minus 0.4em\relax Elsevier, 2018,
  pp. 317--363.

\bibitem{wang2019joint}
F.~Wang, J.~Feng, Y.~Zhao, X.~Zhang, S.~Zhang, and J.~Han, ``Joint activity
  recognition and indoor localization with wifi fingerprints,'' \emph{IEEE
  Access}, vol.~7, pp. 80\,058--80\,068, 2019.

\bibitem{he2015wi}
S.~He and S.-H.~G. Chan, ``Wi-fi fingerprint-based indoor positioning: Recent
  advances and comparisons,'' \emph{IEEE Communications Surveys \& Tutorials},
  vol.~18, no.~1, pp. 466--490, 2015.

\bibitem{liu2019wireless}
J.~Liu, H.~Liu, Y.~Chen, Y.~Wang, and C.~Wang, ``Wireless sensing for human
  activity: A survey,'' \emph{IEEE Communications Surveys \& Tutorials}, 2019.

\bibitem{Tan2014Areal}
B.~Tan, K.~Woodbridge, and K.~Chetty, ``A real-time high resolution passive
  wifi doppler-radar and its applications,'' in \emph{2014 IEEE International
  Radar Conference}.\hskip 1em plus 0.5em minus 0.4em\relax IEEE, 2014, pp.
  1--6.

\bibitem{Wenda2020Passive}
L.~Wenda, R.~Piechocki, K.~Woodbridge, C.~Tang, and K.~Chetty, ``Passive wifi
  radar for human sensing using a stand-alone access point,'' \emph{IEEE
  Transactions on Geoscience and Remote Sensing}, 2020.

\bibitem{Kevin2011Through}
C.~Kevin, G.~Smith, and K.~Woodbridge, ``Through-the-wall sensing of personnel
  using passive bistatic wifi radar at standoff distances,'' \emph{IEEE
  Transactions on Geoscience and Remote Sensing}, vol.~50, no.~4, pp.
  1218--1226, 2011.

\bibitem{MR72}
\BIBentryALTinterwordspacing
77ghz collision avoidance radar mr72. [Online]. Available:
  \url{http://en.nanoradar.cn/Article/detail/id/488.html}
\BIBentrySTDinterwordspacing

\bibitem{NRA15}
\BIBentryALTinterwordspacing
24ghz alimeter radar nra15. [Online]. Available:
  \url{http://en.nanoradar.cn/Article/detail/id/477.html}
\BIBentrySTDinterwordspacing

\bibitem{mizutani2001inter}
K.~Mizutani and R.~Kohno, ``Inter-vehicle spread spectrum communication and
  ranging system with concatenated eoe sequence,'' \emph{IEEE Transactions on
  Intelligent Transportation Systems}, vol.~2, no.~4, pp. 180--191, Dec. 2001.

\bibitem{sturm2009ofdm}
C.~Sturm, T.~Zwick, and W.~Wiesbeck, ``An ofdm system concept for joint radar
  and communications operations,'' in \emph{VTC Spring 2009-IEEE 69th Vehicular
  Technology Conference}.\hskip 1em plus 0.5em minus 0.4em\relax IEEE, 2009,
  pp. 1--5.

\bibitem{sit2011ofdm}
Y.~L. Sit, C.~Sturm, L.~Reichardt, T.~Zwick, and W.~Wiesbeck, ``The ofdm joint
  radar-communication system: An overview,'' in \emph{Proc. Int. Conf. Advances
  in Satellite and Space Communications (SPACOMM 2011)}, 2011, pp. 69--74.

\bibitem{sit2011doppler}
Y.~L. Sit, C.~Sturm, and T.~Zwick, ``Doppler estimation in an ofdm joint radar
  and communication system,'' in \emph{2011 German Microwave Conference}.\hskip
  1em plus 0.5em minus 0.4em\relax IEEE, 2011, pp. 1--4.

\bibitem{liang2019design}
T.~Liang, Z.~Li, M.~Wang, and X.~Fang, ``Design of radar-communication
  integrated signal based on ofdm,'' in \emph{International Conference on
  Artificial Intelligence for Communications and Networks}.\hskip 1em plus
  0.5em minus 0.4em\relax Springer, 2019, pp. 108--119.

\bibitem{han2019artificial}
S.~Han, L.~Ye, and W.~Meng, \emph{Artificial Intelligence for Communications
  and Networks}.\hskip 1em plus 0.5em minus 0.4em\relax Springer, 2019, vol.
  287.

\bibitem{tian2017radar}
X.~Tian and Z.~Song, ``On radar and communication integrated system using ofdm
  signal,'' in \emph{2017 IEEE Radar Conference (RadarConf)}.\hskip 1em plus
  0.5em minus 0.4em\relax IEEE, 2017, pp. 0318--0323.

\bibitem{tedesso2018code}
T.~W. Tedesso and R.~Romero, ``Code shift keying based joint radar and
  communications for emcon applications,'' \emph{Digital Signal Processing},
  vol.~80, pp. 48--56, 2018.

\bibitem{davis1999peak}
J.~A. Davis and J.~Jedwab, ``Peak-to-mean power control in ofdm, golay
  complementary sequences, and reed-muller codes,'' \emph{IEEE Transactions on
  Information Theory}, vol.~45, no.~7, pp. 2397--2417, 1999.

\bibitem{li2019waveform}
W.~Li, Z.~Xiang, and P.~Ren, ``Waveform design for dual-function
  radar-communication system with golay block coding,'' \emph{IEEE Access},
  vol.~7, pp. 184\,053--184\,062, 2019.

\bibitem{podkurkov2019efficient}
I.~Podkurkov, J.~Zhang, A.~F. Nadeev, and M.~Haardt, ``Efficient
  multidimensional wideband parameter estimation for ofdm based joint radar and
  communication systems,'' \emph{IEEE Access}, vol.~7, pp. 112\,792--112\,808,
  2019.

\bibitem{zhang2017efficient}
J.~Zhang, I.~Podkurkov, M.~Haardt, and A.~Nadeev, ``Efficient multidimensional
  parameter estimation for joint wideband radar and communication systems based
  on ofdm,'' in \emph{IEEE ICASSP}, 2017, pp. 3096--3100.

\bibitem{oppenheim1999discrete}
A.~V. Oppenheim, \emph{Discrete-Time Signal Processing}.\hskip 1em plus 0.5em
  minus 0.4em\relax Pearson Education India, 1999.

\bibitem{paredes2015problem}
M.~C.~P. Paredes and M.~Garcia, ``The problem of peak-to-average power ratio in
  ofdm systems,'' \emph{arXiv preprint arXiv:1503.08271}, 2015.

\bibitem{bae2009adaptive}
K.~Bae, J.~G. Andrews, and E.~J. Powers, ``Adaptive active constellation
  extension algorithm for peak-to-average ratio reduction in ofdm,'' \emph{IEEE
  Communications Letters}, vol.~14, no.~1, pp. 39--41, 2009.

\bibitem{krongold2004active}
B.~S. Krongold and D.~L. Jones, ``An active-set approach for ofdm par reduction
  via tone reservation,'' \emph{IEEE transactions on Signal Processing},
  vol.~52, no.~2, pp. 495--509, 2004.

\bibitem{lv2018joint}
X.~Lv, J.~Wang, Z.~Jiang, and W.~Wu, ``A joint radar-communication system based
  on ocdm-ofdm scheme,'' in \emph{2018 International Conference on Microwave
  and Millimeter Wave Technology (ICMMT)}.\hskip 1em plus 0.5em minus
  0.4em\relax IEEE, 2018, pp. 1--3.

\bibitem{lv2018novel}
X.~Lv, J.~Wang, Z.~Jiang, and W.~Jiao, ``A novel papr reduction method for
  ocdm-based radar-communication signal,'' in \emph{2018 IEEE MTT-S
  International Microwave Workshop Series on 5G Hardware and System
  Technologies (IMWS-5G)}.\hskip 1em plus 0.5em minus 0.4em\relax IEEE, 2018,
  pp. 1--3.

\bibitem{ouyang2015discrete}
X.~Ouyang, C.~Antony, F.~Gunning, H.~Zhang, and Y.~L. Guan, ``Discrete fresnel
  transform and its circular convolution,'' \emph{arXiv preprint
  arXiv:1510.00574}, 2015.

\bibitem{braun2009parametrization}
M.~Braun, C.~Sturm, A.~Niethammer, and F.~K. Jondral, ``Parametrization of
  joint ofdm-based radar and communication systems for vehicular
  applications,'' in \emph{IEEE International Symposium on Personal, Indoor and
  Mobile Radio Communications}, 2009, pp. 3020--3024.

\bibitem{gaudio2019effectiveness}
L.~Gaudio, M.~Kobayashi, G.~Caire, and G.~Colavolpe, ``On the effectiveness of
  otfs for joint radar and communication,'' \emph{arXiv preprint
  arXiv:1910.01896}, 2019.

\bibitem{gaudio2019performance}
L.~Gaudio, M.~Kobayashi, B.~Bissinger, and G.~Caire, ``Performance analysis of
  joint radar and communication using ofdm and otfs,'' in \emph{IEEE ICC
  Workshops}.\hskip 1em plus 0.5em minus 0.4em\relax IEEE, 2019, pp. 1--6.

\bibitem{hadani2017orthogonal}
R.~Hadani, S.~Rakib, M.~Tsatsanis, A.~Monk, A.~J. Goldsmith, A.~F. Molisch, and
  R.~Calderbank, ``Orthogonal time frequency space modulation,'' in \emph{IEEE
  WCNC}, 2017, pp. 1--6.

\bibitem{hassanien2018dual}
A.~Hassanien, E.~Aboutanios, M.~G. Amin, and G.~A. Fabrizio, ``A dual-function
  mimo radar-communication system via waveform permutation,'' \emph{Digital
  Signal Processing}, vol.~83, pp. 118--128, 2018.

\bibitem{hassanien2017dual}
A.~Hassanien, B.~Himed, and B.~D. Rigling, ``A dual-function mimo
  radar-communications system using frequency-hopping waveforms,'' in
  \emph{IEEE Radar Conference}, 2017, pp. 1721--1725.

\bibitem{costas1984study}
J.~P. Costas, ``A study of a class of detection waveforms having nearly ideal
  rangeÑdoppler ambiguity properties,'' \emph{Proceedings of the IEEE},
  vol.~72, no.~8, pp. 996--1009, 1984.

\bibitem{wang2019co}
X.~Wang and J.~Xu, ``Co-design of joint radar and communications systems
  utilizing frequency hopping code diversity,'' in \emph{2019 IEEE Radar
  Conference (RadarConf)}.\hskip 1em plus 0.5em minus 0.4em\relax IEEE, 2019,
  pp. 1--6.

\bibitem{golomb1984constructions}
S.~W. Golomb and H.~Taylor, ``Constructions and properties of costas arrays,''
  \emph{Proceedings of the IEEE}, vol.~72, no.~9, pp. 1143--1163, 1984.

\bibitem{nusenu2018dual}
S.~Y. Nusenu and W.-Q. Wang, ``Dual-function fda mimo radar-communications
  system employing costas signal waveforms,'' in \emph{2018 IEEE Radar
  Conference (RadarConf18)}.\hskip 1em plus 0.5em minus 0.4em\relax IEEE, 2018,
  pp. 0033--0038.

\bibitem{hassanien2015dual}
A.~Hassanien, M.~G. Amin, Y.~D. Zhang, and F.~Ahmad, ``Dual-function
  radar-communications using phase-rotational invariance,'' in \emph{IEEE
  EUSIPCO}, 2015, pp. 1346--1350.

\bibitem{euziere2014dual}
J.~Euziere, R.~Guinvarc'h, M.~Lesturgie, B.~Uguen, and R.~Gillard, ``Dual
  function radar communication time-modulated array,'' in \emph{2014
  International Radar Conference}.\hskip 1em plus 0.5em minus 0.4em\relax IEEE,
  2014, pp. 1--4.

\bibitem{hassanien2016non}
A.~Hassanien, M.~G. Amin, Y.~D. Zhang, F.~Ahmad, and B.~Himed, ``Non-coherent
  psk-based dual-function radar-communication systems,'' in \emph{2016 IEEE
  Radar Conference (RadarConf)}.\hskip 1em plus 0.5em minus 0.4em\relax IEEE,
  2016, pp. 1--6.

\bibitem{sahin2017novel}
C.~Sahin, J.~Jakabosky, P.~M. McCormick, J.~G. Metcalf, and S.~D. Blunt, ``A
  novel approach for embedding communication symbols into physical radar
  waveforms,'' in \emph{IEEE Radar Conference (RadarConf)}, 2017, pp.
  1498--1503.

\bibitem{wang2018poster}
C.-H. Wang, C.~D. Ozkaptan, E.~Ekici, and O.~Altintas, ``Poster: Multi-carrier
  modulation on fmcw radar for joint automotive radar and communication,'' in
  \emph{2018 IEEE Vehicular Networking Conference (VNC)}.\hskip 1em plus 0.5em
  minus 0.4em\relax IEEE, 2018, pp. 1--2.

\bibitem{dou2017radar}
Z.~Dou \emph{et~al.}, ``Radar-communication integration based on msk-lfm spread
  spectrum signal,'' \emph{International Journal of Communications, Network and
  System Sciences}, vol.~10, no.~08, p. 108, 2017.

\bibitem{chen2017energy}
X.~Chen, Z.~Liu, Y.~Liu, and Z.~Wang, ``Energy leakage analysis of the radar
  and communication integrated waveform,'' \emph{IET Signal Processing},
  vol.~12, no.~3, pp. 375--382, 2017.

\bibitem{zhipeng2015communication}
L.~Zhipeng, C.~Xingbo, W.~Xiaomo, S.~Xu, and F.~Yuan, ``Communication analysis
  of integrated waveform based on lfm and msk,'' 2015.

\bibitem{zhang2017modified}
Y.~Zhang, Q.~Li, L.~Huang, C.~Pan, and J.~Song, ``A modified waveform design
  for radar-communication integration based on lfm-cpm,'' in \emph{2017 IEEE
  85th Vehicular Technology Conference (VTC Spring)}.\hskip 1em plus 0.5em
  minus 0.4em\relax IEEE, 2017, pp. 1--5.

\bibitem{ni2019high}
Y.~Ni, Z.~Wang, Q.~Huang, and M.~Zhang, ``High throughput rate-shift integrated
  system for joint radar-communications,'' \emph{IEEE Access}, 2019.

\bibitem{li2013integrated}
X.-B. Li, R.-J. Yang, and W.~Cheng, ``Integrated radar and communication based
  on multicarrier frequency modulation chirp signal,'' \emph{Journal of
  Electronics and Information Technology}, vol.~35, no.~2, pp. 406--412, 2013.

\bibitem{zhang2017waveform}
Y.~Zhang, Q.~Li, L.~Huang, and J.~Song, ``Waveform design for joint
  radar-communication system with multi-user based on mimo radar,'' in
  \emph{2017 IEEE Radar Conference (RadarConf)}.\hskip 1em plus 0.5em minus
  0.4em\relax IEEE, 2017, pp. 0415--0418.

\bibitem{moon2005error}
T.~K. Moon, \emph{Error correction coding: mathematical methods and
  algorithms}.\hskip 1em plus 0.5em minus 0.4em\relax John Wiley \& Sons, 2005.

\bibitem{li2019integrated}
Q.~Li, K.~Dai, Y.~Zhang, and H.~Zhang, ``Integrated waveform for a joint
  radar-communication system with high-speed transmission,'' \emph{IEEE
  Wireless Communications Letters}, 2019.

\bibitem{bahl1974optimal}
L.~Bahl, J.~Cocke, F.~Jelinek, and J.~Raviv, ``Optimal decoding of linear codes
  for minimizing symbol error rate (corresp.),'' \emph{IEEE Transactions on
  Information Theory}, vol.~20, no.~2, pp. 284--287, 1974.

\bibitem{kumari2015investigating}
P.~{Kumari}, N.~{Gonzalez-Prelcic}, and R.~W. {Heath}, ``Investigating the ieee
  802.11ad standard for millimeter wave automotive radar,'' in \emph{Vehicular
  Technology Conference (VTC2015-Fall)}, Sep. 2015, pp. 1--5.

\bibitem{kishida201579}
M.~Kishida, K.~Ohguchi, and M.~Shono, ``79 ghz-band high-resolution
  millimeter-wave radar,'' \emph{Fujitsu Scient. and Tech. J}, vol.~51, no.~4,
  pp. 55--59, 2015.

\bibitem{kumari2017ieee}
P.~Kumari, J.~Choi, N.~Gonz{\'a}lez-Prelcic, and R.~W. Heath, ``Ieee 802.11
  ad-based radar: An approach to joint vehicular communication-radar system,''
  \emph{IEEE Transactions on Vehicular Technology}, vol.~67, no.~4, pp.
  3012--3027, 2017.

\bibitem{kumari2017performance}
P.~Kumari, D.~H. Nguyen, and R.~W. Heath, ``Performance trade-off in an
  adaptive ieee 802.11 ad waveform design for a joint automotive radar and
  communication system,'' in \emph{IEEE ICASSP}, 2017, pp. 4281--4285.

\bibitem{kumari2019adaptive}
P.~Kumari, S.~A. Vorobyov, and R.~W. Heath~Jr, ``Adaptive virtual waveform
  design for millimeter-wave joint communication-radar,'' \emph{arXiv preprint
  arXiv:1904.05516}, 2019.

\bibitem{kumari2018virtual}
P.~Kumari, R.~W. Heath, and S.~A. Vorobyov, ``Virtual pulse design for ieee
  802.11 ad-based joint communication-radar,'' in \emph{2018 IEEE International
  Conference on Acoustics, Speech and Signal Processing (ICASSP)}.\hskip 1em
  plus 0.5em minus 0.4em\relax IEEE, 2018, pp. 3315--3319.

\bibitem{petrov2019unified}
V.~Petrov, G.~Fodor, J.~Kokkoniemi, D.~Moltchanov, J.~Lehtomaki, S.~Andreev,
  Y.~Koucheryavy, M.~Juntti, and M.~Valkama, ``On unified vehicular
  communications and radar sensing in millimeter-wave and low terahertz
  bands,'' \emph{IEEE Wireless Communications}, to appear.

\bibitem{ahmedoptimized}
A.~Ahmed, S.~Zhang, and Y.~D. Zhang, ``Optimized sensor selection for joint
  radar-communication systems.''

\bibitem{khawar2014mimo}
A.~Khawar, A.~Abdel-Hadi, and T.~C. Clancy, ``Mimo radar waveform design for
  coexistence with cellular systems,'' in \emph{2014 IEEE International
  Symposium on Dynamic Spectrum Access Networks (DYSPAN)}.\hskip 1em plus 0.5em
  minus 0.4em\relax IEEE, 2014, pp. 20--26.

\bibitem{shajaiah2014impact}
H.~Shajaiah, A.~Abdelhadi, and C.~Clancy, ``Impact of radar and communication
  coexistence on radar's detectable target parameters,'' \emph{arXiv preprint
  arXiv:1408.6736}, 2014.

\bibitem{noam2013blind}
Y.~Noam and A.~J. Goldsmith, ``Blind null-space learning for mimo underlay
  cognitive radio with primary user interference adaptation,'' \emph{IEEE
  Transactions on Wireless Communications}, vol.~12, no.~4, pp. 1722--1734,
  Apr. 2013.

\bibitem{ma2018novel}
D.~Ma, T.~Huang, Y.~Liu, and X.~Wang, ``A novel joint radar and communication
  system based on randomized partition of antenna array,'' in \emph{2018 IEEE
  International Conference on Acoustics, Speech and Signal Processing
  (ICASSP)}.\hskip 1em plus 0.5em minus 0.4em\relax IEEE, 2018, pp. 3335--3339.

\bibitem{wang2018sparse}
X.~Wang, A.~Hassanien, and M.~G. Amin, ``Sparse transmit array design for
  dual-function radar communications by antenna selection,'' \emph{Digital
  Signal Processing}, vol.~83, pp. 223--234, 2018.

\bibitem{wang2018dual}
X.~Wang, A.~Hassanien, and M.~G. Amin, ``Dual-function mimo radar
  communications system design via sparse array optimization,'' \emph{IEEE
  Transactions on Aerospace and Electronic Systems}, vol.~55, no.~3, pp.
  1213--1226, 2018.

\bibitem{wang2012generalised}
J.~Wang, S.~Jia, and J.~Song, ``Generalised spatial modulation system with
  multiple active transmit antennas and low complexity detection scheme,''
  \emph{IEEE Transactions on Wireless Communications}, vol.~11, no.~4, pp.
  1605--1615, 2012.

\bibitem{F.2019Liu}
F.~Liu and C.~Masouros, ``Hybrid beamforming with sub-arrayed mimo radar:
  Enabling joint sensing and communication at mmwave band,'' in \emph{ICASSP
  2019-2019 IEEE International Conference on Acoustics, Speech and Signal
  Processing (ICASSP)}.\hskip 1em plus 0.5em minus 0.4em\relax IEEE, 2019, pp.
  7770--7774.

\bibitem{X.2020Yuan}
X.~Yuan, Z.~Feng, J.~A. Zhang, W.~Ni, R.~P. Liu, Z.~Wei, and C.~Xu, ``Waveform
  optimization for mimo joint communication and radio sensing systems with
  training overhead,'' \emph{arXiv preprint arXiv:2002.00338}, 2020.

\bibitem{F.2018Liu}
F.~Liu, C.~Masouros, A.~Li, H.~Sun, and L.~Hanzo, ``Mu-mimo communications with
  mimo radar: From co-existence to joint transmission,'' \emph{IEEE
  Transactions on Wireless Communications}, vol.~17, no.~4, pp. 2755--2770,
  2018.

\bibitem{ChengchengMar2020Xu}
C.~Xu, B.~Clerckx, S.~Chen, Y.~Mao, and J.~Zhang, ``Rate-splitting multiple
  access for multi-antenna joint communication and radar transmissions,''
  \emph{arXiv preprint arXiv:2002.00407}, 2020.

\bibitem{A.2009Absil}
P.-A. Absil, R.~Mahony, and R.~Sepulchre, \emph{Optimization Algorithms on
  Matrix Manifolds}.\hskip 1em plus 0.5em minus 0.4em\relax Princeton
  University Press, 2009.

\bibitem{P.2006Petersen}
P.~Petersen, S.~Axler, and K.~Ribet, \emph{Riemannian Geometry}.\hskip 1em plus
  0.5em minus 0.4em\relax Springer, 2006, vol. 171.

\bibitem{S.2011Boyd}
S.~Boyd and J.~Mattingley, ``Branch and bound methods,'' \emph{Notes for
  EE364b, Stanford University}, pp. 2006--07, 2007.

\bibitem{orn1995Ottersten}
B.~orn Ottersten, ``Spatial division multiple access (sdma) in wireless
  communications,'' in \emph{Proceedings Nordic Radio Symposium},
  vol.~95.\hskip 1em plus 0.5em minus 0.4em\relax Citeseer, 1995.

\bibitem{A.2018Hassanien}
A.~Hassanien, B.~Himed, and M.~G. Amin, ``Transmit/receive beamforming design
  for joint radar and communication systems,'' in \emph{2018 IEEE Radar
  Conference (RadarConf18)}.\hskip 1em plus 0.5em minus 0.4em\relax IEEE, 2018,
  pp. 1481--1486.

\bibitem{P.2018Kumari}
P.~Kumari, M.~E. Eltayeb, and R.~W. Heath, ``Sparsity-aware adaptive
  beamforming design for ieee 802.11 ad-based joint communication-radar,'' in
  \emph{2018 IEEE Radar Conference (RadarConf18)}.\hskip 1em plus 0.5em minus
  0.4em\relax IEEE, 2018, pp. 0923--0928.

\bibitem{Liuarxiv2020F}
F.~Liu, W.~Yuan, C.~Masouros, and J.~Yuan, ``Radar-assisted predictive
  beamforming for vehicular links: Communication served by sensing,''
  \emph{arXiv preprint arXiv:2001.09306}, 2020.

\bibitem{pcrb}
Y.~Shen and M.~Z. Win, ``Fundamental limits of wideband localizationÑpart i: A
  general framework,'' \emph{IEEE Transactions on Information Theory}, vol.~56,
  no.~10, pp. 4956--4980, 2010.

\bibitem{A.2018HassanienSPAWC}
A.~Hassanien, C.~Sahin, J.~Metcalf, and B.~Himed, ``Uplink signaling and
  receive beamforming for dual-function radar communications,'' in \emph{2018
  IEEE 19th International Workshop on Signal Processing Advances in Wireless
  Communications (SPAWC)}.\hskip 1em plus 0.5em minus 0.4em\relax IEEE, 2018,
  pp. 1--5.

\bibitem{Wolfel2005M}
M.~Wolfel and J.~McDonough, ``Minimum variance distortionless response spectral
  estimation,'' \emph{IEEE Signal Processing Magazine}, vol.~22, no.~5, pp.
  117--126, 2005.

\bibitem{A.2019Ahmed}
A.~Ahmed, Y.~D. Zhang, and B.~Himed, ``Distributed dual-function
  radar-communication mimo system with optimized resource allocation,'' in
  \emph{Proc. IEEE Radar Conference}, Boston, MA, Apr. 2019.

\bibitem{ZhouY2018}
Y.~Zhou, H.~Zhou, F.~Zhou, Y.~Wu, and V.~C. Leung, ``Resource allocation for a
  wireless powered integrated radar and communication system,'' \emph{IEEE
  Wireless Communications Letters}, vol.~8, no.~1, pp. 253--256, 2018.

\bibitem{X.SecondQuarter2015Lu}
X.~Lu, P.~Wang, D.~Niyato, D.~I. Kim, and Z.~Han, ``Wireless networks with rf
  energy harvesting: A contemporary survey,'' \emph{IEEE Communications Surveys
  \& Tutorials}, vol.~17, no.~2, pp. 757--789, 2014.

\bibitem{C2017Shi}
C.~Shi, F.~Wang, S.~Salous, and J.~Zhou, ``Optimal power allocation strategy in
  a joint bistatic radar and communication system based on low probability of
  intercept,'' \emph{Sensors}, vol.~17, no.~12, p. 2731, 2017.

\bibitem{A.G.2004Stove}
A.~Stove, A.~Hume, and C.~Baker, ``Low probability of intercept radar
  strategies,'' \emph{IEE Proceedings-Radar, Sonar and Navigation}, vol. 151,
  no.~5, pp. 249--260, 2004.

\bibitem{B.2017Li}
B.~Li and A.~P. Petropulu, ``Joint transmit designs for coexistence of mimo
  wireless communications and sparse sensing radars in clutter,'' \emph{IEEE
  Transactions on Aerospace and Electronic Systems}, vol.~53, no.~6, pp.
  2846--2864, 2017.

\bibitem{J.2018LiuGlobalSIP}
J.~Liu and M.~Saquib, ``Transmission design for a joint mimo radar and mu-mimo
  downlink communication system,'' in \emph{2018 IEEE Global Conference on
  Signal and Information Processing (GlobalSIP)}.\hskip 1em plus 0.5em minus
  0.4em\relax IEEE, 2018, pp. 196--200.

\bibitem{F.WCLLiu2017}
F.~Liu, C.~Masouros, A.~Li, and T.~Ratnarajah, ``Robust mimo beamforming for
  cellular and radar coexistence,'' \emph{IEEE Wireless Communications
  Letters}, vol.~6, no.~3, pp. 374--377, 2017.

\bibitem{F.Liu2017}
F.~Liu, C.~Masouros, A.~Li, T.~Ratnarajah, and J.~Zhou, ``Interference
  exploitation for radar and cellular coexistence-the power-efficient
  approach,'' \emph{IET}, Jun. 2017.

\bibitem{D.2006P}
D.~P. Palomar and M.~Chiang, ``A tutorial on decomposition methods for network
  utility maximization,'' \emph{IEEE Journal on Selected Areas in
  Communications}, vol.~24, no.~8, pp. 1439--1451, 2006.

\bibitem{S1999Wright}
J.~Nocedal and S.~Wright, \emph{Numerical Optimization}.\hskip 1em plus 0.5em
  minus 0.4em\relax Springer Science \& Business Media, 2006.

\bibitem{S.2004Boyd}
S.~Boyd, S.~P. Boyd, and L.~Vandenberghe, \emph{Convex Optimization}.\hskip 1em
  plus 0.5em minus 0.4em\relax Cambridge University Press, 2004.

\bibitem{Chen2004S}
S.~Chen and C.~Zhu, ``Ici and isi analysis and mitigation for ofdm systems with
  insufficient cyclic prefix in time-varying channels,'' \emph{IEEE
  Transactions on Consumer Electronics}, vol.~50, no.~1, pp. 78--83, 2004.

\bibitem{S.2011Sen}
S.~Sen and A.~Nehorai, ``Adaptive ofdm radar for target detection in multipath
  scenarios,'' \emph{IEEE Transactions on Signal Processing}, vol.~59, no.~1,
  pp. 78--90, 2010.

\bibitem{K.2008Fazel}
K.~Fazel and S.~Kaiser, \emph{Multi-carrier and Spread Spectrum Systems: From
  OFDM and MC-CDMA to LTE and WiMAX}.\hskip 1em plus 0.5em minus 0.4em\relax
  John Wiley \& Sons, 2008.

\bibitem{M.2016Bica}
M.~Bica, K.-W. Huang, V.~Koivunen, and U.~Mitra, ``Mutual information based
  radar waveform design for joint radar and cellular communication systems,''
  in \emph{IEEE ICASSP}, 2016, pp. 3671--3675.

\bibitem{T.2019Tian}
T.~Tian, T.~Zhang, G.~Li, and T.~Zhou, ``Mutual information-based power
  allocation and co-design for multicarrier radar and communication systems in
  coexistence,'' \emph{IEEE Access}, vol.~7, pp. 159\,300--159\,312, 2019.

\bibitem{F2019WangTSP}
F.~Wang, H.~Li, and M.~A. Govoni, ``Power allocation and co-design of
  multicarrier communication and radar systems for spectral coexistence,''
  \emph{IEEE Transactions on Signal Processing}, vol.~67, no.~14, pp.
  3818--3831, 2019.

\bibitem{Fukushima1992M}
M.~Fukushima, ``Application of the alternating direction method of multipliers
  to separable convex programming problems,'' \emph{Computational Optimization
  and Applications}, vol.~1, no.~1, pp. 93--111, 1992.

\bibitem{P-SCP}
M.~Fardad and M.~R. Jovanovi{\'c}, ``On the design of optimal structured and
  sparse feedback gains via sequential convex programming,'' in \emph{2014
  American Control Conference}, 2014, pp. 2426--2431.

\bibitem{Bica2019M}
M.~Bic{\u{a}} and V.~Koivunen, ``Multicarrier radar-communications waveform
  design for rf convergence and coexistence,'' in \emph{IEEE ICASSP}, 2019, pp.
  7780--7784.

\bibitem{Ahmed2019}
A.~Ahmed, Y.~D. Zhang, A.~Hassanien, and B.~Himed, ``Ofdm-based joint
  radar-communication system: Optimal sub-carrier allocation and power
  distribution by exploiting mutual information,'' in \emph{Asilomar Conference
  on Signals, Systems, and Computers}, 2019.

\bibitem{C.2019Shi}
C.~Shi, F.~Wang, S.~Salous, and J.~Zhou, ``Joint subcarrier assignment and
  power allocation strategy for integrated radar and communications system
  based on power minimization,'' \emph{IEEE Sensors Journal}, to appear.

\bibitem{M.May2019Bica}
M.~Bic{\u{a}} and V.~Koivunen, ``Multicarrier radar-communications waveform
  design for rf convergence and coexistence,'' in \emph{IEEE ICASSP}, 2019, pp.
  7780--7784.

\bibitem{C.May2019Kai}
C.~Kai, H.~Li, L.~Xu, Y.~Li, and T.~Jiang, ``Joint subcarrier assignment with
  power allocation for sum rate maximization of d2d communications in wireless
  cellular networks,'' \emph{IEEE Transactions on Vehicular Technology},
  vol.~68, no.~5, pp. 4748--4759, 2019.

\bibitem{B.2016LiTSP}
B.~Li, A.~P. Petropulu, and W.~Trappe, ``Optimum co-design for spectrum sharing
  between matrix completion based mimo radars and a mimo communication
  system,'' \emph{IEEE Transactions on Signal Processing}, vol.~64, no.~17, pp.
  4562--4575, 2016.

\bibitem{B.ICASSPLi2015}
B.~Li and A.~Petropulu, ``Spectrum sharing between matrix completion based mimo
  radars and a mimo communication system,'' in \emph{IEEE ICASSP}, 2015, pp.
  2444--2448.

\bibitem{L.2017ZhengJSSP}
L.~Zheng, M.~Lops, and X.~Wang, ``Adaptive interference removal for
  uncoordinated radar/communication coexistence,'' \emph{IEEE Journal of
  Selected Topics in Signal Processing}, vol.~12, no.~1, pp. 45--60, 2017.

\bibitem{Y.2019LiarXiv}
Y.~Li, L.~Zheng, M.~Lops, X.~Wang \emph{et~al.}, ``Interference removal for
  radar/communication co-existence: the random scattering case,'' \emph{arXiv
  preprint arXiv:1902.03436}, 2019.

\bibitem{B2019Jain}
H.~B. Jain, I.~P. Roberts, and S.~Vishwanath, ``Enabling in-band coexistence of
  millimeter-wave communication and radar,'' \emph{arXiv preprint
  arXiv:1911.11283}, 2019.

\bibitem{S.2003Burer}
S.~Burer and R.~D. Monteiro, ``A nonlinear programming algorithm for solving
  semidefinite programs via low-rank factorization,'' \emph{Mathematical
  Programming}, vol.~95, no.~2, pp. 329--357, 2003.

\bibitem{L.2011SitEuropean}
A.~Munari, N.~Grosheva, L.~Simi{\'c}, and P.~M{\"a}h{\"o}nen, ``Performance of
  radar and communication networks coexisting in shared spectrum bands,''
  \emph{arXiv preprint arXiv:1902.01359}, 2019.

\bibitem{L.2011SitRADAR}
Y.~L. Sit, L.~Reichardt, C.~Sturm, and T.~Zwick, ``Extension of the ofdm joint
  radar-communication system for a multipath, multiuser scenario,'' in
  \emph{IEEE RadarCon (RADAR)}, 2011, pp. 718--723.

\bibitem{Y.2018Zeng}
Y.~Zeng, Y.~Ma, and S.~Sun, ``Joint radar-communication: Low complexity
  algorithm and self-interference cancellation,'' in \emph{2018 IEEE Global
  Communications Conference (GLOBECOM)}.\hskip 1em plus 0.5em minus 0.4em\relax
  IEEE, 2018, pp. 1--7.

\bibitem{L.2012SitMicrowave}
Y.~L. Sit, L.~Reichardt, C.~Sturm, and T.~Zwick, ``Extension of the ofdm joint
  radar-communication system for a multipath, multiuser scenario,'' in
  \emph{2011 IEEE RadarCon (RADAR)}.\hskip 1em plus 0.5em minus 0.4em\relax
  IEEE, 2011, pp. 718--723.

\bibitem{A.Deligiannis2018}
A.~Deligiannis, A.~Daniyan, S.~Lambotharan, and J.~A. Chambers, ``Secrecy rate
  optimizations for mimo communication radar,'' \emph{IEEE Transactions on
  Aerospace and Electronic Systems}, vol.~54, no.~5, pp. 2481--2492, 2018.

\bibitem{N.2019Su}
N.~Su, F.~Liu, and C.~Masouros, ``Enhancing the physical layer security of
  dual-functional radar communication systems,'' \emph{arXiv preprint
  arXiv:1906.09284}, 2019.

\bibitem{S.2008Goel}
S.~Goel and R.~Negi, ``Guaranteeing secrecy using artificial noise,''
  \emph{IEEE transactions on Wireless Communications}, vol.~7, no.~6, pp.
  2180--2189, Jun. 2008.

\bibitem{K2018Chalise}
B.~K. Chalise and M.~G. Amin, ``Performance tradeoff in a unified system of
  communications and passive radar: A secrecy capacity approach,''
  \emph{Digital Signal Processing}, vol.~82, pp. 282--293, 2018.

\bibitem{Q.2010Luo}
Z.-Q. Luo, W.-K. Ma, A.~M.-C. So, Y.~Ye, and S.~Zhang, ``Semidefinite
  relaxation of quadratic optimization problems,'' \emph{IEEE Signal Processing
  Magazine}, vol.~27, no.~3, pp. 20--34, 2010.

\bibitem{A.2018Garnaev}
A.~Garnaev, W.~Trappe, and A.~Petropulu, ``Optimal design of a dual-purpose
  communication-radar system in the presence of a jammer,'' in
  \emph{International Workshop on Signal Processing Advances in Wireless
  Communications (SPAWC)}, 2018, pp. 1--5.

\bibitem{salberg2005doppler}
A.-B. Salberg and A.~Swami, ``Doppler and frequency-offset synchronization in
  wideband ofdm,'' \emph{IEEE Transactions on Wireless Communications}, vol.~4,
  no.~6, pp. 2870--2881, 2005.

\bibitem{nguyen2012time}
C.~L. Nguyen, A.~Mokraoui, P.~Duhamel, and N.~Linh-Trung, ``Time
  synchronization algorithm in ieee 802.11 a communication system,'' in
  \emph{2012 Proceedings of the 20th European Signal Processing Conference
  (EUSIPCO)}.\hskip 1em plus 0.5em minus 0.4em\relax IEEE, 2012, pp.
  1628--1632.

\bibitem{ElSawy2016H}
H.~ElSawy, A.~Sultan-Salem, M.-S. Alouini, and M.~Z. Win, ``Modeling and
  analysis of cellular networks using stochastic geometry: A tutorial,''
  \emph{IEEE Communications Surveys \& Tutorials}, vol.~19, no.~1, pp.
  167--203, 2016.

\bibitem{wu2019intelligent}
Q.~Wu and R.~Zhang, ``Intelligent reflecting surface enhanced wireless network
  via joint active and passive beamforming,'' \emph{IEEE Transactions on
  Wireless Communications}, vol.~18, no.~11, pp. 5394--5409, Nov. 2019.

\bibitem{shi2016edge}
W.~Shi, J.~Cao, Q.~Zhang, Y.~Li, and L.~Xu, ``Edge computing: Vision and
  challenges,'' \emph{IEEE internet of Things Journal}, vol.~3, no.~5, pp.
  637--646, Oct. 2016.

\end{thebibliography}


\begin{thebibliography}{}
\providecommand{\url}[1]{#1}
\csname url@samestyle\endcsname
\providecommand{\newblock}{\relax}
\providecommand{\bibinfo}[2]{#2}
\providecommand{\BIBentrySTDinterwordspacing}{\spaceskip=0pt\relax}
\providecommand{\BIBentryALTinterwordstretchfactor}{4}
\providecommand{\BIBentryALTinterwordspacing}{\spaceskip=\fontdimen2\font plus
\BIBentryALTinterwordstretchfactor\fontdimen3\font minus
  \fontdimen4\font\relax}
\providecommand{\BIBforeignlanguage}[2]{{%
\expandafter\ifx\csname l@#1\endcsname\relax
\typeout{** WARNING: IEEEtran.bst: No hyphenation pattern has been}%
\typeout{** loaded for the language `#1'. Using the pattern for}%
\typeout{** the default language instead.}%
\else
\language=\csname l@#1\endcsname
\fi
#2}}
\providecommand{\BIBdecl}{\relax}
\BIBdecl

\end{thebibliography}

\end{document}